\documentclass[preprint,aps,amssymb,showpacs,superscriptaddress,nofootinbib]{revtex4}
\usepackage{amsmath}
\usepackage{accents}
\usepackage{color}
\usepackage{graphicx}
\usepackage{bbold}
\definecolor{DarkRed}{rgb}{0.545098,0.000000,0.000000}
\definecolor{blue1}{rgb}{0.000000,0.000000,1.000000}
\definecolor{red1}{rgb}{1.000000,0.000000,0.000000}
\definecolor{indianred}{rgb}{0.803922,0.360784,0.360784}
\newcommand{\Case}[2]{{\textstyle \frac{#1}{#2}}}

\newcommand{\U}[1]{\underline{#1}}

\begin{document}
\preprint{\today \hspace{11cm} IMSc/2015/10/07}

\title{Gravitational Waves from Compact Sources in de Sitter Background} 

\author{Ghanashyam Date}
\email{shyam@imsc.res.in}
\affiliation{The Institute of Mathematical Sciences\\
CIT Campus, Chennai-600 113, INDIA.}

\author{Sk Jahanur Hoque}
\email{jahanur@imsc.res.in}
\affiliation{The Institute of Mathematical Sciences\\
CIT Campus, Chennai-600 113, INDIA.}

\begin{abstract} 
The concordance model of cosmology favours a universe with a tiny
positive cosmological constant. A tiniest positive constant curvature,
profoundly alters the asymptotic structure, forcing a re-look at a
theory of gravitational radiation. Even for compact astrophysical
sources, the intuition from Minkowski background is challenged at every
step. Nevertheless, at least for candidate sources such as compact
binaries, it is possible to quantify influence of the cosmological
constant, as small corrections to the leading order Minkowski background
results.  Employing suitably chosen Fermi normal coordinates in the
static patch of the de Sitter background, we compute the field due to a
compact source to first order in $\Lambda$.  For contrast, we also
present the field in the Poincare patch where the leading correction is
of order $\sqrt{\Lambda}$. {We introduce a gauge invariant quantity, {\em
deviation scalar}, containing polarization information and compute it in
both charts for a comparison.}
\end{abstract}


\maketitle

\section{Introduction}

Asymptotically flat space-times as a model for space-times with
compactly supported sources, {are fashioned after the choice of the
Minkowski space-time as the background space-time. This choice
constitutes a special case of maximally symmetric background
space-times}.  We could also have a cosmological constant, $\Lambda$, in
the Einstein equation and take the {\em de Sitter} ( $\Lambda > 0$) or
the anti-de Sitter ($\Lambda < 0$) solutions as background. The
conformal completion a la Penrose, immediately reveals the {\em
qualitatively} different structure of the infinity. In particular, {\em
irrespective of the non-zero value} of the cosmological constant, the
null infinity - the set consisting of the beginnings and the ends of all
in-extendible null curves - is {\em space-like} for de Sitter and {\em
time-like} for Anti de Sitter \cite{Penrose, HawkingEllis}. This has
drastic effect on the kinds of fluxes that can be used as measures of
radiation at infinity in these backgrounds. The asymptotic symmetry
groups are different too \cite{Strominger, AshtekarBongaKesavanI}. It is
important to note that {\em these {qualitative}} differences are
independent of the numerical value (in any suitably chosen units) of the
cosmological constant. {The quantitative estimates of} the
deviations from Minkowski background {\em are} sensitive to the
numerical value.  This raises the question that if we choose a
background space-time with a non-zero cosmological constant, how does
the linearized theory work out? In particular, what are the
modifications to the ``quadrupole formula(e)''?  Can the modifications
be obtained as `small' corrections in powers of the cosmological
constant?

At this stage, it is worth noting the different facets of the
gravitational fields far away from dynamical sources such as
astrophysical bodies. The most basic question is: what is the field due
to a source at large separations? The very characterization of compact
sources, presumes a source free region where vacuum equations, possibly
including the cosmological term, hold. Thus at large separations we have
a natural split of the field into a background and a small deviation
caused by the source. The simplest approach is then to linearize the
Einstein equation about a background and study its solutions, keeping in
mind the inherent non-linear nature of the theory and hoping for
reliable estimates. The linearized equation is a wave equation with a
finite propagation speed. Among these linear waves, are also the fields
due to sources which are computed from the retarded Green function. The
Green functions of course depend on the choice of `gauge conditions' on
the linear fields and their explicit form depends on the choice of
coordinate chart on the background space-time. 

The next level of physical questions relate to physicality of the wave
solutions. The general covariance of the theory manifests as a {\em
gauge equivalence} at the linearized level and this complicates
identification of physical (gauge invariant) attributes of the wave
solutions. In the Minkowski background, the {linearised} Riemann
tensor is gauge invariant and consequently the induced geodesic
deviation or tidal distortion is a physical effect of the waves. In the
de Sitter background, the {linearised} Riemann tensor itself is
{\em not} gauge invariant but thanks to its conformal flatness, certain
{\em deviation scalar} can be constructed which is gauge invariant.
{It too is related to tidal distortions and contains information
about physical attributes of the waves.} To the extent that there exists
fully gauge fixed solutions with a non-zero tidal distortion, the
gravitational waves are `real'. All the interferometric detectors
measure these distortions in some form or other.  Since the waves are
capable of doing work, we could ask for a measure of the {\em energy}
carried by the waves. 

The natural strategy for defining a measure of energy through a stress
tensor, does not work for gravity.  There is simply no gauge invariant,
tensorial definition of a gravitational stress tensor. There are two
approaches taken for a measure of the {\em flux of gravitational
energy}. One is based on an {\em effective gravitational stress tensor}
tailored for the context wherein there are two widely separated scales,
$\lambda \ll L$, of spatio-temporal variations of the metric which are
used to identify the $L-$scale component of the metric as a {\em
background metric} and $\lambda-$scale component as a small {\em ripple}
\cite{Isaacson}. The other approach directly defines the  {\em flux of
gravitational radiation} in reference to the {\em null infinity} using
the the canonical structure of the space of asymptotically flat/de
Sitter solutions of the Einstein equation. This is applicable for all
spatially compact sources \cite{AshtekarBongaKesavanII}. 

A spatially compact source has two natural scales - its physical size
$R$ and the scale of its time variation $T$. For $R$ sufficiently small
compared to the distance to the source, ${d}$, it is essentially
the scale $T$ that is relevant for gravitational radiation and we may
take the corresponding equivalent length scale as, {$\lambda \sim
T$. ($c = 1$ units)} On the other hand, the curvature scale of the
ambient geometry sufficiently far away from the source, provides the
scale $L$.  For Minkowski space-time background, $L = \infty$ whereas
for non-zero cosmological constant, $L \sim |\Lambda|^{-1/2}$.  A
sufficiently rapidly varying source is one which has its time scale of
variation or equivalent spatial scale $\lambda \ll L${ while a
source is distant if $\lambda/d \ll 1$.}
%
%

Our focus in this work is on sufficiently rapidly varying, distant,
spatially compact sources. For current interferometric detectors, the
scale $\lambda \sim 10^{4} - 10^{5}$ meters, the distances ${d}$,
are in the range of kilo to hundreds of mega parsecs {($\sim
10^{19} - 10^{24}$ meters)} while the spatial extents, $R$, vary over
light seconds or less {($\lesssim 10^8$ meters)}.  We would like to
note that induced tidal distortions are needed in the direct detection
of gravitational waves, regardless of a measure of the energy carried,
while for indirect detection based on energy loss due to gravitational
radiation, reliable flux measures are crucial. In this work, we focus on
the gravitational field and the induced tidal distortion. Computation of
flux(es) will be presented in a separate publication. The quadrupole
flux based on the canonical approach is already available in
\cite{AshtekarBongaKesavanIII}.

In obtaining the field due to a compact source, we follow the basic
steps which are well known and well understood for Minkowski background:
(a) set up the linearized equations, (b) choose a suitable gauge and
obtain a retarded Green function, (c) identify the physical solutions
for subsequent computation of geodesic deviation and power radiated and
(d) relate the physical field to appropriate source multipole moments.
At each of these steps, we encounter new features compared to the
computations in the Minkowski background. 

\begin{figure}[htb]
\includegraphics[scale=0.5]{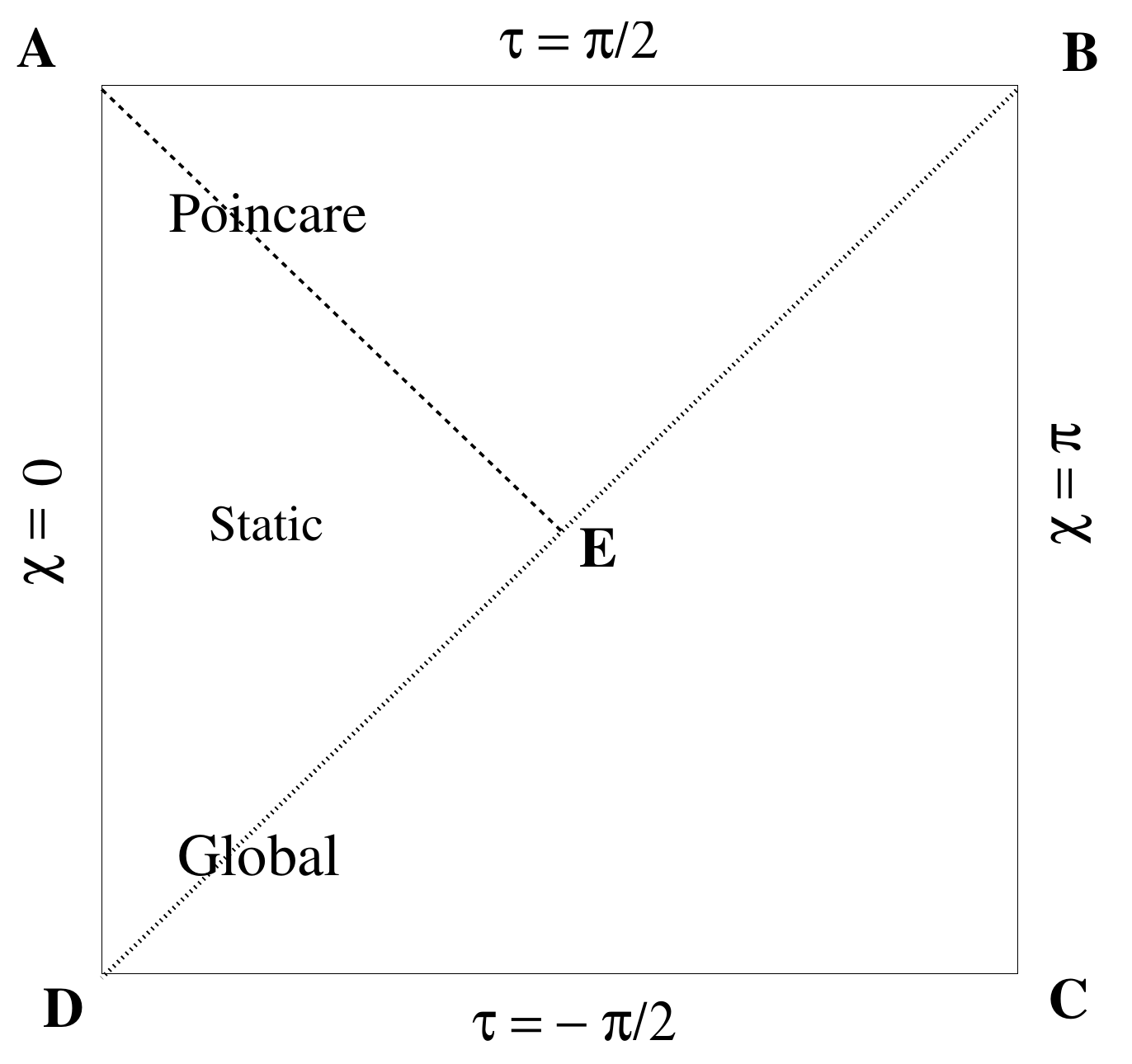} 
\caption{ABCD denotes the Global Chart, ABD is a Poincare patch while
AED is a static patch.  The angular coordinates, $\theta, \phi$ are
suppressed.  The metric in global chart is given by: \centerline{$ ds^2
= \Case{3}{\Lambda}\mathrm{sec}^2\tau\left[ - d\tau^2 + d\chi^2 +
\mathrm{sin}^2\chi \left(d\theta^2 + \mathrm{sin}^2\theta d\phi^2\right)
\right]\ . $ }  \label{GlobalChartFig}}
\end{figure}

Unlike the Minkowski space-time which admits a natural, global Cartesian
chart, de Sitter space-time has several charts appropriate for different
situations. The de Sitter space-time defined as the hyperboloid in five
dimensional Minkowski space-time, has a `{\em global chart}' of
coordinates $(\tau, \chi, \theta, \phi)$, as shown in figure
\ref{GlobalChartFig}. There are natural `Poincare patches' which
constitute the causal future (past) of observers and cover `half' of the
global chart.  For instance, an observer represented by the world line
DA, has its causal future $J^+$ spanning the region DBA and is one of
the {\em Poincare patches}. Being appropriate for the cosmological
context we focus on this Poincare patch. Its boundary denoted by the
line AB, is the {\em future null infinity}, ${\cal J}^+$. There are two
natural coordinate charts for the Poincare patch eg. a {\em conformal
chart}: $(\eta, x^i)$ and a {\em cosmological chart}: $(t, x^i)$. A
`half' of the Poincare patch admits a time-like Killing vector and is
referred to as a {\em static patch}.  This is a natural patch for an
isolated body or a black hole with a stationary neighbourhood.  We
present computations in two different charts: suitably defined {\em
Fermi Normal Coordinates} (FNC) covering the static patch and a
conformal chart {covering the Poincare patch}, see figure
\ref{GlobalChartFig}.

{

While physical implications should not depend on choice of charts, their
explicit computations do depend on the chosen chart. For convenience as
well as for building up intuition, different charts could have different
advantages. For instance, the time coordinate of the FNC chart is the
Killing parameter of the stationary Killing vector. This reduces the
Lie derivative with respect to the Killing vector, to a simple coordinate
derivative. The metric too is obtained as a Taylor series in the
curvature and hence effects due to the cosmological constant can
naturally be expected to appear as a power series in $\Lambda$. However,
this advantage is not available outside the static patch. By contrast,
in the conformal chart the metric is conformal to the Minkowski metric
which considerably simplify the computations. It is possible to scale
out $\Lambda$ by a suitable choice of variables. However, to get
corrections in terms of $\Lambda$ one needs to go to the cosmological
chart. A priori, it is not clear which chart(s) are convenient for what
aspect and we present computations for two choices of charts - the FNC
and the conformal chart. 

}

After obtaining the linearized equation, the next step is to choose `a
gauge'. The natural choice (also used in the Minkowski background) is
the {\em transverse, traceless (TT) gauge}. But there has been another
gauge choice \cite{deVega}, which in the conformal chart simplifies the
linearized equations as well as subsequent analysis due to its
similarity with the Minkowski space-time. This is a gauge which
{imposes a variant of the transversality condition.}
We present the solutions in both gauges. The wave propagation has a {\em
tail} term in both gauges. The TT gauge computations are performed in a
FNC system and are restricted to order $\Lambda$. The tail term is of
order $\Lambda^2$. {In the second gauge}, in a large separation
regime, the tail integral can be computed explicitly. 

Next, to identify the physical fields, one chooses the so-called {\em
synchronous gauge} which sets all fields with {at least} one
temporal index to zero. This steps needs a generalization when the
background has a curvature and needs a suitable time-like vector field.
Fortunately such a generalization is available \cite{AA-Hunt} in a
neighbourhood of a Cauchy surface. 

In a curved space-time, the notion of {\em source multipole moments}
needs to be defined appropriately. In the Minkowski background, the
coordinates of the global chart are {\em vectors} under spatial
rotations on a constant $t$ hypersurface. In a curved background, the
local coordinates have no such property.  A suitable definition can be
constructed by setting-up Fermi normal coordinates.  We show that in the
FNC chart with TT gauge, the physical fields and the source moments can
be obtained as the Minkowski background results with corrections in
powers of $\Lambda\times (\mbox{distant to the source})^2$ and present
the first correction. The computations are useful and reliable at best
up to a distance of about $\Lambda^{-1/2}$ and certainly not up to the
null infinity, ${\cal J}^+$. The FNC chart is contained within a static
patch. For the subset of compact sources we limit ourselves to, this is
adequate.  Unlike the Minkowski background, the correction terms contain
additional types of moments as well as lower order time derivatives of
the moments.

The paper is organized as follows. In the section \ref{Linearization},
we recall the linearization procedure together with the associated
notion of gauge freedom. We collect the expression for the Ricci tensor
up to the quadratic order and give the linearized wave equation for the
metric perturbations. We discuss the gauge choices and residual gauge
invariance.  The section \ref{RNC} is divided in two sub-sections. In
the first subsection we choose the usual transverse, traceless gauge. We
present the Hadamard form of the retarded Green function and simplify
the expression for field due to a localized source, using the Fermi
Normal Coordinates (FNC). The leading contribution of the order
$\Lambda^0$ to the quadrupole field is the same as that in the Minkowski
background and we present the order $\Lambda$ contributions.  Here
appropriate source moments are defined and the solution in a synchronous
gauge is presented. For contrast, in the second sub-section we summarize
the computation of the quadrupole field in an alternative gauge
\cite{deVega}.  The solution in synchronous gauge is presented in terms
of analogously defined source moments. Here, using the cosmological
chart, the corrections appear in powers of $\sqrt{\Lambda}$. In the
section \ref{TidalDistortion}, we present a suitably defined, gauge
invariant deviation scalar and compute it for the suitably projected
fields. In the final section \ref{FinalSec}, we summarize and discuss
our results.  Some of the technical details are given in {the}
three appendices.

\section{Linearization about de Sitter background} \label{Linearization}

As noted above, there are several natural patches and charts available
in the de Sitter space-time.  To introduce perturbations without
referring to coordinates\footnote{Some times a coordinate system is
presumed in which the metric is split into a background plus small
perturbations. This obscures the tensorial nature of the perturbation
and is avoided as discussed, for example, in \cite{Wald}.}, consider a
one parameter family of metrics, $g_{\mu\nu}(\epsilon)$ which is
differentiable with respect to $\epsilon$ at $\epsilon = 0$ and let
$\bar{g}_{\mu\nu} := g_{\mu\nu}(0)$ be a given solution of the exact
Einstein equation. Define {\em a perturbation} of the exact solution as:
$h_{\mu\nu} := \Case{dg_{\mu\nu}(\epsilon)}{d\epsilon}|_{\epsilon = 0}$.
As the one parameter families of metrics are varied, we generate the
space of perturbations from the corresponding $h_{\mu\nu}$. If every
member of the family of metrics solves Einstein equation (with sources
and cosmological constant), then the perturbation satisfies a {\em
linear equation} obtained by differentiating the exact equation with
respect to $\epsilon$ and setting $\epsilon$ to zero. Thus every one
parameter family of exact solutions of the Einstein equation gives a
solution of the linearized equation. The converse is not always true and
is known as the linearization instability problem. In our context, this
is not a concern. The general covariance of the Einstein equation
implies that every one parameter family of metrics, obtained by
diffeomorphisms generated by a vector field {on a solution to
Einstein equation}, also solves the equation and leads to a
corresponding perturbation satisfying the linearized equation. However,
these families give the {\em same physical space-time}. The
corresponding perturbations do {\em not} give physically distinct,
nearby space-times and therefore do not represent {\em physical
perturbations}. These perturbations have the form: $h_{\mu\nu} = {\cal
L}_{\xi}\bar{g}_{\mu\nu}$ where ${\cal L}_{\xi}$ denotes the Lie
derivative. To identify the physical perturbations, we have to `mod out'
these perturbations, generated by diffeomorphisms. In other words,
physical perturbations are equivalence classes of perturbations:
\[
[h_{\mu\nu}] := \left\{ h'_{\mu\nu}/ h'_{\mu\nu} = h_{\mu\nu} + {\cal
L}_{\xi}\bar{g}_{\mu\nu}\ \forall \mbox{~vector fields~} \xi\ \right\}\
.
\] 
More commonly, the expression $h'_{\mu\nu} = h_{\mu\nu} + {\cal
L}_{\xi}\bar{g}_{\mu\nu}$ is referred to as a {\em gauge transformation}
and the equivalence classes are of course the physical
perturbations. Thus, by definition of gauge transformations, the
linearized equation is gauge invariant. While the perturbations are
subjected to these gauge transformations, it should be borne in mind
that they are tensors with respect to general coordinate
transformations.

While the linearization can be specified in a coordinate free manner,
explicit computations of solutions needs coordinates to be introduced.
In practice, one begins by writing $g_{\mu\nu}(\epsilon, x) \approx
\bar{g}_{\mu\nu}(x) + \epsilon h_{\mu\nu}(x)$ and obtains the linearized
equation by substituting this in the full equation and keeping terms to
order $\epsilon$. Since we consider perturbations of the source free de
Sitter solution, the matter stress tensor is of order $\epsilon$ while
the cosmological constant is of order $\epsilon^0$.  Under an
infinitesimal diffeomorphism generated by a vector field $\xi^{\mu}(x),~
x'^{\mu} = x^{\mu} - \epsilon\xi^{\mu}(x)$, the Lie derivative of
the background metric, $\bar{g}_{\mu\nu}(x)$, is given by ${\cal
L}_{\xi}\bar{g}_{\mu\nu} = \bar{\nabla}_{\mu}\xi_{\nu} +
\bar{\nabla}_{\nu}\xi_{\mu}$. Here the $\bar{\nabla}$ denotes the
covariant derivative with the Riemann-Christoffel connection of
$\bar{g}$ and $\xi_{\mu} := \bar{g}_{\mu\nu}\xi^{\nu}$.  The gauge
transformations thus take the form: $h'_{\mu\nu}(x) = h_{\mu\nu}(x) +
\bar{\nabla}_{\mu}\xi_{\nu} + \bar{\nabla}_{\nu}\xi_{\mu} $.

We begin by summarizing the expansions of the connection and Ricci
tensor to $o(h^2)$.

In the following, the indices are raised and lowered using the
background metric which is taken to be a maximally symmetric one.
Background quantities carry an overbar.
\begin{eqnarray}
g^{\mu\nu} & = & \bar{g}^{\mu\nu} - \epsilon h^{\mu\nu} +
\epsilon^2 h^{\mu}_{~\alpha} h^{\alpha\nu} \\
\Gamma^{\lambda}_{~\mu\nu} & = & \bar{\Gamma}^{\lambda}_{~\mu\nu} +
\epsilon\left[\frac{1}{2}\bar{g}^{\lambda\alpha}(\bar{\nabla}_{\nu}h_{\alpha\mu}
+ \bar{\nabla}_{\mu}h_{\alpha\nu} -
\bar{\nabla}_{\alpha}h_{\mu\nu})\right] \nonumber \\
& & \hspace{0.9cm} -
\epsilon^2\left[\frac{1}{2}h^{\lambda\alpha}(\bar{\nabla}_{\nu}h_{\alpha\mu}
+ \bar{\nabla}_{\mu}h_{\alpha\nu} -
\bar{\nabla}_{\alpha}h_{\mu\nu})\right] 
\end{eqnarray}
\begin{eqnarray}
R_{\mu\nu} & = & \bar{R}_{\mu\nu} + \epsilon R^{(1)}_{\mu\nu} +
\epsilon^2 R^{(2)}_{\mu\nu} \nonumber \\
R^{(1)}_{\mu\nu} & = & - \frac{1}{2} \bar{\Box}h_{\mu\nu} -
\frac{1}{2}\bar{\nabla}_{\mu}\bar{\nabla}_{\nu}h +
\frac{1}{2}\left(\bar{\nabla}_{\mu}\bar{\nabla}_{\alpha}
h^{\alpha}_{~\nu} + \bar{\nabla}_{\nu}\bar{\nabla}_{\alpha}
h^{\alpha}_{~\mu} \right) \nonumber \label{OrderOneRicci} \\
& & \hspace{1.7cm} + \frac{1}{2}\left(
\bar{R}_{\mu\alpha}h^{\alpha}_{~\nu} +
\bar{R}_{\nu\alpha}h^{\alpha}_{~\mu} \right) +
\bar{R}_{\mu\alpha\beta\nu}h^{\alpha\beta} ~~,\hspace{1.0cm} h :=
\bar{g}^{\alpha\beta} h_{\alpha\beta} \ ; \\
R^{(2)}_{\mu\nu} & = & \frac{1}{2}h^{\alpha\beta}\left[
\bar{\nabla}_{\nu}\bar{\nabla}_{\mu}h_{\alpha\beta} +
\bar{\nabla}_{\alpha}\bar{\nabla}_{\beta}h_{\mu\nu} -
\bar{\nabla}_{\alpha}\bar{\nabla}_{\mu} h_{\beta\nu} -
\bar{\nabla}_{\alpha}\bar{\nabla}_{\nu}h_{\beta\mu} \right] \nonumber \\
& & - \frac{1}{4}\left\{2\bar{\nabla}_{\alpha}h^{\alpha}_{~\beta} -
\bar{\nabla}_{\beta}h\right\}\left\{ \bar{\nabla}_{\mu}h_{\nu}^{~\beta}
+ \bar{\nabla}_{\nu}h_{\mu}^{~\beta} -
\bar{\nabla}^{\beta}h_{\mu\nu}\right\} \nonumber \\ 
& & + \frac{1}{4}\left( \bar{\nabla}_{\mu}h^{\alpha\beta} +
\bar{\nabla}^{\alpha}h^{\beta}_{~\mu} -
\bar{\nabla}^{\beta}h^{\alpha}_{~\mu} \right) \left(
\bar{\nabla}_{\nu}h_{\alpha\beta} + \bar{\nabla}_{\alpha}h_{\beta\nu} -
\bar{\nabla}_{\beta}h_{\alpha\nu} \right) \label{OrderTwoRicci} 
\end{eqnarray}
\begin{eqnarray}
\bar{g}^{\mu\nu}R^{(1)}_{\mu\nu} & = & - \bar{\Box} h +
\bar{\nabla}_{\mu}\bar{\nabla}_{\nu} h^{\mu\nu} ~ ,
\label{OrderOneRicciScalar} \\
G_{\mu\nu} + \Lambda g_{\mu\nu} & = & [\bar{G}_{\mu\nu} +
\Lambda\bar{g}_{\mu\nu}] + [G^{(1)}_{\mu\nu} + \Lambda h_{\mu\nu}]
\nonumber \\
G^{(1)}_{\mu\nu} + \Lambda h_{\mu\nu} & = & -\frac{1}{2}\bar{\Box}
h_{\mu\nu} -\frac{1}{2}\left(\bar{\nabla}_{\mu}\bar{\nabla}_{\nu} -
\bar{g}_{\mu\nu}\bar{\Box} \right) h + h_{\mu\nu}\left(\Lambda
-\frac{1}{2}\bar{R}\right) \nonumber \\
& &  + \frac{1}{2}\left(\bar{\nabla}_{\mu}\bar{\nabla}_{\alpha}
h^{\alpha}_{~\nu} + \bar{\nabla}_{\nu}\bar{\nabla}_{\alpha}
h^{\alpha}_{~\mu} - \bar{g}_{\mu\nu}\left(
\bar{\nabla}_{\alpha}\bar{\nabla}_{\beta}h^{\alpha\beta}\right) \right)
\nonumber \\
&  & + \bar{R}_{\mu\alpha\beta\nu}h^{\alpha\beta} + \frac{1}{2}\left(
\bar{R}_{\mu\alpha} h^{\alpha}_{~\nu} + \bar{R}_{\nu\alpha}
h^{\alpha}_{~\mu} + \bar{g}_{\mu\nu}
\bar{R}^{\alpha\beta}h_{\alpha\beta} \right) 
\end{eqnarray}

The expressions simplify further for the {\em maximally symmetric}
solution of the background equation, $\bar{G}_{\mu\nu} +
\Lambda\bar{g}_{\mu\nu} = 0$.  Maximal symmetry implies
$\bar{R}_{\mu\alpha\beta\nu} = {K}
(\bar{g}_{\mu\beta}\bar{g}_{\nu\alpha}
-~\bar{g}_{\mu\nu}\bar{g}_{\alpha\beta})$ while the background equation
fixes ${K} = \Lambda/3$ and the linearized equation
becomes\footnote{From now on, the background is taken to be the de
Sitter space-time with $\Lambda > 0$ and the units are chosen so that $G
= 1 = c$.},
\begin{eqnarray}
-\frac{1}{2}\bar{\Box} h_{\mu\nu}
-\frac{1}{2}\left(\bar{\nabla}_{\mu}\bar{\nabla}_{\nu}
-~\bar{g}_{\mu\nu}\bar{\Box} \right) h + \frac{\Lambda}{3} h_{\mu\nu} +
\frac{\Lambda}{6}\bar{g}_{\mu\nu} h & & \nonumber \\
+ \frac{1}{2}\left(\bar{\nabla}_{\mu}\bar{\nabla}_{\alpha}
h^{\alpha}_{~\nu} + \bar{\nabla}_{\nu}\bar{\nabla}_{\alpha}
h^{\alpha}_{~\mu} -
\bar{g}_{\mu\nu}\left(\bar{\nabla}_{\alpha}\bar{\nabla}_{\beta}
h^{\alpha\beta}\right) \right) & = & 8\pi T_{\mu\nu} .
\end{eqnarray}
It is customary and convenient to use the trace-reversed combination:
$\tilde{h}_{\mu\nu} := h_{\mu\nu} - \Case{1}{2}\bar{g}_{\mu\nu} h$.
Denoting, $B_{\mu} := \bar{\nabla}_{\alpha}\tilde{h}^{\alpha}_{~\mu}$,
in terms of the tilde variables, the linearized equation takes the form,
\begin{equation}
\frac{1}{2}\left[ - \bar{\Box} \tilde{h}_{\mu\nu} + \left\{
\bar{\nabla}_{\mu}B_{\nu} + \bar{\nabla}_{\nu}B_{\mu} -
\bar{g}_{\mu\nu}(\bar{\nabla}^{\alpha}B_{\alpha})\right\}\right] +
\frac{\Lambda}{3}\left[\tilde{h}_{\mu\nu} -
\tilde{h}\bar{g}_{\mu\nu}\right] ~ = ~ 8\pi T_{\mu\nu} \label{LinEqn}
\end{equation}
The divergence of the left hand side, $\bar{\nabla}^{\mu}
[LHS]_{\mu\nu}$ is identically zero and thus source tensor is conserved
automatically as it should be. For $\Lambda = 0$ the equation goes over
to the flat background equation. Under the gauge transformations,
$\tilde{h}_{\mu\nu}$ transforms as,
\[
\delta \tilde{h}_{\mu\nu}(x) = \bar{\nabla}_{\mu}\xi_{\nu} +
\bar{\nabla}_{\nu}\xi_{\mu} -
\bar{g}_{\mu\nu}\bar{\nabla}_{\alpha}\xi^{\alpha} ,
\]
and the linearized equation (\ref{LinEqn}) is explicitly invariant under
these as it should be. It is well known that availing this freedom, it
is possible to impose the {\em transversality condition},
$\bar{\nabla}_{\alpha}\tilde{h}^{\alpha}_{~\mu} = 0$.  The trace can be
further gauged away \cite{Higuchi91} in the absence of sources (or for
traceless stress tensor).  The particular choice of arranging
$\bar{\nabla}_{\alpha}\tilde{h}^{\alpha}_{~\mu} = 0 = h$, is the {\em
transverse, traceless gauge} or TT gauge for short. It simplifies the
equation (\ref{LinEqn}) to (for traceless stress tensor),
\begin{equation} \label{TTLinEqn}
-~\frac{1}{2} \bar{\Box} \tilde{h}_{\mu\nu} +
\frac{\Lambda}{3}\tilde{h}_{\mu\nu} ~ = ~ 8\pi T_{\mu\nu}
\end{equation}
The transversality condition still allows {\em residual gauge
transformations} generated by vector fields $\xi^{\mu}$ satisfying,
\begin{equation} \label{ResidualGauge}
\delta (\bar{\nabla}_{\mu}\tilde{h}^{\mu}_{~\nu}) =
\bar{\nabla}_{\mu}(\delta \tilde{h}^{\mu}_{~\nu}) = \bar{\Box}\xi_{\nu}
+ \bar{R}_{\alpha\nu}\xi^{\alpha} = (\bar{\Box} + \Lambda)\xi_{\nu} = 0
\end{equation}
If, in addition, the trace (zero or non-zero) is to be preserved, then
$\xi^{\mu}$ must further satisfy, $\bar{\nabla}_{\alpha}\xi^{\alpha} =
0$ and this is consistent with the above equation. 

While it is common to choose the TT gauge,  it is also possible to make
a different choice of gauge \cite{deVega} in the Poincare patch of the
de Sitter space-time.  This will be done in the subsection
\ref{PoincareChart} below. 

The task now is to obtain the particular solution of the linearized,
inhomogeneous equation (\ref{LinEqn}), and extract the {\em physical
solutions} i.e. solutions satisfying conditions which leaves {\em no}
gauge transformations possible, in the source free region. Within the
perturbative framework, this is obtained at the leading order by using a
suitable Green function for the linearized equation on the de Sitter
background. The {\em retarded Green functions} will be determined after
some gauge fixing simplifying the equation (\ref{LinEqn}). 

\section{The Retarded Green function} \label{RNC}

There have been several computations of two point functions for scalar,
vector and tensor fields on de Sitter background \cite{AllenTuryn,
Higuchi91, TsamisWoodard, deVega}.  We will consider two retarded Green
functions. In the subsection \ref{CovariantGauge}, we impose first the
transversality condition and then also the tracelessness condition.  We
refer to these as the {\em transverse gauge} and the {\em TT gauge}
respectively.  In the subsection \ref{PoincareChart}, following
\cite{deVega}, we choose a gauge which changes the transversality
condition by making its right hand side non-zero. We refer to it as {\em
generalized transverse gauge}. With the tracelessness condition imposed,
we refer to it as {\em generalized-TT} gauge.  The two computations will
provide different views of the physical solutions, in particular the
form of the manifestation of the $\Lambda$ dependence.  The computations
in the transverse gauge, employing the Hadamard construction
\cite{Friedlander}, follow reference \cite{Poisson} while the
generalized transverse gauge computations are based on \cite{deVega}.

\subsection{The Transverse and the TT gauge}
\label{CovariantGauge}
%

It turns out to be convenient to separate the trace part of the equation
and construct the retarded Green function in the transverse, traceless
(TT) gauge directly with a source which is traceless.

Imposing the transversality condition, $B_{\mu} = 0$ in eqn.
(\ref{LinEqn}) gives,
\begin{equation}
\bar{\Box} \tilde{h}_{\mu\nu} -
\frac{2\Lambda}{3}\left[\tilde{h}_{\mu\nu} -
\tilde{h}\bar{g}_{\mu\nu}\right] ~ = ~ -16\pi T_{\mu\nu}
\label{T-LinEqn}
\end{equation}
and taking the trace of the above equation, gives an equation for the
trace, $\tilde{h}$,
\begin{equation}\label{TraceEqn}
(\bar{\Box} + 2\Lambda)\tilde{h} = - 16\pi T ~~,~~~ T :=
\bar{g}_{\mu\nu}T^{\mu\nu} \ .
\end{equation}
Subtracting $\Case{1}{4}\bar{g}_{\mu\nu} \times$ eqn.(\ref{TraceEqn})
from eqn.(\ref{T-LinEqn}), we get
\begin{equation}\label{TraceFreeEqn}
\bar{\Box} \tilde{h}'_{\mu\nu} - \frac{2\Lambda}{3}\tilde{h}'_{\mu\nu} ~
= ~ -16 \pi T'_{\mu\nu} ~~~,~~~ \tilde{h}'_{\mu\nu} := \tilde{h}_{\mu\nu}
- \frac{1}{4}\tilde{h}\bar{g}_{\mu\nu}~,~ T'_{\mu\nu} := T_{\mu\nu} -
\frac{1}{4}T\bar{g}_{\mu\nu}~.
\end{equation}

The equation (\ref{TraceEqn}) for $\tilde{h}$ is a scalar equation and
its solution is determined by a corresponding Green function with a
source which is the trace of the stress tensor. However we know that in
the source free region, we {\em can} make a gauge transformation to set
the $\tilde{h}$ to zero. Hence, in the region of observational interest,
we can gauge away the effect of the trace $T$. With this understood, we
take $\tilde{h} = 0$ which gives $\tilde{h}'_{\mu\nu} =
\tilde{h}_{\mu\nu}$ and use the traceless $T'_{\mu\nu}$ as the source.
For notational simplicity, we drop the prime from the stress tensor.
Thus, we focus on the TT gauge equation (\ref{TTLinEqn}) with a {\em
trace-free stress tensor} as the source.

The equation for the Green function is,
\begin{equation} \label{GreenFnEqn}
\bar{\Box} G^{\alpha\beta}_{~~\mu'\nu'}(x, x') - \frac{2\Lambda}{3}
G^{\alpha\beta}_{~~\mu'\nu'}(x, x') ~ = ~ - 4\pi
J^{\alpha\beta}_{~~\mu'\nu'} \delta_4(x, x') ~~,~~ \mbox{where,}
\end{equation}
{
\begin{equation}
J^{\alpha\beta}_{~~\mu'\nu'}(x, x') ~ := ~
\frac{g^{\alpha}_{~\mu'}g^{\beta}_{~\nu'} +
g^{\alpha}_{~\nu'}g^{\beta}_{~\mu'}}{2} -
\frac{1}{4}\bar{g}^{\alpha\beta}(x)\bar{g}_{\mu'\nu'}(x') ~~,~~
\mbox{and,}
\end{equation}
}
$g^{\alpha}_{~\mu'}(x, x')$ denotes the {\em parallel propagator} along
the geodesic connecting $x, x'$. The tensor
$J^{\alpha\beta}_{~~\mu'\nu'}$ is symmetric and traceless in the pairs
of indices $\alpha\beta$ and $\mu'\nu'$. {The Green's function is
obtained using the {\em Hadamard ansatz}}.

The Hadamard ansatz for the retarded Green function for {{\em a
general}} wave equation is \cite{Friedlander},
\begin{equation}
G^{\alpha\beta}_{~~\mu'\nu'}(x, x') ~ = ~
U^{\alpha\beta}_{~~\mu'\nu'}(x, x')\delta_+(\sigma + \epsilon) +
V^{\alpha\beta}_{~~\mu'\nu'}(x, x')\theta_+(-\sigma -
\epsilon)~,~\mbox{where}
\end{equation}
the space-time points $x, x'$ belong to a {\em convex normal
neighbourhood} with $x$ in the chronological future of $x'$; $\sigma(x,
x')$ is the Synge world function which is half the geodesic distance
squared between $x$ and $x'$ \cite{Synge, Poisson}; $\theta_+, \delta_+$
are distributions, viewed as functions of $x$, having support in the
chronological future and future light cone of $x'$ respectively. The
small parameter $\epsilon$ is introduced to permit differentiation of
the distribution and is to be taken to zero in the end. The bi-tensors
$U, V$ are determined by inserting the ansatz in the equation
(\ref{GreenFnEqn}).

Using the relation $\bar{g}^{\alpha\beta}\bar{\nabla}_{\alpha}\sigma
\bar{\nabla}_{\beta}\sigma = 2 \sigma$ and the distributional identities
\cite{Poisson}:
\begin{eqnarray}
(\sigma + \epsilon)\delta'(\sigma + \epsilon) = - \delta(\sigma +
\epsilon) & , & (\sigma + \epsilon)\delta''(\sigma + \epsilon) = -
2\delta'(\sigma + \epsilon) \nonumber \\
\mbox{as } \epsilon \to 0 : ~\epsilon \delta'(\sigma + \epsilon) \to 0 &
, & \epsilon \delta''(\sigma + \epsilon) \to 2\pi\delta_4(x, x') ~~,  
\end{eqnarray}
leads to four equations by equating the coefficients of
$\theta(-\sigma), \delta(\sigma), \delta'(\sigma)$ and $\delta_4(x, x')$
to zero.  The respective equations are:
\begin{eqnarray}
\bar{\Box}V^{\alpha\beta}_{~~\mu'\nu'}(x, x') - \frac{2\Lambda}{3}
V^{\alpha\beta}_{~~\mu'\nu'}(x,x') \hspace{0.3cm}  = ~~ 0 \hspace{2.4cm}
& ~~,~~ & \sigma(x, x') < 0 ; \label{ThetaEqn} \\
2 \sigma^{\lambda}\bar{\nabla}_{\lambda}V^{\alpha\beta}_{~~\mu'\nu'} +
(~\bar{\Box}\sigma - 2)V^{\alpha\beta}_{~~\mu'\nu'} =
\bar{\Box}U^{\alpha\beta}_{~~\mu'\nu'} -
\frac{2\Lambda}{3}U^{\alpha\beta}_{~~\mu'\nu'}
& ~~,~~ & \sigma(x, x') = 0  ; \label{DeltaEqn}\\
\left(2 \sigma^{\lambda}\bar{\nabla}_{\lambda}  + (\bar{\Box}\sigma -
4)\right) U^{\alpha\beta}_{~~\mu'\nu'} \hspace{0.0cm}  = ~ 0
\hspace{2.5cm} & ~~,~~ & \sigma(x, x') = 0 ; \label{DeltaPrimeEqn} \\
\left[U^{\alpha\beta}_{~~\mu'\nu'}\right] \hspace{0.3cm}  = ~
\left[J^{\alpha\beta}_{~~\mu'\nu'}\right] ~ = ~
\delta^{(\alpha'}_{\mu'}\delta^{\beta')}_{\nu'} -
\frac{1}{4}\bar{g}^{\alpha'\beta'}\bar{g}_{\mu'\nu'} \hspace{1.1cm} &
~~,~~ & x = x' \ . \label{Delta4Eqn}
\end{eqnarray}
In the above, the quantity enclosed within square brackets denotes its
{\em coincidence limit} - evaluation for $ x = x'$ and super(sub)script
on $\sigma$ denotes its covariant derivative.

The last two equations uniquely determine
$U^{\alpha\beta}_{~~\mu'\nu'}(x,x')$ on the light cone through $x'$
while the first two equations uniquely determine
$V^{\alpha\beta}_{~~\mu'\nu'}(x,x')$ inside and on the light cone
through $x'$. The cosmological constant appears explicitly in these two
equations. 

{\em Determination of $U^{\alpha\beta}_{~~\mu'\nu'}$:} Equation
(\ref{DeltaPrimeEqn}) is a homogeneous, first order, linear differential
equation and its solution is completely determined by the initial
condition provided by eqn. (\ref{Delta4Eqn}).  Noting that
$\sigma^{\lambda}\bar{\nabla}_{\lambda}$ on the parallel propagator and
the metric gives zero, we get $\sigma^{\lambda}\bar{\nabla}_{\lambda}
J^{\alpha\beta}_{~~\mu'\nu'} = 0$.

Hence the ansatz $U^{\alpha\beta}_{~~\mu'\nu'}(x, x') ~ := ~
J^{\alpha\beta}_{~~\mu'\nu'}\tilde{U}(x, x')$ in
eqs.(\ref{DeltaPrimeEqn}, \ref{Delta4Eqn}) leads to 
\begin{equation}
\left(2 \sigma^{\alpha}\nabla_{\alpha}  + (\bar{\Box}\sigma - 4)\right)
\tilde{U} ~ = ~ 0 ~~,~~ [\tilde{U}] = 1 ~~~~ \Rightarrow ~~~~
\tilde{U}(x, x') ~ := ~ \sqrt{\Delta(x, x')} \ ,
\end{equation}
where $\Delta(x, x')$ is the {(scalarised)} Van Vleck
determinant or Van Vleck bi-scalar defined as, {$\Delta(x, x') := -
\mathrm{det}(-\sigma_{\alpha\beta'}(x, x'))/\sqrt{g(x)g(x')}$, with $g$
in the denominator denoting the modulus of the determinant of the
metric} \cite{Poisson}. The bi-scalar $\tilde{U}$, being de Sitter
invariant, depends on $x, x'$ only through the world function
$\sigma(x,x')$ which means that value of $\tilde{U}$ along the light
cone is same as its value in the coincidence limit i.e.
$\tilde{U}|_{\sigma = 0} = [\tilde{U}] = 1 (= \Delta(x, x')|_{\sigma =
0})$ and we need the solution only on the light cone.  Thus, 
\[
U^{\alpha\beta}_{~~\mu'\nu'}(x, x')|_{\sigma = 0} ~ := ~
J^{\alpha\beta}_{~~\mu'\nu'}|_{\sigma = 0}.
\]

We {\em cannot} similarly factor out $J^{\alpha\beta}_{~~\mu'\nu'}$ from
$V^{\alpha\beta}_{~~\mu'\nu'}(x, x')$. The reason is that equation
(\ref{DeltaEqn}) is an {\em inhomogeneous} equation and the tensor
structure of its right hand side is {\em not} the same as that of
$U^{\alpha\beta}_{~~\mu'\nu'}$. Indeed, to order $(\sigma)^2$, we find
\cite{Poisson},
\begin{eqnarray}
\left(\bar{\Box} - \frac{2\Lambda}{3}\right)U^{\alpha\beta}_{~~\mu'\nu'}
& = & \left\{ - \frac{\Lambda}{6}\left(4 - \bar{\Box}\sigma\right) -
\frac{\Lambda^2\sigma}{9}\right\} J^{\alpha\beta}_{~~\mu'\nu'} \\
& & + \frac{\Lambda^2}{18}\left\{
\bar{g}^{\alpha\beta}\sigma_{\mu'}\sigma_{\nu'} +
\sigma^{\alpha}\sigma^{\beta}\bar{g}_{\mu'\nu'} -
\sigma(g^{\alpha}_{~\mu'} g^{\beta}_{~\nu'} + g^{\alpha}_{~\nu'}
g^{\beta}_{~\mu'}) \right. \nonumber \\
& & \hspace{1.3cm} +
\left.\left(g^{\alpha}_{~\mu'}\sigma^{\beta}\sigma_{\nu'} +
g^{\beta}_{~\mu'}\sigma^{\alpha}\sigma_{\nu'} +
g^{\alpha}_{~\nu'}\sigma^{\beta}\sigma_{\mu'} +
g^{\beta}_{~\nu'}\sigma^{\alpha}\sigma_{\mu'}\right) \right\} \nonumber
\\
& := & \Phi(\sigma)J^{\alpha\beta}_{~~\mu'\nu'} +
\frac{\Lambda^2}{18}K^{\alpha\beta}_{~~\mu'\nu'} + o(\sigma^3)
\end{eqnarray}
Note that the bi-tensor $K^{\alpha\beta}_{~~\mu'\nu'}$ is traceless and
just as the bi-tensor $J^{\alpha\beta}_{~~\mu'\nu'}$, it too is
annihilated by $\sigma^{\lambda}\bar{\nabla}_{\lambda}$.

Noting the coincidence limits: $[\bar{\Box}\sigma] = 4, [\sigma] = 0,
[\sigma^{\alpha}] = 0$, we see that, $[\Phi] = 0 =
[K^{\alpha\beta}_{~~\mu'\nu'}]$ and hence the coincidence limit of the
left hand side vanishes.

The coincidence limit of the equation (\ref{DeltaEqn}) then implies
$[V^{\alpha\beta}_{~~\mu'\nu'}(x,x')] = 0$. However, this does {\em not}
imply $V^{\alpha\beta}_{~~\mu'\nu'}(x,x')|_{\sigma = 0} = 0$. To order
$\sigma^2$, we can write, 
\[
V^{\alpha\beta}_{~~\mu'\nu'}(x,x') := \tilde{V}_1(\sigma)
J^{\alpha\beta}_{~~\mu'\nu'} +
\tilde{V}_2(\sigma)K^{\alpha\beta}_{~~\mu'\nu'} \ .
\]

This leads to two inhomogeneous differential equations for the
bi-scalars $\tilde{V}_1, \tilde{V}_2$. The coincidence limits of these
equations, combined with $[\Phi] = 0$ leads to $[\tilde{V}_1] = 0$ and
$[\tilde{V}_2] = \frac{\Lambda^2}{108}$. Once again, these values
determine these bi-scalars everywhere on the light cone. Hence, to order
$\sigma^2$,
\begin{equation}
V^{\alpha\beta}_{~~\mu'\nu'}(x, x')|_{\sigma = 0} ~ = ~
\frac{\Lambda^2}{108}K^{\alpha\beta}_{~~\mu'\nu'}|_{\sigma = 0} +
o(\sigma^3)
\end{equation}

This shows clearly that the data for characteristic evolution off the
light cone is non-zero and hence the tail term is non-zero as well.
Equally well, it also shows that the tail term is {\em at least of order
$\Lambda^2$}.  The Green function is then given by,
\[
G^{\alpha\beta}_{~~\mu'\nu'}(x, x') ~ = ~
J^{\alpha\beta}_{~~\mu'\nu'}(x, x') \delta_+(\sigma) +
V^{\alpha\beta}_{~~\mu'\nu'} \theta_+(-\sigma) .
\]

We will be computing corrections to order $\Lambda$ and hence {\em we do
not compute the effect of the tail term in this work}. From now on, we
restrict to the sharp propagation term only and only the {\em
trace-free} part of the source stress tensor contributes.

Using the sharp term of the Green function above, the solution to the
inhomogeneous equation becomes,
\begin{eqnarray}
\tilde{h}^{\alpha\beta}(x) & = & 4 \int_{\mathrm{source}}d^4x'
\sqrt{-g(x')} \delta_+(\sigma) J^{\alpha\beta}_{~~\mu'\nu'}(x, x')
T^{\mu'\nu'}(x') \\
& = & 4 \int_{\mathrm{source}}d^4x'\sqrt{- g(x')}\delta_+(\sigma)
g^{\alpha}_{~\mu'}(x, x')g^{\beta}_{~\nu'}(x, x')T^{\mu'\nu'}(x')
\label{InhomogeneousSoln}
\end{eqnarray}
In the second line we have substituted for
$J^{\alpha\beta}_{~~\mu'\nu'}$ and used the fact that the stress tensor
is trace-free and symmetric. 

\begin{figure}[htb] 
\includegraphics[scale=0.55]{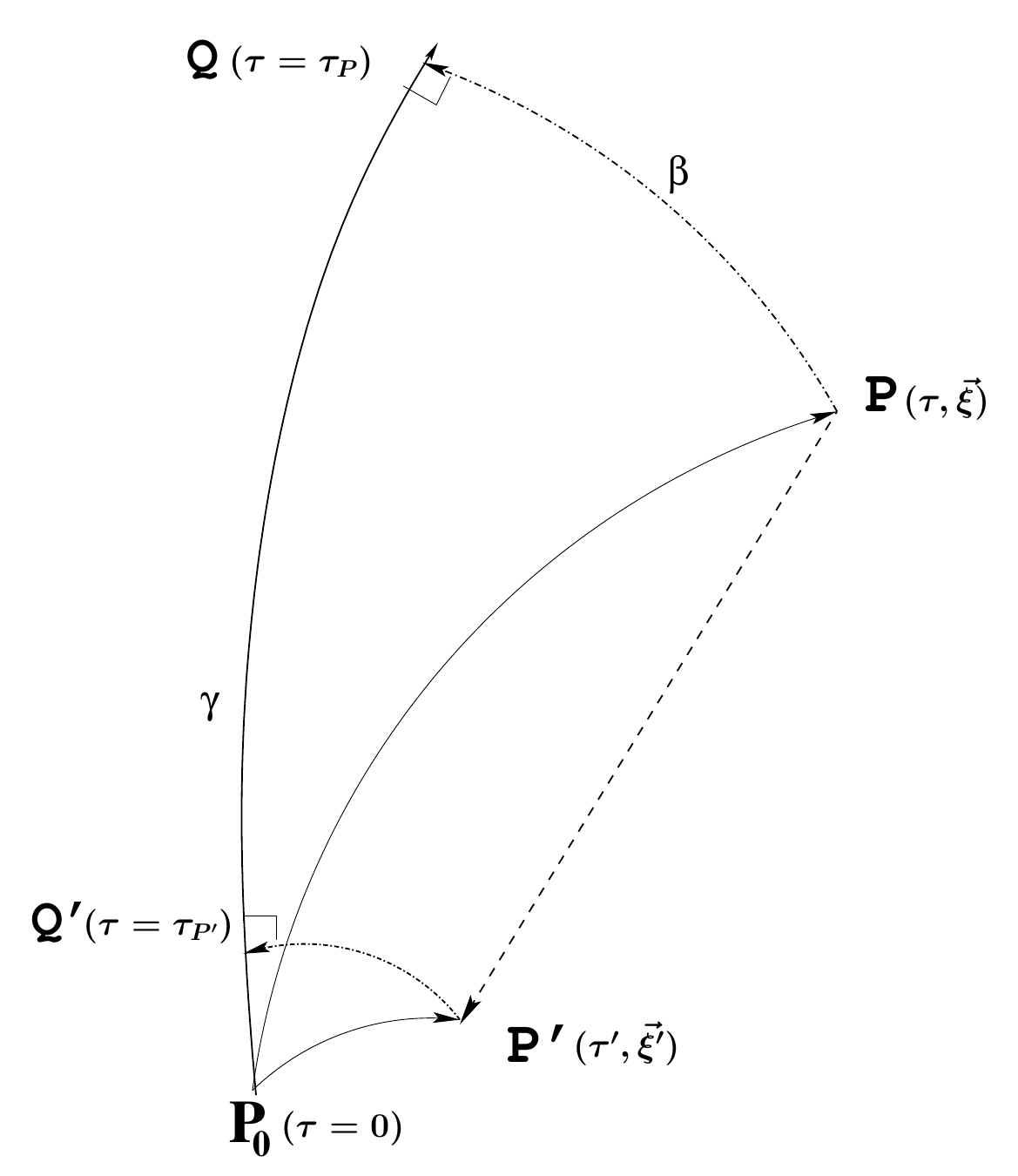}
\caption{The definition of Fermi Normal Coordinates. The dotted line
from $P$ to $P'$ is the unique null geodesic for which the parallel
propagator is computed in appendix \ref{ParallelPropagatorCaln}. The
geodesics $P_0P,\ P_0 P'$ are used in setting up the Riemann Normal
Coordinates. $P$ and $P'$ denote the observation and the source points
respectively. \label{FNCfig}}
\end{figure}

To proceed further, we employ {\em Fermi Normal Coordinates}
(FNC) and {\em Riemann Normal Coordinates} (RNC). These coordinate
charts are based on the choice of a time-like reference curve $\gamma$,
a reference point $P_0$ on it, and an orthonormal tetrad $E^{\alpha}_a$
at $P_0$ such that $E^{\alpha}_0$ equals the normalised tangent to
$\gamma$, at $P_0$. To be definite, let us take the world tube of the
spatially compact source to be around the line $AD$ of the figure
\ref{GlobalChartFig}. The line $AD$ is a time-like geodesic and we
naturally choose the reference curve, $\gamma$, to be this line.
Denoting the proper time along $\gamma$ by $\tau$, we choose $P_0 =
\gamma(\tau = 0)$, as the reference point.  Let $E^{\alpha}_a$ denote an
orthonormal tetrad at $P_0$ chosen such that $E^{\alpha}_0$ is the
normalized, geodesic tangent to $\gamma$.  Fermi transport the tetrad
along $\gamma$ (which is same as parallel transport since $\gamma$ is a
geodesic).  Thus we have an orthonormal tetrad,
$e^{\alpha}_{~a}e^{\beta}_{~b}\bar{g}_{\alpha\beta} = \eta_{ab}\ ,$ with
$e^{\alpha}_{~0} = $ the geodesic tangent to $\gamma$, all along
$\gamma(\tau)$. The corresponding orthonormal co-tetrad is denoted as
$e^{a}_{~\alpha}$. It follows that, all along $\gamma(\tau)$,
$\bar{g}_{\alpha\beta} = \eta_{\alpha\beta}$ and the Christoffel
connection is zero. With these choices, the {\em Fermi Normal
Coordinates} (FNC) and the {\em Riemann Normal Coordinates} (RNC) are
set up as follows (see \ref{FNCfig}).


%

To define the Fermi coordinates of a point $P$ off $\gamma$, let $\beta$
be the unique (space-like) geodesic from $P$, {\em orthogonally} meeting
$\gamma$ at a point $Q = \gamma(\tau_P)$, {with a unit affine
parameter interval}. Its tangent vector, $n^{\alpha}$ at $Q$ can be
resolved along the triad of space-like vectors at $Q$ as: $n^{\alpha} :=
\xi^i e^{\alpha}_{~i}$. Its norm gives the proper distance between $P$
and $Q$, $s^2 := n^{\alpha}n^{\beta}\eta_{\alpha\beta} =
\xi^i\xi^j\delta_{ij}$. The FNC of $P$ are then defined to be $(\tau_P,
\xi^i)$. Evidently, for points along $\gamma$, the spatial coordinates
$\xi^i$, are zero.  To define the RNC for the same point $P$ as above,
construct the unique geodesic starting from $P_0$ and reaching $P$ in a
unit affine parameter interval. This fixes the geodesics tangent vector
$N^{\alpha}$ at $P_0$. The normal coordinates of $P$, $X^a$, are then
defined through $N^{\alpha} := X^aE^{\alpha}_{~a}$. We will use them in
intermediate computations. 

Generally, the FNC and the RNC have a domain consisting of points
$P$ which have the required unique geodesics from the reference
curve/point.  By examining the geodesic equation in the global chart it
is easy to see that the RNC's and the FNC's would be valid in the static
patch (see also \cite{FNCDomain}).  {\em In effect, the computations of
this subsection are restricted to the static patch.}

Our task is to evaluate the terms in the integrand of eqn.
(\ref{InhomogeneousSoln}). The final answer will be expressed in terms
of the FNC introduced above.

{\em Computation of $\sigma(x, x')$}: Let $P, P'$ denote the observation
point and a source point respectively.  With the base point $P_0$, we
get a geodesic triangle $P_0 P P'$ with the $P' P$ geodesic being null
and future directed. Let $X^a, X'^a$ denote the RNCs of $P$ and $P'$
respectively.  In terms of the RNC set up in this manner, we have to
obtain $\sigma(P', P)$. For this we follow chapter II of \cite{Synge}. 
\begin{figure}[htb]
\includegraphics[width=7.0cm,height=5.2cm]{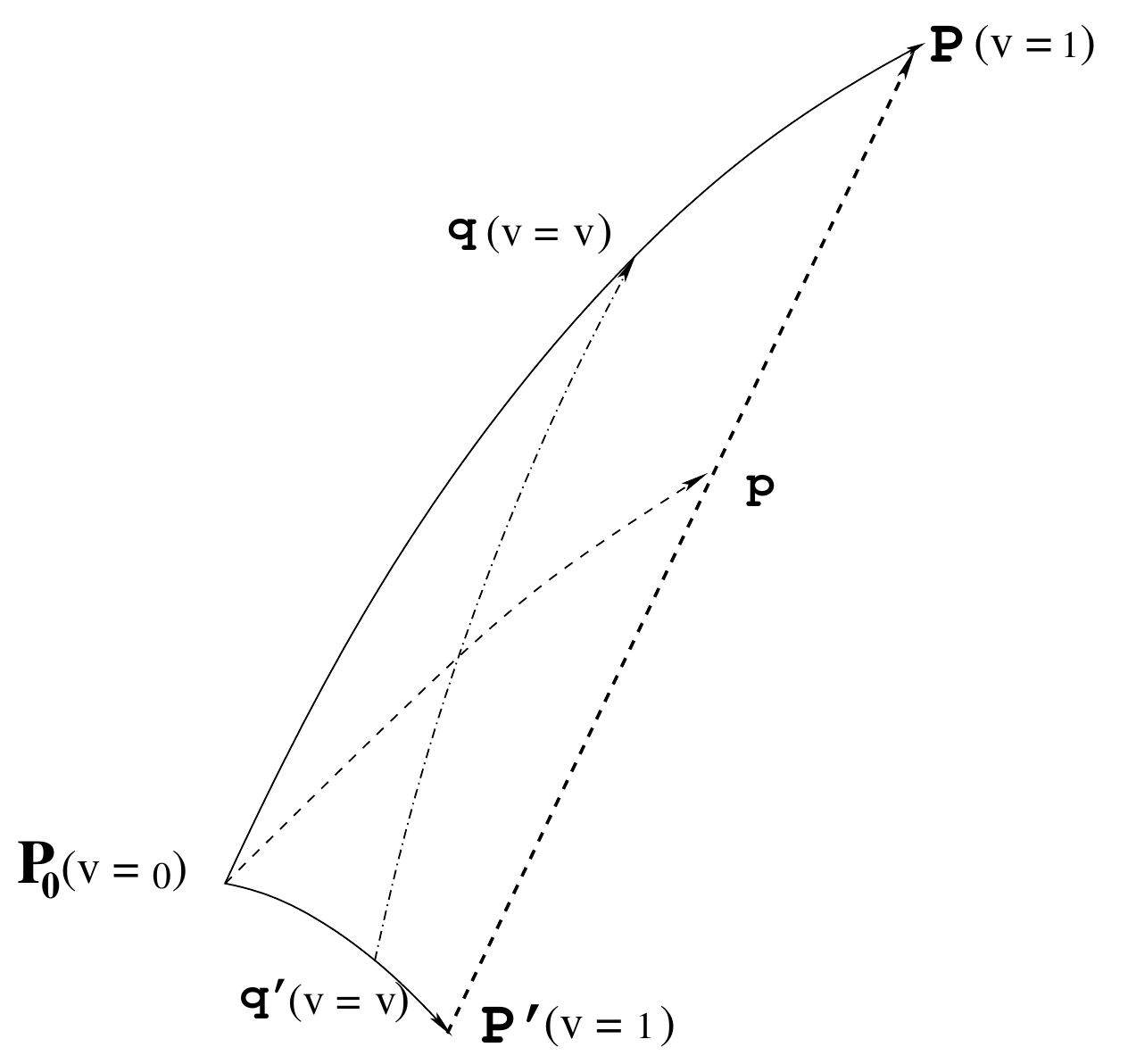} 
\caption{All lines are the unique geodesics in the Riemann normal
neighbourhood. As the point $p$ slides between $P'$ and $P$, a two
dimensional surface is generated.\label{RNCfig}}
\end{figure}

The idea is to construct a surface spanning family of geodesics (figure
\ref{RNCfig}), interpolating between the geodesics $P_0 P, P_0 P'$, all
originating at $P_0$ and ending on a point $p$ on the geodesic
connecting $P' P_0$.  Each of these have their affine parameters, $v$'s,
running from $0$ to $1$.  Choose points $q'$ and $q$ on the geodesics
$P_0 P'$ and $P_0 P$ respectively and having the same value of affine
parameter, $0 \le v \le 1$. The world function $\sigma(q', q)$ depends
{\em only} on $v$ and gives the desired answer for $v = 1$. When the
Riemann tensor is ``small'' i.e. can be treated as order 1 (different
from the orders used in the metric expansion), the $\sigma(q', q)$ is
expressed as a Taylor expansion, in $v$, to third order together with
the remainder. This gives,
\begin{eqnarray} \label{RNCDistance}
\sigma(P',P) & = & \sigma(P_0, P') + \sigma(P_0, P) - \left. \left(
g^{\alpha\beta}\frac{\partial \sigma(y, P')}{\partial y^{\alpha}}
\frac{\partial \sigma(y, P)}{\partial y^{\beta}}\right)\right|_{P_0}
\nonumber \\
& &  + \frac{1}{6}\int_0^1 dv (1 - v)^3 \frac{D^4 \sigma(q', q)}{Dv^4} \
; \hspace{1.0cm} \leftarrow ~~ \mbox{vanishes for flat space.} 
\end{eqnarray}

The term in the second line comes from the remainder in the Taylor
expansion and contains the modifications due to non-zero Riemann tensor.
This is computed to the first order in the curvature. For maximally
symmetric space-time, the computation simplifies. The steps are sketched
in the appendix (\ref{TriangleLaw}) and here is the final result
expressed in terms of the RNCs of $P, P'$:
\begin{eqnarray} \label{LeadingOrderDistance}
2 \sigma(P, P') & = & (X - X')\cdot (X - X') - \frac{\Lambda}{9}\left\{
(X\cdot X)(X'\cdot X') - (X\cdot X')^2\right\} + o(\Lambda^2) ~.
\end{eqnarray}
Here the dot product is the Minkowski dot product, $X\cdot Y :=
\eta_{ab}X^a Y^b$ etc. 

At this stage we convert the above expression from RNC to FNC. The
coordinate transformation between the RNC and the FNC is given by
\cite{WillPoisson},
\begin{eqnarray} \label{RNC2FNC}
X^0(\tau, \vec{\xi}) & = & \tau + \tau \frac{R^0_{~ij0} +
R^{0}_{~ji0}}{6} \xi^i\xi^j + \cdots \nonumber \\
& = & \tau\left(1 - \frac{\Lambda s^2}{9}\right) \\
X^i(\tau, \vec{\xi}) & = & \xi^i + \frac{R^i_{~0j0}}{6}\xi^j\tau^2 +
\frac{R^i_{~jk0}}{3}\xi^j\xi^k\tau \nonumber \\
& = & \xi^i\left(1 - \frac{\Lambda\tau^2}{18}\right)
\end{eqnarray}
In the second lines we have used the de Sitter curvature.

Substitution in (\ref{LeadingOrderDistance}) leads to,
\begin{eqnarray}
2\sigma(\tau, \vec{\xi}, \tau', \vec{\xi}') & = & \left\{- (\tau -
\tau')^2 + (\vec{\xi} - \vec{\xi}')^2\right\} +
\frac{\Lambda}{9}\left\{(\tau - \tau')^2(\vec{\xi}^2 + \vec{\xi}'^2 +
\vec{\xi}\cdot\vec{\xi}') - (\tau\vec{\xi}' + \tau'\vec{\xi})^2
\right. \nonumber \\
& & ~ - \left. ( - \tau^2 + \vec{\xi}^2)(- \tau'^2 + \vec{\xi}'^2) + (
-\tau\tau' + \vec{\xi}\cdot\vec{\xi}')^2 \right\}
\end{eqnarray}

{\em Solving the $\delta_+(\sigma)$:} We have to solve the
$\delta_+(\sigma(P, P'))$ for $\tau'$ and eliminate the $d\tau'$
integration. The solution is sought in the form of $\tau' = \tau_0 +
\Lambda\tau_1$. The $\tau_0$ is determined by vanishing of the first
braces and the retarded condition picks out one solution, namely,
$\tau_0 = \tau - |\vec{\xi} - \vec{\xi}'|$. The full solution is
obtained as,
\begin{eqnarray}
\tau'_{\mathrm{ret}} & := & \tau_0 + \Lambda\tau_1 ~~
\hspace{3.0cm} \mbox{where,}\nonumber \\
\tau_0 & = & \tau - |\vec{\xi} - \vec{\xi}'| \\
\tau_1 & = & - \frac{1}{18}\frac{1}{|\vec{\xi} -
\vec{\xi}'|}\left\{|\vec{\xi} - \vec{\xi}'|^2(\vec{\xi}^2 + \vec{\xi}'^2
+ \vec{\xi}\cdot\vec{\xi}') - (\tau(\vec{\xi} + \vec{\xi}') - |\vec{\xi}
- \vec{\xi}'|\vec{\xi})^2 \right. \nonumber \\
& & \left. - (-\tau^2 + \vec{\xi}^2)\left(\vec{\xi}'^2 - (\tau -
|\vec{\xi} - \vec{\xi}'|)^2\right) + (- \tau^2 + \tau|\vec{\xi} -
\vec{\xi}'| + \vec{\xi}\cdot\vec{\xi}')^2 \right\}
\end{eqnarray}

Now we introduce the approximation that the source size is much smaller
than its distance from observers i.e. $\vec{\xi}'^2 \ll \vec{\xi}^2
\leftrightarrow s' \ll s$ \footnote{The spatial coordinates are
proportional to the {\em proper distance} along the corresponding
spatial geodesics. This distance is related to but not equal to the
`physical distance' equaling the scale factor times the co-moving
distance.  Explicit relation is given in eqn. (\ref{CosmoToFNC}).
Nevertheless, $s' \ll s$, reflects the assumption of source size being
much smaller than the distance to the observer. The cosmological horizon
bounding the static chart is at a `physical distance' of
$\sqrt{3/\Lambda}$ and all our $s', s$ are within the static chart.}.
%
With this assumption, 
\[
|\vec{\xi}' - \vec{\xi}| ~ \approx ~ s\sqrt{1 + \frac{s'^2}{s^2} - 2
\hat{\xi}\cdot\hat{\xi'}\frac{s'}{s}} ~ \approx ~ s \ ,
\]
and keeping only the leading term in powers of $s$, we get, 
\begin{equation}\label{RetTauDefn}
\tau'_{\mathrm{ret}} ~ := ~ \tau - \left(s + \frac{\Lambda}{18}
s^3\right) ~~ =: \tau - \bar{s}(s) ~ =: \tau_{\mathrm{ret}}\ .
\end{equation}

From this, it follows that,
\begin{equation} \label{DeltaDet}
\left.\frac{\partial\sigma(\tau, \vec{\xi}, \tau',
\vec{\xi'})}{\partial\tau'}\right|_{\tau' = \tau - s - \frac{\Lambda
s^3}{18}} \approx - s\left(1 - \frac{\Lambda s^2}{18}\right) ~~
\Longrightarrow ~~ \left|\frac{\partial\sigma}{\partial\tau'}\right|^{-1}
\approx \frac{1}{s}\left(1 + \frac{\Lambda}{18}s^2\right) \ .
\end{equation}

{\em Note:} The $\tau_{\mathrm{ret}}(\tau, s)$ defined above in terms of
$\bar{s}$ reflects the non-Minkowskian metric and exactly corresponds to
the light-cone. 

{\em Metric and its determinant in FNC}: In terms of the FNC, the metric
to first order in the curvature, is given as \cite{Poisson},
\begin{equation}\label{FNCMetric}
g_{00}(\tau,\vec{\xi}) ~ = ~ - 1 + \frac{\Lambda s^2}{3} ~~,~~ g_{0i}
~=~ 0 ~~,~~ g_{ij} ~ = ~ \delta_{ij} - \frac{\Lambda}{9}(\delta_{ij} s^2
- \xi_i\xi_j) \ .
\end{equation}
The metric is {\em static} and its determinant is given by,
\begin{equation}\label{MetricDetFNC}
\sqrt{- g}|_{\mathrm{FNC}} ~ \approx ~ 1 - \frac{5}{18}\Lambda s^2 ~ = ~
\left(1 - \frac{\Lambda s^2}{6}\right)\left(1 - \frac{\Lambda
s^2}{9}\right)\ . 
\end{equation}
The second factor is the square root of the determinant of the induced
metric on a constant $\tau$ hypersurface. The metric being static
(independent of $\tau$ with $g_{0i} = 0$) also means that
$\partial_{\tau}$ is the stationary Killing vector in the FNC chart. 

{\em Riemann-Christoffel connection in FNC}: We compute this from the
metric. Noting that the metric is of the same form as the perturbation
about flat metric, $g_{\mu\nu} = \eta_{\mu\nu} + \delta g_{\mu\nu}$ with
$\delta g_{00} = \Lambda s^2/3, \delta g_{0i} = 0, \delta g_{ij} =
-~\Case{\Lambda}{9}(\delta_{ij}\vec{\xi}^2 - \xi_i \xi_j)$, we obtain,
\begin{equation}
\Gamma^{\mu}_{~\alpha\beta} ~ = ~
\frac{1}{2}\left(\partial_{\alpha}\delta g^{\mu}_{~\beta} +
\partial_{\beta}\delta g^{\mu}_{~\alpha} - \partial^{\mu}\delta
g_{\alpha\beta}\right) \ .
\end{equation}
Using,
\[
\delta g^{\mu}_{~\alpha} ~ = ~ - \frac{\Lambda s^2}{3}\delta^{\mu}_0
\delta^0_{\alpha} -
\frac{\Lambda}{9}(\delta^{\mu}_i\delta^i_{\alpha})(\xi^j\xi_j) +
\frac{\Lambda}{9}(\delta^{\mu}_i\xi^i)(\delta^j_{\alpha}\xi_j) \ ,
\]
we get,
\begin{eqnarray} \label{ConnectionFNC}
\Gamma^{\mu}_{~\alpha\beta} & = & \frac{\Lambda}{18}\left[ -
6\left\{\delta^{\mu}_0(\delta^0_{\alpha}\delta^i_{\beta}\xi_i +
\delta^0_{\beta}\delta^i_{\alpha}\xi_i) +
\delta^0_{\alpha}\delta^0_{\beta}\delta^{\mu}_i\xi^i\right\} \right.  \\
& & \left.
-~2\left\{\delta^{\mu}_i(\delta^i_{\alpha}\delta^j_{\beta}\xi_j +
\delta^i_{\beta}\delta^j_{\alpha}\xi_j) - 2
\delta^i_{\alpha}\delta^i_{\beta}\delta^{\mu}_j\xi^j \right\}\right]
\nonumber
\end{eqnarray}
For future reference, we also give the derivative of the connection.
\begin{eqnarray} \label{ConnectionDerivativeFNC}
\partial_{\gamma}\Gamma^{\mu}_{~\alpha\beta} & = &
\frac{\Lambda}{18}\left[ -
6\left\{\delta^{\mu}_0(\delta^0_{\alpha}\delta^i_{\beta}\delta_{\gamma
i} + \delta^0_{\beta}\delta^i_{\alpha}\delta_{\gamma i}) +
\delta^0_{\alpha}\delta^0_{\beta}\delta^{\mu}_i\delta_{\gamma}^i\right\}
\right.  \\
& & \left.
-~2\left\{\delta^{\mu}_i(\delta^i_{\alpha}\delta^j_{\beta}\delta_{\gamma
j} + \delta^i_{\beta}\delta^j_{\alpha}\delta_{\gamma j}) - 2
\delta^i_{\alpha}\delta^i_{\beta}\delta^{\mu}_j\delta_{\gamma}^j\right\}
\right] \nonumber
\end{eqnarray}

The parallel propagator $g^{\mu}_{~\alpha'}(P, P')$ is given in the
equation (\ref{ParallelPropagator}). It involves the coordinate
differences $(x' - x)^{\beta}$ while the coefficients are evaluated at
$x^{\alpha}$. We need the parallel propagator at the
retarded time and in the regime of $s \gg s'$.  The coordinate
differences are then given as,
\[
(x' - x)^0 ~ = ~ \tau' - \tau ~ \approx ~ - s ~~~,~~~ (x' - x)^i ~ = ~
\xi'^i - \xi^i \ .
\]
Thus, the parallel propagator depends only on $(\tau, \vec{\xi})$.

At this stage we recall that in the Minkowski background, a
simplification is achieved by further imposing the {\em synchronous
gauge} condition, $\tilde{h}_{0\alpha} = 0$ which removes the residual
gauge freedom of the TT gauge completely and we are left with only the
physical solution: the components $\tilde{h}_{ij}$ satisfying
$\partial^i\tilde{h}_{ij} = 0 = \delta^{ij}\tilde{h}_{ij}$. Is such a
simplification available in the de Sitter background?

As a matter of fact, it is a general result \cite{AA-Hunt}, that in a
globally hyperbolic space-time, given any Cauchy surface, $\Sigma$, the
normalised, time-like geodesic vector field, $\eta^{\alpha}$ orthogonal
to the Cauchy surface allows us to impose the synchronous gauge
condition $\tilde{h}_{\alpha\beta}\eta^{\beta} = 0$ in a {\em normal
neighbourhood} of $\Sigma$.  The vector field also provides us with a
convenient way to identify the physical components of the solution. 

For the static patch we are working in, the hypersurface of constant
$\tau$ corresponding to the horizontal line through the point `E' of
figure \ref{GlobalChartFig} is a Cauchy surface and the required
$\eta^{\alpha}$ field can be constructed  easily to order $\Lambda$. For
instance, let $\tau = \tau_0$ be the surface $\Sigma_0$, with a
normalized normal given by $\hat{n}^{\alpha} := (1 +
\Case{\Lambda}{6}s^2)\delta^{\alpha}_0$. Then the vector field $\eta$ is
determined as the solution of an initial value problem:
\begin{eqnarray}
0 & = & \eta^{\beta}(\partial_{\beta} \eta^{\alpha} +
\Gamma^{\alpha}_{~~\beta\gamma}\eta^{\gamma})  ~~ , ~~ \left.
\eta^{\alpha}\right|_{\Sigma_0} = \hat{n}^{\alpha} ~~~ \Longrightarrow
\\
\eta^{\alpha} & = & \delta^{\alpha}_{~0} +
\frac{\Lambda}{3}\left(\frac{s^2}{2}\delta^{\alpha}_{~0} + (\tau -
\tau_0)~ \xi^{i} \delta^{\alpha}_{~i} \right) + o(\Lambda^2)
\end{eqnarray}

From this it follows that in the synchronous gauge,
\begin{equation}
\tilde{h}^{\alpha\beta}\eta_{\beta} = 0 ~~ \Rightarrow ~~ \tilde{h}^{00}
= \frac{\Lambda}{3}(\tau - \tau_0)\tilde{h}^{0i}\xi_i ~~,~~
\tilde{h}^{0i} ~ = ~ \frac{\Lambda}{3}(\tau - \tau_0)\tilde{h}^{ij}\xi_j
\ .
\end{equation}
Clearly, $\tilde{h}^{00} \sim o(\Lambda^2)$ and can be set to be zero
while $\tilde{h}^{0i}$ is completely determined by $\tilde{h}^{ij}$. It
will turn out in the next section that {\em for `TT-projected'}
$\mathit{\tilde{h}^{ij}}$, $\tilde{h}^{0i} = 0$. Therefore, we now
specialise to the spatial components, $\mu = m, \nu = n$. 

Keeping only the leading powers in $s'/s$, the expressions simplify and
we obtain the parallel propagator as,
\begin{equation}
g^m_{~\alpha'}(\tau, \vec{\xi}, \tau'_{\mathrm{ret}}, \vec{\xi}') ~
\approx ~ \delta^m_{~\alpha'} + \frac{\Lambda
s^2}{18}\left[\delta^m_{~\alpha'} + 3\delta^0_{~\alpha'}\frac{\xi^m}{s}
-~\delta^j_{~\alpha'}\frac{\xi_j\xi^m}{s^2}\right] \ .
\end{equation}
Note that it is independent of the source point $(\tau', \vec{\xi}')$,
thanks to the leading $s'/s$ approximation. It is also independent of
$\tau$.

Now we have assembled all the terms in equation
(\ref{InhomogeneousSoln}). The $\tau'$ integration exhausts the
first factor in the $\sqrt{-g}$ and we get, 
\begin{eqnarray}\label{InhomoSoln0}
\tilde{h}^{\mu\nu}(\tau, \vec{\xi}) & = & \frac{4}{s}\left(1 +
\frac{\Lambda s^2}{18}\right)
g^{\mu}_{~\alpha'}(\vec{\xi})g^{\nu}_{~\beta'}(\vec{\xi}) \int d^3\xi'
\sqrt{g_3(\xi')} T^{\alpha'\beta'}(\tau_{\mathrm{ret}}, \vec{\xi}') \ .
\end{eqnarray}

The integral over the source is usually expressed in terms of time
derivatives of moments, using the conservation of the stress tensor. To
make these integrals well defined, it is convenient and transparent to
introduce suitable orthonormal tetrad and convert the coordinate
components to frame components. The frame components are coordinate
scalars (although they change under Lorentz transformations) and their
integrals are well defined. In the FNC chart, there is a natural choice
provided by the $\tau'$ = constant hypersurface passing through the
source world tube.  At any point on this hypersurface, we have a unique
orthonormal triad obtained from the triad on the reference curve by
parallel transport along the spatial geodesic. The unit normal,
$n^{\alpha}$, together with this triad, $e^{\alpha}_{~m}\, ,  m =
1,2,3$, provide the frame, $e^{\alpha}_{~a}$.  Explicitly, to order
$\Lambda$,
\begin{equation*}
n^{\tau}(\tau, \vec{\xi}') ~ = ~ 1 + \frac{\Lambda s'^2}{6}~,~ n^i = 0
~~~,~~~ e^{\tau}_{m}(\tau, \vec{\xi}') ~ = ~ 0~,~ e^{i}_{m}(\tau,
\vec{\xi}') ~ = ~ \left(1 + \frac{\Lambda s'^2}{18}\right)\delta^i_{m} -
\frac{\Lambda}{18}\xi'^i\xi'_m ~~,
\end{equation*}
In more compact form ({\em underlined indices denote frame indices}),
\begin{eqnarray} \label{FNCTetrad}
e^{\alpha'}_{~\U{a}} & := & \left(1 + \frac{\Lambda
s'^2}{6}\right)\delta^{\alpha'}_{~\tau}\delta^{\U{0}}_{\U{a}} +
\delta^{\alpha'}_{i}\delta^{\U{j}}_{\U{a}}\left\{\delta^i_{\U{j}}\left(1
+ \frac{\Lambda s'^2}{18}\right) -
\frac{\Lambda}{18}\xi'^i\xi'_{\U{j}}\right\}
\end{eqnarray}
It is easy to check that
$e^{\alpha'}_{\U{a}}e^{\beta'}_{\U{b}}g_{\alpha'\beta'} =
\eta_{\U{a}\U{b}}$. It follows that,  
\begin{equation} \label{ContractedPropagator}
g^m_{~\alpha'}(x)e^{\alpha'}_{~\U{a}}(x') ~ \simeq ~
\delta^{m}_{~\U{a}}\left(1 + \frac{\Lambda s^2}{18}\right) +
\frac{\Lambda}{6}\delta^{\U{0}}_{~\U{a}} s\ \xi^{m} -
\frac{\Lambda}{18}\delta^{\U{j}}_{~\U{a}}\xi_{\U{j}}\xi^{m} .
\end{equation} 

Defining the frame components of the stress tensor through the relation,
$T^{\mu\nu} := e^{\mu}_{~\U{a}}e^{\nu}_{~\U{b}}\Pi^{\U{ab}}$.  and
substituting for $g^m_{\alpha'}e^{\alpha'}_{\U{a}}(\tau,\vec{\xi},
\tau',\vec{\xi}')$, we obtain the final expression for the solution in
the synchronous gauge, to leading order in $s'/s$ and to $o(\Lambda)$
as,
\begin{eqnarray} \label{InhomoSoln}
\tilde{h}^{mn}(\tau, \vec{\xi}\ ) & = & \frac{4}{s}\left(1 +
\frac{\Lambda s^2}{18}\right)\left[\left(1 + \frac{\Lambda
s^2}{9}\right)\delta^{m}_{~\U{m}}\delta^n_{~\U{n}}\int\sqrt{g_3(\vec{\xi'})}
\Pi^{\U{mn}}(\tau_{\mathrm{ret}}, \vec{\xi}')\right.  \nonumber \\
& & + \frac{\Lambda
s}{6}\left\{\xi^m\delta^n_{~\U{n}}\int\sqrt{g_3(\vec{\xi}')}\Pi^{\U{0n}}
+ \xi^n\delta^m_{~\U{m}}\int\sqrt{g_3(\vec{\xi}')}\Pi^{\U{0m}} \right\}
\\
& & \left. - \frac{\Lambda}{18}\left\{
\xi^m\delta^n_{~\U{n}}\xi_{\U{k}}\int\sqrt{g_3(\vec{\xi}')}\Pi^{\U{kn}}
+
\xi^n\delta^m_{~\U{m}}\xi_{\U{k}}\int\sqrt{g_3(\vec{\xi}')}\Pi^{\U{km}}
\right\} \right] \nonumber 
\end{eqnarray}

The stress tensor is a function of ($\tau_{\mathrm{ret}}, \vec{\xi}'$),
$\tau_{\mathrm{ret}}$ being defined in equation (\ref{RetTauDefn}). The
terms in the second and the third line will drop out when a suitable TT
(transverse, traceless) projection is applied to the above solution to
extract its {\em gauge invariant} content, in section
\ref{TidalDistortion}.  Each of these integrals over the source on
$\tau' =$ constant hypersurface, are well defined and give a quantity
which is a function of the retarded time and carry only the frame
indices. The explicit factors of the mixed-indexed $\delta$'s are
constant triad which serve to convert the integrated quantities from
frame indices to coordinate indices.

To express the source integrals in terms of moments, we have to consider
the conservation equation. 

{\em Conservation equation}: The conservation equation is
$\partial_{\mu}T^{\mu\nu} = - \Gamma^{\mu}_{~\mu\lambda}T^{\lambda\nu} -
\Gamma^{\nu}_{~ \mu\lambda}T^{\mu\lambda}$  and we have computed the
connection in FNC in equation (\ref{ConnectionFNC}).  Recalling that the
stress tensor is trace free, $T^{\mu\nu}\bar{g}_{\mu\nu} =
{T^{\mu\nu}(\eta_{\mu\nu} + \delta g_{\mu\nu}) = } (- T^{00} +
T^j_{~j}) + T^{\mu\nu}\delta g_{\mu\nu} = 0$, we eliminate the spatial
trace by using $T^j_{~j} = T^{00} - T^{\mu\nu}\delta g_{\mu\nu}$. The
second term is order $\Lambda$.  To within our approximation and {\em
momentarily suppressing the primes on the coordinates}, we find,
\begin{eqnarray}\label{ConversationEqns}
\partial_0 T^{00} + \partial_i T^{i0} & = & \frac{11
\Lambda}{9}T^{0i}\xi_i \\
\partial_0T^{0i} + \partial_j T^{ji} & = & \frac{\Lambda}{9}( 7
T^{ij}\xi_j ~+~ T^{00} \xi^i)
\end{eqnarray}

Taking second derivatives and eliminating $T^{0i}$ we get,
\begin{equation}
\partial^2_{ij} T^{ij} ~ = ~ \partial^2_0 T^{00} +
\frac{\Lambda}{9}\left\{ 10 T^{00} + \xi^j\partial_j T^{00} + 18
\xi_i\partial_j T^{ij}\right\}
\end{equation}

Introducing the notation, $\rho := \Pi^{\U{00}}, \pi :=
\Pi^{\U{ij}}\delta_{\U{ij}}$, we express the the coordinate components
of the stress tensor in terms of the frame components as,
\begin{eqnarray} \label{FrameComponents}
T^{00} & = & \delta^0_{~\U{0}}\delta^0_{~\U{0}}\left(1 +
\frac{\Lambda s^2}{3}\right)\rho ; \\
T^{0i} & = & \delta^0_{~\U{0}}\left[\left(1 + \frac{2\Lambda
s^2}{9}\right)\delta^i_{~\U{j}}
- \frac{\Lambda}{18}\xi^i\xi_{\U{j}}\right]\Pi^{\U{0j}} \nonumber \\
T^{ij} & = & \left[\left(1 + \frac{\Lambda
s^2}{9}\right)\delta^i_{~\U{k}}\delta^j_{~\U{l}} -
\frac{\Lambda}{18}\left(\delta^i_{~\U{k}}\xi^j\xi_{\U{l}} +
\delta^j_{~\U{l}}\xi^i\xi_{\U{k}}\right)\right]\Pi^{\U{kl}} \nonumber 
\end{eqnarray} 

In terms of the frame components, the conservation equations take the
form (the constant tetrad are suppressed),
\begin{eqnarray}
\partial_{\tau}\Pi^{\U{00}} & = & - \left(1 -~\frac{\Lambda
s^2}{9}\right)\partial_j\Pi^{\U{0j}} +
\frac{\Lambda}{18}\xi_j\xi^i\partial_i\Pi^{\U{0j}} + \Lambda
\Pi^{\U{0j}}\xi_j \label{DelTauRhoEqn} \\
\partial_{\tau}\Pi^{\U{0i}} & = & -~\left(1 - \frac{\Lambda
s^2}{9}\right)\partial_j\Pi^{\U{ji}} +
\frac{\Lambda}{18}\left\{\xi_j\xi\cdot\partial\ \Pi^{\U{ji}} + 15
\Pi^{\U{ij}} \xi_j + 3{\rho}\xi^{i} \right\} \label{DelTauPiEqn}
\end{eqnarray}

Eliminating $\Pi^{\U{0i}}$ and using $\pi = \rho$ thanks to the trace
free stress tensor, we get the second order conservation equation as,
\begin{eqnarray} \label{SecondOrderConservationEqn}
\partial_{\tau}^2\rho & = & \left(1 - \frac{2\Lambda
s^2}{9}\right)\partial^2_{ij}\Pi^{\U{ij}} -
\frac{\Lambda}{9}\left[\xi^i\xi_j\partial^2_{ik}\Pi^{\U{jk}} + 19
\xi_i\partial_j\Pi^{\U{ij}} + 2 \xi^j\partial_j\rho + 12\rho\right]
\end{eqnarray}

The usual strategy is to define suitable moments of energy
density/pressures and taking moments of the above equation, express the
integral of $\Pi^{\U{ij}}$ in terms of the moments and its time
derivatives. To maintain coordinate invariance, the moment variable
(analogue of $x^i$ in the Minkowski background) must also be a {\em
coordinate scalar}. Note that in FNC (as in RNC), $\xi^i$ is a
contravariant vector.  Its frame components naturally provide coordinate
scalars. We still have the freedom to multiply these by suitable scalar
functions. It is easy to see that the frame components of $\xi$ are the
same as the coordinate components at best up to permutations i.e.
$\xi^{\U{i}} := e^{\U{i}}_j\xi^j = \delta^{\U{i}}_{~j}\xi^j$. It is also
true that $g_{ij}\xi^i\xi^j = \xi^{\U{i}}\xi^{\U{j}}\delta_{ij} = s^2$.
Hence suitable functions of $s^2$ would qualify to be considered as
coordinate scalars.

In equation (\ref{InhomoSoln}), we need $\int
d^3\xi'\sqrt{g_3(\vec{\xi}')}$. To get this from the equation
(\ref{SecondOrderConservationEqn}), we introduce a moment variable
$\zeta^i(\vec{\xi})$ and {\em define} moments of $\rho$ as,
\begin{eqnarray}\label{MomentsDefn}
{\cal M}^{\U{i_1i_2}\ldots \U{i_n}}(\tau) & := & \int_{\mathrm{source}}
d^3\xi\sqrt{g_3(\vec{\xi})} \zeta^{\U{i_1}}\cdots \zeta^{\U{i_n}} \
\rho(\tau, \vec{\xi}) ~~,~~ \zeta^{\U{i}}(\vec{\xi}) := \left(1 +
\frac{\Lambda s^2}{9}\right)\xi^{\U{i}}\ ,  
\end{eqnarray}
where the integration is over the support of the source on the
constant$-\tau$ hypersurface.

{Multiplying the equation (\ref{SecondOrderConservationEqn}) by
$\sqrt{g_3(\vec{\xi}')}\zeta^{\U{i_1}}\ldots\zeta^{\U{i_n}}$, 
and integrating over the source, we get, }
\begin{eqnarray}
\ddot{\cal M}^{\U{i_1}\ldots \U{i_n}} & = & \int d^3\xi\
\Pi^{\U{ij}}\partial^2_{ij}\left(\left(1 + \frac{(n - 3)\Lambda
s^2}{9}\right)\xi^{\U{i_1}\ldots \U{i_n}}\right) - \frac{\Lambda}{9}\left[\int
d^3\xi\
\Pi_{\U{j}}^{~\U{k}}\partial^2_{ik}\left(\xi^{\U{i}}\xi^{\U{j}}\xi^{\U{i_1}\ldots
\U{i_n}}\right) \right. \nonumber \\
& & \left.\hspace{0.0cm}+ 19\int d^3\xi\
\Pi^{\U{ij}}\partial_j\left(\xi_{\U{i}}\xi^{\U{i_1}\ldots \U{i_n}}\right) + 2\int d^3\xi\
\rho\partial_j\left(\xi^{\U{j}}\xi^{\U{i_1}\ldots \U{i_n}}\right) + 12\int d^3\xi\
\rho\xi^{\U{i_1}\ldots \U{i_n}} \right]
\end{eqnarray}
There are no factors of $(1 - \Lambda s^2/9)$ in the terms enclosed by
the square brackets since there is already an explicit pre-factor of
$\Lambda$. 

The first few moments {satisfy},
\begin{eqnarray}
\partial^2_{\tau}{{\cal M}} & = & \frac{\Lambda}{3}{\cal M}  \ ,
\hspace{4.5cm}\mbox{(`Mass conservation')}
\label{MassConservation}\\
\partial^2_{\tau}{{\cal M}}^{\U{i}} & = & \frac{2\Lambda}{3}{\cal
M}^{\U{i}} + \frac{2\Lambda}{3}\int d^3\xi\ \Pi^{\U{ij}}\xi_{\U{j}} \ ,
\hspace{1.0cm}\mbox{(`Momentum conservation')}
\label{MomentumConservation}\\
\partial^2_{\tau}{{\cal M}}^{\U{ij}} & = & 2
\int_{\mathrm{source}}d^3\xi\sqrt{g_3(\vec{x})}\ \Pi^{\U{ij}} +
\Lambda{\cal M}^{\U{ij}} + \Lambda\int
d^3\xi\xi_{\U{k}}\left(\Pi^{\U{ki}}\xi^{\U{j}} +
\Pi^{\U{kj}}\xi^{\U{i}}\right) \label{QuadrupoleConservation}
\end{eqnarray}

There are additional types of integrals over $\Pi^{\U{0n}}$ and
$\xi\cdot\Pi\cdot\xi$ in equation (\ref{InhomoSoln}). But these come
with an explicit factor of $\Lambda$ which simplifies the calculation.
These can be expressed in terms of different moments using both the
second order conservation equation, (\ref{SecondOrderConservationEqn})
and the first order one, (\ref{DelTauRhoEqn}). In particular, taking the
fourth moment and tracing over a pair gives (to order $(\Lambda)^0$),
\begin{eqnarray}
\int d^3\xi\ \xi_{\U{k}}(\Pi^{\U{km}}\xi^{\U{n}} +
\Pi^{\U{kn}}\xi^{\U{m}}) & = &
\frac{1}{4}\left[\delta_{\U{rs}}\partial_{\tau}^2{\cal M}^{\U{mnrs}} - 2
{\cal M}^{\U{mn}} - 2{\cal N}^{\U{mn}}\right]
\hspace{0.5cm}\mbox{where,} \nonumber \\ 
{\cal N}^{\U{mn}} & := & \int d^3\xi\sqrt{g_3(\vec{\xi})} \
\Pi^{\U{mn}}s^2 \, .
\end{eqnarray}
Likewise, {taking the first moment of equation
(\ref{DelTauRhoEqn})}, we get
\begin{equation}
\int d^3\xi \sqrt{g_3(\vec{\xi})} \Pi^{\U{0n}} ~ = ~
\partial_{\tau}{\cal M}^{\U{n}}  \ .
\end{equation}

Collecting all these, we write the solution in the form,
\begin{eqnarray} \label{FNCSoln}
\tilde{h}^{mn}(\tau, \vec{\xi}) & = &
\delta^m_{~\U{m}}\delta^n_{~\U{n}}
\left[\left(\frac{2}{s}\partial^2_{\tau}{\cal M}^{\U{mn}}\right) -
\frac{\Lambda}{3
s}\left(\xi_{\U{k}}\frac{\xi^{\U{m}}\partial^2_{\tau}{\cal M}^{\U{kn}} +
\xi^{\U{n}}\partial^2_{\tau}{\cal M}^{\U{km}}}{3} -
s^2\partial^2_{\tau}{\cal M}^{\U{mn}}\right)\right. \\
& & \left. + \frac{\Lambda}{s}\left(- {\cal M}^{\U{mn}} + {\cal
N}^{\U{mn}} -~\frac{1}{2}\delta_{\U{rs}}\partial^2_{\tau}{\cal
M}^{\U{mnrs}}\right) +
\frac{2\Lambda}{3}\left(\xi^{\U{m}}\partial_{\tau}{\cal M}^{\U{n}} +
\xi^{\U{n}}\partial_{\tau}{\cal M}^{\U{m}} \right) \right]\nonumber
\end{eqnarray}
The moments on the right hand side are all evaluated at the retarded
$\tau$ and we have displayed the constant triad. The constant triad
plays no role here but a similar one in the next subsection is
important.

There are several noteworthy points. 

(1) The leading term has exactly the same form as for the usual flat
space background. The correction terms involve the first, the second and
the fourth moments as well as a new type of moment ${\cal N}^{\U{mn}}$.
We will see in the next section that the term involving $\xi^{i}$ will
drop out in a TT projection.

(2) There are terms which have no time derivative of any of the moments
and hence can have constant (in time) field.  This is a new feature not
seen in the Minkowski background.  A priori, such a term is permitted
even in the Minkowski background. For instance, if
$\partial_{\tau}T^{\alpha\beta} = 0$ i.e. the source is {\em static},
then eqn. (\ref{InhomoSoln0}) or eqn. (\ref{InhomoSoln}) imply that
$\partial_{\tau}\tilde{h}^{mn} = 0$ and hence the solution {\em can}
have a $\tau-$independent piece. However, {in this case ($\Lambda =
0$), the conservation equation (\ref{QuadrupoleConservation}) equation
relates the field to double $\tau-$ derivative of the quadrupole moment
which vanishes for a static source.} It reflects the physical
expectation that a static source does not radiate. Does this expectation
change in a curved background?

{In a general curved background, `staticity' could be defined in a
coordinate invariant manner only if there is a time-like Killing vector,
say, $T$. A source would then be called static if the Lie derivative of
the stress tensor vanishes, ${\cal L}_TT^{\alpha\beta} = 0$. In the de
Sitter background, in the static patch we are working in, the stationary
Killing vector is precisely $\partial_{\tau}$. Hence, from the
definition of moments (\ref{MomentsDefn}, \ref{FNCTetrad}), it follows
that for a static source, $\partial_{\tau}T^{\alpha\beta} = 0$, all its
moments would be $\tau-$independent. However, the conservation equations
(\ref{MassConservation}) for the zeroth moment\footnote{ The non-zero
curvature always does `work' on the test matter and the `mass of the
matter' alone is not conserved.} contradicts this, unless ${\cal M}$
itself vanishes. Hence we {\em cannot} even have strictly static (test)
sources in a curved background. Thus, in the specific case of the de
Sitter background, the non-derivative terms in the equation
(\ref{QuadrupoleConservation}), do not indicate the possibility of time
independent field $\tilde{h}^{mn}$.}

For a very slowly varying source - so that we can neglect the derivative
terms - we can have a left over, slowly varying field, falling off as
$\sim \Lambda/s$.  Such a field has a very long wavelength and is not
`radiative' in the static patch. To isolate radiative fields, one should
probe the vicinity of the null infinity which is beyond the extent of
the static patch.  For typical rapidly changing sources, ($\lambda \ll
s$) the $\tau-$derivative terms dominate over these terms and in the
context of present focus, we {\em drop them hereafter}. 

The remaining terms that survive the TT projection, all have second
order $\tau-$derivative. Similar features also arise in the Cosmological
chart in the next subsection.

(3) The mass conservation equation can be immediately integrated and
have exponentially growing and decaying components. The scale of this
time variation is $\sim (\Lambda)^{-1/2}$ which is extremely slow, about
the age of the universe. These equations do {\em not} depend on the
Green's function at all and are just consequences of the matter
conservation equation for {\em small} curvature. We are working in a
static patch of the space-time, so the time variation is not driven by
the time dependence of the background geometry. It is the background
{\em curvature} that is responsible for the changes in the matter
distribution and hence its moments. In effect, this confirms that test
matter {\em cannot} remain static in a {\em curved} background even if
the background is static. In a flat background, there is no work done on
the test matter and hence the sources' mass and linear momenta are
conserved (the zeroth and the first moment are time independent).

For contrast, in the next subsection, we recall the computation in the
generalized transverse gauge \cite{deVega}. This subsection has the tail
contribution explicitly available and the correction terms are in powers
of $\sqrt{\Lambda}$.

\subsection{Generalized Transverse Gauge in Poincare
Patch}\label{PoincareChart}

The computation takes advantage of the conformally flat form of the
metric in the conformal chart and makes a choice of a generalized
transverse gauge to simplify the linearized equation. We summarize them
for convenience and present the radiative solution. 

\begin{figure}[htb]
\begin{center}
\includegraphics[scale=0.62]{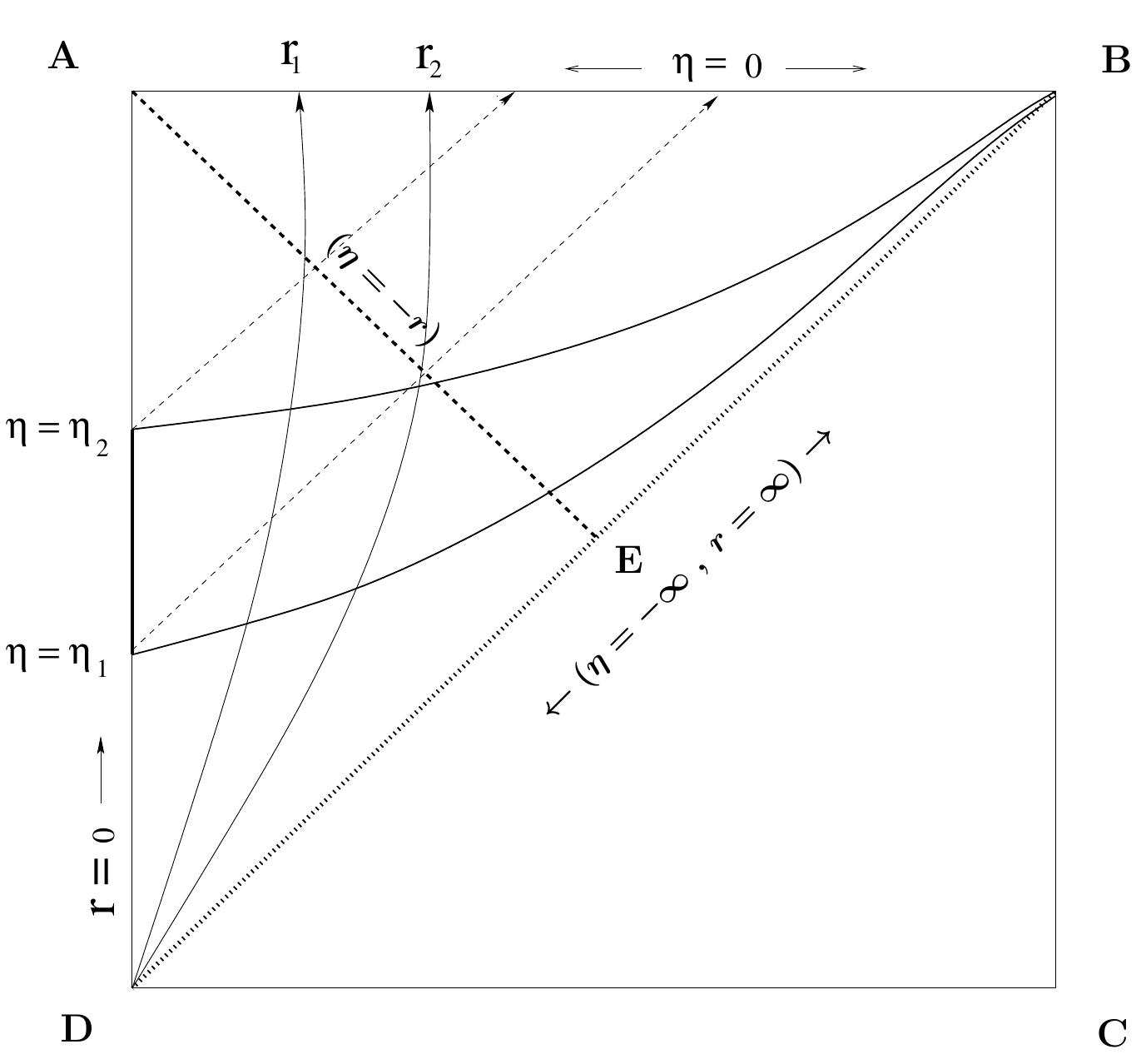}
\caption{The full square is the Penrose diagram of de Sitter space-time
with generic point representing a 2-sphere.  The Poincare patch labeled
ABD, is covered by the conformal chart $(\eta, r, \theta, \phi)$.  The
line BD does not belong to the chart. The line AB is the {\em future
null infinity}, ${\cal J}^+$ and the line AE is the {\em
cosmological horizon}. Two constant $\eta$ space-like hypersurfaces are
shown with $\eta_2 > \eta_1$. The two constant $r$, time-like
hypersurfaces have $r_2 > r_1$.  The two dotted lines at 45 degrees,
denote the paths of gravitational waves emitted at $\eta = \eta_1,
\eta_2$ on the world line at $r = 0$, through the source. During the
interval ($\eta_1 , \eta_2$), the source is `active' i.e. varying
rapidly enough to be in the detectable range of frequencies.  The region
AED is a static patch.  \label{DeSitterPenrose}}
\end{center} 
\end{figure}

In the conformal chart (see figure \ref{DeSitterPenrose}), the
coordinates and the metric take the form,
\begin{eqnarray}
z^0 & = & H^{-1}\, \mathrm{sinh}(Ht) + Hr^2 \frac{e^{Ht}}{2} ~~,~~ z^1 ~
= ~ H^{-1}\, \mathrm{cosh}(Ht) - Hr^2 \frac{e^{Ht}}{2} ~~,~~ (\therefore
z^0 + z^1 > 0) , \nonumber \\ 
z^i & = & e^{Ht} x^i ~~~,~~~ i = 2, 3, 4~~,~~ r^2 := \sum_i (x^i)^2
\hspace{2.7cm},~~~ t, ~ x^i \in \mathbb{R} ~; \\ 
ds^2 & = & - dt^2 + e^{2Ht}\sum_{i = 2}^4 (dx^i)^2 \ .  ~~~\mbox{The
substitution, $\eta := - H^{-1}\ e^{-Ht} ~~ \Rightarrow~~ $,}
\label{ConformalTime}\\
ds^2 & = & \frac{1}{H^2\eta^2}\left[ - d\eta^2 + \sum_i (dx^i)^2
\right]\ , ~~ \eta \in ( - \infty , 0 )~~,~~ H :=
\sqrt{\frac{\Lambda}{3}} . \label{ConformallyFlat}
\end{eqnarray}
The conformally flat form leads to a great deal of simplification. The
${\cal J}^+$ is approached as $\eta \to 0_-$ while the $\eta \to
-~\infty$ corresponds to the FLRW singularity.

In this chart, the de Sitter d'alembertian can be conveniently expressed
in terms of the Minkowski d'alembertian leading to, 
\begin{eqnarray}
0 & = & \Omega^{-2}\left[ \Box \tilde{h}_{\mu\nu} +
\frac{2}{\eta}\left\{\left(
\delta_{\mu}^0\partial^{\sigma}\tilde{h}_{\sigma\nu} +
\delta_{\nu}^0\partial^{\sigma}\tilde{h}_{\sigma\mu}\right) + \left( -
\partial_0 \tilde{h}_{\mu\nu} + \partial_{\mu}\tilde{h}_{0\nu} +
\partial_{\nu}\tilde{h}_{0\mu} \right) \right\} \right. \nonumber \\
& & \left. + \frac{2}{\eta^2}\left\{ \delta_{\mu}^0\delta_{\nu}^0
\tilde{h}_{\alpha\beta}\eta^{\alpha\beta} + \eta_{\mu\nu}\tilde{h}_{00}
+ 2 \left( \delta_{\mu}^0\tilde{h}_{0\nu} +
\delta_{\nu}^0\tilde{h}_{0\mu} \right) \right\} \right] 
-~\left(\frac{2\Lambda}{3}\right)\left[\tilde{h}_{\mu\nu} -
\eta_{\mu\nu} \tilde{h}_{\alpha\beta}\eta^{\alpha\beta}\right] \nonumber
\\ 
& & -~ \left\{\left(\partial_{\mu}B_{\nu} + \partial_{\nu}B_{\mu} -
\eta_{\mu\nu}\partial^{\alpha}B_{\alpha}\right) +
\frac{2}{\eta}\left(\delta_{\mu}^0B_{\nu} +
\delta_{\nu}^0B_{\mu}\right)\right\} ~~,~~ \Omega^2 :=
\frac{1}{H^2\eta^2} = \frac{3}{\Lambda\eta^2} \ .
\label{LinEqnGaugeFixed}
\end{eqnarray}
The left hand side will be $- 16\pi T_{\mu\nu}$ in presence of
matter.

While the transverse gauge will eliminate the $B_{\mu}$ terms, it still
keeps the linearized equation in a form that mixes different components
of $\tilde{h}_{\mu\nu}$.  A different choice of $B_{\mu}$ achieves
decoupling of these components.  Taking $B_{\mu}$ of the form
$f(\eta)\tilde{h}_{0\mu}$, shows that for the choice $f(\eta) :=
\Case{2\Lambda}{3}\eta$, the equation (with source included) simplifies
to \cite{deVega},
\begin{eqnarray}
-16\pi T_{\mu\nu}\Omega^2 & = & \Box\tilde{h}_{\mu\nu} -
\frac{2}{\eta}\partial_0\tilde{h}_{\mu\nu} -
\frac{2}{\eta^2}\left\{\delta_{\mu}^0\delta_{\nu}^0\tilde{h}_{\alpha}^{~\alpha}
-~\tilde{h}_{\mu\nu} + \delta_{\mu}^0\tilde{h}_{0\nu} +
\delta_{\nu}^0\tilde{h}_{0\mu} \right\} ~~ , ~~ \mbox{with}
\label{NonCovariantGaugeEqn} \\
0 & = & \partial^{\alpha}\tilde{h}_{\alpha\mu} +
\frac{1}{\eta}\delta_{\mu}^0 \tilde{h}_{\alpha}^{~\alpha} ~~,~~
\tilde{h}_{\alpha}^{~\alpha} :=
\tilde{h}_{\alpha\beta}\eta^{\alpha\beta}\ \hspace{1.0cm} \mbox{(gauge
fixing condition)} \label{GaugeCondn}
\end{eqnarray}
{\em From now on in this subsection, the tensor indices are
raised/lowered with the Minkowski metric.}

It turns out to be convenient to work with new variables, $\chi_{\mu\nu}
:= \Omega^{-2}\tilde{h}_{\mu\nu}$. All factors of $\Omega^2$ {\em and}
$\Lambda$ drop out of the equations and $\chi_{\mu\nu}$ satisfies
\cite{deVega},
\begin{eqnarray}
-~16 \pi G T_{\mu\nu} & = & \Box \chi_{\mu\nu} +
\frac{2}{\eta}\partial_0\chi_{\mu\nu} -
\frac{2}{\eta^2}\left(\delta_{\mu}^0\delta_{\nu}^0\chi_{\alpha}^{~\alpha}
+ \delta_{\mu}^0\chi_{0\nu} + \delta_{\nu}^0\chi_{0\mu} \right) \ .
\label{LinEqnChi} \\ 
0 & = & \partial^{\alpha}\chi_{\alpha\mu} + \frac{1}{\eta}\left(2
\chi_{0\mu} + \delta_{\mu}^0 \chi_{\alpha}^{~\alpha} \right)
\hspace{2.5cm} (\mbox{gauge condition}). \label{ChiGauge}
\end{eqnarray}

Under the gauge transformations generated by a vector field $\xi^{\mu}$,
the $\chi_{\mu\nu}$ transform as, 
\begin{equation} 
\delta\chi_{\mu\nu} ~ = ~ (\partial_{\mu}\underline{\xi}_{\nu} +
\partial_{\nu}\underline{\xi}_{\mu} -
\eta_{\mu\nu}\partial^{\alpha}\underline{\xi}_{\alpha}) -
\frac{2}{\eta}\eta_{\mu\nu}\underline{\xi}_0 ~~,~~ \underline{\xi}_{\mu}
:= \Omega^{-2}\xi_{\mu} = \eta_{\mu\nu}\xi^{\nu}.  \label{ChiGaugeTrans}
\end{equation}

The gauge condition (\ref{ChiGauge}) is preserved by the transformation
generated by a vector field $\xi^{\mu}$ satisfying, 
\begin{equation}
{\Box\underline{\xi}_{\mu} +
\frac{2}{\eta}\partial_0\underline{\xi}_{\mu} -
\frac{2}{\eta^2}\delta_{\mu}^0\underline{\xi}_0  ~ = ~ 0 }
\label{ResidualChiGauge}
\end{equation}
and the equation (\ref{LinEqnChi}) is invariant under the gauge
transformations generated by these restricted vector fields.

It is further shown in \cite{deVega} that the residual invariance is
exhausted by setting $\chi_{0i} = 0 = \hat{\chi} (:= \chi_{00} +
\chi_{i}^{~i})$. The gauge condition (\ref{ChiGauge}) then implies
$\partial^0\chi_{00} = 0$ and by choosing it to be zero at some
initial $\eta = $constant hypersurface we can take $\chi_{00} = 0$ as
well. Thus the {\em physical solutions}, satisfy conditions:
$\partial^{i}\chi_{ij} = 0 = \chi^{i}_{~i}$ and it suffices to focus on
the equation (\ref{LinEqnChi}) for $\mu, \nu = i, j$. 

To obtain the inhomogeneous solution, we return to the equations
satisfied by the $\chi_{00}, \chi_{0i}$ and $\chi_{ij}${, which}
are decoupled and we are interested only in the $\chi_{ij}$ equation:
\[
\Box \chi_{ij} + \frac{2}{\eta}\partial_0 \chi_{ij} = - 16\pi T_{ij}
~~~~~,~~~~~ \partial_i \chi^{i}_{~j} = 0 = \chi_i^{~i} \ .
\]

The corresponding, retarded Green function is defined by 
\begin{equation}
\left(\Box + \frac{2}{\eta}\partial_0\right) G_R(\eta, x; \eta', x') = -
\frac{\Lambda}{3}\eta^2\delta^4(x - x') ~ ,
\end{equation}
and is given by \cite{deVega},
\begin{equation} \label{NonCovariantGreenFn}
G_R(\eta, x; \eta' x') ~ = ~ \frac{\Lambda}{3}\eta \eta' \frac{1}{4\pi}
\frac{ \delta(\eta - \eta' - |x - x'|) }{|x - x'|} ~+~
\frac{\Lambda}{3}\frac{1}{4\pi}\theta(\eta - \eta' - |x - x'|) ~ .
\end{equation}

The particular solution is given by,
\begin{eqnarray}
\chi_{ij}(\eta, x) & = & 16\pi \int_{\mathrm{source}} \frac{d\eta'
d^3x'}{\Case{\Lambda}{3}\eta'^2} G_R(\eta, x; \eta' x') T_{ij}(\eta',
x') \\
& = & 4\int d\eta'd^3x' \frac{\eta}{\eta'}\frac{\delta(\eta - \eta' - |x
- x'|) }{|x - x'|} T_{ij}(\eta', x') \nonumber \\
& & \hspace{1.0cm} ~ + ~ 4\int d\eta'd^3x' \frac{1}{\eta'^2}\theta(\eta
- \eta' - |x - x'|) T_{ij}(\eta', x') \label{RadiativeSoln0} \\
%
%
& = & 4 \int d^3x' \frac{\eta}{|x - x'|(\eta - |x - x'|)} \left.
T_{ij}(\eta', x') \right|_{\eta' = \eta - |x - x'|} \nonumber \\
& & \hspace{0.5cm} 
+ ~ 4 \int d^3x' \int_{- \infty}^{\eta - |x
-x'|}d\eta'\frac{T_{ij}(\eta', x')}{\eta'^2} \label{RadiativeSoln}
\end{eqnarray}
The spatial integration is over the matter source confined to a compact
region and is finite. {The second term in the eqns.
(\ref{RadiativeSoln0}, \ref{RadiativeSoln}) is the {\em tail term}.} 



It is possible to put the solution in the same form as in the case of
flat background, in terms of suitable Fourier transforms with respect to
$\eta$ \cite{deVega}. However, we work with the $(\eta, \vec{x})$-space. 

For $|\vec{x}| \gg |\vec{x}'|$, we can approximate $|\vec{x} - \vec{x}'|
\approx r := |\vec{x}|$. This allows us separate out the $\vec{x}'$
dependence from the $\eta - |x - x'|$. In the first term, this leads to
the spatial integral over $T_{ij}(\eta - r, x')$ while in the second
term, we can interchange the order of integration again leading to the
same spatial integral. The spatial integral of $T_{ij}$ can be
simplified using moments.  This is done through the matter conservation
equation using the conformally flat form of the metric,
\begin{equation}\label{ConservationEqnPoincare}
\partial^{\mu}T_{\mu 0} + \frac{1}{\eta}\left(T_{00} + T_i^{~i}\right) ~
= ~ 0 ~~~,~~~ \partial^{\mu}T_{\mu i} + \frac{2}{\eta}T_{0i} ~ = ~ 0 ~.
\end{equation}

Taking derivatives of these equations to eliminate $T_{0i}$, we get,
\begin{equation} \label{ConservePoincareEqn}
\partial^i\partial^j T_{ij} ~ = ~ \partial^2_{\eta}T_{00} -
\frac{1}{\eta}\partial_{\eta}(T_{00} + T_i^{~i}) +
\frac{3}{\eta^2}(T_{00} + T_i^{~i}) -
\frac{2}{\eta}\partial_{\eta}T_{00}
\end{equation}
As in the previous section, we introduce a tetrad to define the
frame components of the stress tensor. The conformal form of the metric
suggests a natural choice: ($\sqrt{\Lambda/3} =: H$), 
\begin{equation}
f^{\alpha}_{~\U{0}} ~ := ~ - H\eta(1, \vec{0}) ~~,~~ f^{\alpha}_{~\U{m}}
~ := ~  - H\eta\ \delta^{\alpha}_{~\U{m}} \hspace{0.7cm}
\Longleftrightarrow \hspace{0.7cm} f^{\alpha}_{~\U{a}} := - H\eta
\delta^{\alpha}_{~\U{a}} 
\end{equation}
The corresponding components of the stress tensor are given by,
\begin{eqnarray}
\rho ~:=~ P_{\U{00}} =
T_{\alpha\beta}f^{\alpha}_{~\U{0}}f^{\beta}_{~\U{0}} ~ = ~
H^2\eta^2T_{00}\delta^0_{~\U{0}}\delta^0_{~\U{0}} & , & P_{\U{ij}} ~ :=
~ T_{\alpha\beta}f^{\alpha}_{~\U{i}}f^{\beta}_{~\U{j}} ~ = ~ H^2\eta^2
T_{ij}\delta^i_{~\U{i}}\delta^j_{~\U{j}} \ ; \\
P_{\U{0i}} := T_{\alpha\beta}f^{\alpha}_{~\U{0}}f^{\beta}_{~\U{i}} & , &
\pi := P_{\U{ij}}\delta^{\U{ij}} \ .
\end{eqnarray}

In terms of these, the conservation equations take the form (suppressing
the constant tetrad),
\begin{eqnarray} 
0 & = & \partial_{\eta}P^{\U{00}} + \partial_iP^{\U{i0}} -
\frac{1}{\eta}\left(3P^{\U{00}} + \pi\right) \\
0 & = & \partial_{\eta}P^{\U{0i}} + \partial_jP^{\U{ji}} -
\frac{1}{\eta}4P^{\U{0i}}
\end{eqnarray}
%
It is convenient to go over to the cosmological chart, $(t, \vec{x})$
and convert the $\partial_{\eta}$ to $\partial_t$ using the definitions:
$\eta := - H^{-1} e^{- Ht}$. This leads to $\partial_{\eta} ~ = ~
e^{Ht}\partial_t ~~ := ~~ a(t)\partial_t\ $.
\begin{eqnarray}
0 & = & \partial_t\rho + \frac{1}{a}\partial_i P^{\U{0i}} + H (3\rho +
\pi) \ ; \label{Delt00Eqn}\\
0 & = & \partial_t P^{\U{0i}} + \frac{1}{a}\partial_jP^{\U{ij}} + 4H
P^{\U{0i}} \ \label{Delt0iEqn}; \\
0 & = & \partial^2_t\rho - \frac{1}{a^2}\partial^2_{ij}P^{{\U{ij}}} +
8H\partial_t\rho + H\partial_t\pi + 5H^2(3\rho + \pi)
\end{eqnarray}

As before, we define the moments of the two rotational scalars, $\rho,
\pi$, by integrating over the source distribution at $\eta = $constant
hypersurface. The determinant of the induced metric on these
hypersurfaces is $a^3(\eta)$. The tetrad components of the moment
variable are given by, $\bar{x}^{\U{i}} := f^{\U{i}}_{\alpha} x^{\alpha}
= - (\eta H)^{-1}\delta^{\U{i}}_{~j}x^{j} = a(t)x^{\U{i}}$. The moments
are defined by,
\begin{eqnarray}
{Q}^{\U{i_1}\cdots\U{i_n}}(t) & := & \int_{Source(t)} d^3x \
a^{3}(t)\rho(t,\vec{x}) \bar{x}^{\U{i}_1} \cdots \bar{x}^{\U{i}_n} ~~,
\\
~~,~~ \bar{Q}^{\U{i_1}\cdots\U{i_n}}(t) & := & \int_{Source(t)} d^3x \
a^{3}(t)\pi(t,\vec{x}) \bar{x}^{\U{i}_1} \cdots \bar{x}^{\U{i}_n} ~.~
\end{eqnarray}
Taking second moment of the eqn. (\ref{ConservePoincareEqn}) and
lowering the frame indices we get, 
\begin{equation}
\int d^3x a^3(t) P_{\U{ij}}(t, x) ~ = ~ \frac{1}{2}\left[\partial^2_t
Q_{\U{ij}} - 2H\partial_t Q_{\U{ij}} + H \partial_t\bar{Q}_{\U{ij}}
\right] 
\end{equation} 

Let us write the solution, eq.(\ref{RadiativeSoln}), in terms of the
cosmological chart, incorporating the approximation $|\vec{x}'| \ll
|\vec{x}|$.
\begin{eqnarray}
\chi_{ij}(\eta, x) & = & \left. 4 \frac{\eta}{r(\eta - r)} \int d^3x'
T_{ij}(\eta', x') \right|_{\eta' = \eta - r} 
%
~ + ~ 4 \int d^3x' \int_{- \infty}^{\eta - r }d\eta'\frac{T_{ij}(\eta',
x')}{\eta'^2} \label{RadiativeSoln1}
\end{eqnarray}

Define the {\em retarded time}, $t_{\mathrm{ret}}$, through $(\eta - r)
:= - H^{-1}e^{-Ht_{\mathrm{ret}}}$ and set $\bar{a} :=
a(t_{\mathrm{ret}})$. Then we have, $\eta = - (a H)^{-1} ~,~ (\eta - r)
= - (\bar{a}H)^{-1}$. Using these,
\begin{eqnarray}
\frac{\eta}{\eta - r}T_{ij}(\eta - r, x') & =  &
a(t)^{-1}\bar{a}^3P_{\U{ij}}(t_{\mathrm{ret}}, x') ~,~
%
%
d\eta' \frac{1}{\eta'^2}T_{ij}(\eta', x')  =  H^2 dt'
a^3(t')P_{\U{ij}}(t', x')
\end{eqnarray}
All terms involve only the $\int a^3P_{ij}$ which is obtained above.
With these, the solution takes the form,
\begin{eqnarray}
\chi_{\U{ij}}(t, r) & \approx & \frac{2}{r\
a(t)}\left.\left[\partial^2_{t'} Q_{\U{ij}} - 2H\partial_{t'} Q_{\U{ij}}
+ H \partial_{t'}\bar{Q}_{\U{ij}} \right]\right|_{t_{\mathrm{ret}}} 
%
%
+ 2H^2\left.\left\{ \partial_{t'} Q_{\U{ij}} - 2H Q_{\U{ij}} + H
\bar{Q}_{\U{ij}} \right\}\right|_{t_{\mathrm{ret}}} \nonumber \\
& & \hspace{6.5cm} - 2H^2 ~\left.\left\{\partial_{t'} Q_{\U{ij}} - 2H
Q_{\U{ij}} + H \bar{Q}_{\U{ij}} \right\}\right|_{-\infty}
\label{NonCovariantGaugeGWAmplitude}
\end{eqnarray} 
We have restored the constant triad and used the definition:
$\chi_{\U{ij}} := \delta_{\U{i}}^{~i}\delta_{\U{j}}^{~j}\chi_{ij}$.
{ The first term in the equation
(\ref{NonCovariantGaugeGWAmplitude}) is the contribution of the sharp
term and the remaining terms are from the tail.  The tail contribution
has separated into a term which depends on the retarded time just as the
sharp term does, and the contribution from the history is given by the
limiting value in the last line.}

This expression is valid {\em as a leading term for $|\vec{x}| \gg
|\vec{x}'|$}.  (For Hulse-Taylor system, the physical size is about 3
light-seconds and it is about 20,000 light years away, giving $|x'|/|x|
\sim 10^{-12}$.) We work with this expression in the following and
suppress the $\approx$ sign.

We write $a^{-1} = \bar{a}^{-1}(\Case{\bar{a}}{a}) = \bar{a}^{-1}(1
-~Hr\bar{a}) {= \bar{a}^{-1} - rH}$ in the first term {to make
manifest the dependence on retarded time $t_{\mathrm{ret}}$.} 
The solution is then expressed as,
\begin{eqnarray} \label{NonCovariantGaugeGWAmplitude1}
\chi_{\U{ij}}(t, r) & \approx & \frac{2}{r\ \bar{a}}\left\{\partial^2_t
Q_{\U{ij}} - 2H\partial_t Q_{\U{ij}} + H \partial_t\bar{Q}_{\U{ij}}
\right\} \nonumber\\
& & - 2H\left\{\partial^2_tQ_{\U{ij}} - 3H\partial_t Q_{\U{ij}} +
H\partial_t\bar{Q}_{\U{ij}} + 2H^2Q_{\U{ij}} - H^2\bar{Q}_{\U{ij}}
\right\} \\
& & 
%
-2H^2\left.\left\{\partial_{t'} Q_{\U{ij}} - 2H Q_{\U{ij}} + H
\bar{Q}_{\U{ij}} \right\}\right|_{-\infty} \nonumber
\end{eqnarray} 

{\em Remarks:}

(1) In the conformal chart, there is no explicit dependence on
the cosmological constant and it is not a suitable chart for exploring
the subtle limit of vanishing cosmological constant
\cite{AshtekarBongaKesavanII, AshtekarBongaKesavanIII}. Hence we changed
to the cosmological chart and exhibited the solution with explicit
powers of $H$. 
{Although the solution in eqn. (\ref{RadiativeSoln}) showed the presence
of a tail term as an integral over the history of the source, in the
final expression the field depends only on the properties of the source
at the retarded time $t_{\mathrm{ret}}$ which was defined through $(\eta
-~r)$ {except for the limiting value in the last line.}}

(2) Unlike the FNC chart, here the {\em tail contribution} has moments
without a time derivative which naively indicates that for `static'
sources, there could be a non-zero field. A coordinate invariant way of
specifying staticity of a source is to refer to the Killing parameter of
a stationary Killing vector in its vicinity, eg, $T\cdot\partial :=
-H(\eta\partial_{\eta} + x^i\partial_i) = \partial_t - Hx^i\partial_i$
(This also equals the $\partial_{\tau}$ of the FNC).  A static source
satisfies ${\cal L}_T T_{\mu\nu} = T\cdot\partial T_{\mu\nu} - 2H
T_{\mu\nu} = 0$.  Explicitly, ${\cal L}_T f^{\alpha}_{~\U{a}} = 0$ and
hence for a static source, ${\cal L}_T P_{\U{ab}} = 0$. Furthermore, the
Lie derivative of the moment variable $x^{\U{i}} = a x^i$ also vanishes
as does that of the volume element.  Hence, ${\cal L}_T Q_{\U{ij}} = 0$.
Since the moments are coordinate scalars and independent of spatial
coordinates, their Lie derivative is just $\partial_t$. Hence, for
static sources, $\partial_t Q_{\U{ij}} = 0 = \partial_t
\bar{Q}_{\U{ij}}$ (indeed {\em all} moments will be independent of $t$).
For constant moments, there is a cancellation between the terms in the
second line of equation (\ref{NonCovariantGaugeGWAmplitude}) and the
field vanishes. The boundary term at $t = - \infty$ is essential for
this cancellation.

However, the conservation equations for the zeroth and the first moment
are,
\begin{eqnarray}
\partial_tQ + H\bar{Q} ~ = ~ 0 & , & \partial_t^2Q_{\U{i}} + H
\partial_t \bar{Q}_{\U{i}} - H^2 \left(Q_{\U{i}} - \bar{Q}_{\U{i}}\right)
~ = ~ 0 
\end{eqnarray}
The equation for the zeroth moment can be derived directly from
(\ref{Delt00Eqn}). 
These again show that in a curved background, test matter cannot remain
static.

For very slowly varying moments, the sharp contribution is negligible
while the tail has a contribution, {\em not} falling off as $r^{-1}$. In
FNC, the slowly varying contribution is in the sharp term, but could not
be thought of as `radiation'. The absence of such a contribution in the
sharp term in eqn.  (\ref{NonCovariantGaugeGWAmplitude}), suggests that
the slowly varying sharp term of FNC (eqn. \ref{FNCSoln}), would not
survive as `radiation' at ${\cal J}^+$, though of course this cannot be
analysed within FNC chart. The surviving tail contribution has been
thought of as inducing a linear memory effect in
\cite{Chu-MemoryEffect}.

The contribution from $t = -\infty$ boundary, is in any case a constant
and does not play any role in any physical observables which typically
involve time derivatives. With this understood, we now suppress this
boundary contributions.

(3) To link with \cite{AshtekarBongaKesavanIII}, the final step involves
replacement of $\partial_t$ by the Lie derivative with respect to the
stationary Killing vector. Using ${\cal
L}_{T}Q_{\U{ij}}(t_{\mathrm{ret}}) = (\partial_t -
Hx^i\partial_i)(t_{\mathrm{ret}})\partial_{t_{\mathrm{ret}}}Q_{\U{ij}}(t_{\mathrm{ret}})
= \partial_{t_{\mathrm{ret}}}Q_{\U{ij}}(t_{\mathrm{ret}})$ and ${\cal
L}_T\delta_i^{~\U{i}} = T\cdot\partial\delta_i^{~\U{i}} -
H\delta_i^{~\U{i}}$, we get
\begin{equation}
{\cal L}_TQ_{ij} = {\cal
L}_T(\delta_i^{~\U{i}}\delta_j^{~\U{j}}Q_{\U{ij}}) = ({\cal
L}_T\delta_i^{~\U{i}}\delta_j^{~\U{j}})Q_{\U{ij}} +
\delta_i^{~\U{i}}\delta_j^{~\U{j}}\partial_{t_{\mathrm{ret}}}Q_{\U{ij}}
= \partial_{t_{\mathrm{ret}}}Q_{ij} - 2HQ_{ij}\ .
\end{equation}
This is where the constant triad plays a role, unlike in the FNC chart
where ${\cal L}_T = \partial_{\tau}$ on all tensors.  With these
translations, our solution in equation
(\ref{NonCovariantGaugeGWAmplitude1}) takes the forms,
\begin{eqnarray}
\chi_{ij}(t, r) & = & \frac{2}{r
\bar{a}}\delta_i^{~\U{i}}\delta_j^{~\U{j}} \left[\partial^2_tQ_{\U{ij}}
-~2H \partial_tQ_{\U{ij}} + H\partial_t\bar{Q}_{\U{ij}}\right](t)
\label{CosmoGWAmplitude0}\\
& & -2H\delta_i^{~\U{i}}\delta_j^{~\U{j}} \left[\partial^2_tQ_{\U{ij}} -
3H\partial_tQ_{\U{ij}} + 2H^2Q_{\U{ij}} + H\partial_t\bar{Q}_{\U{ij}} -
H^2\bar{Q}_{\U{ij}}\right](t), \nonumber \\
& = & \frac{2}{r \bar{a}}\left[{\cal L}^2_T{Q}_{ij} + 2H{\cal
L}_T{Q}_{ij} + H{\cal L}_T{\bar{Q}}_{ij} + 2H^2 \bar{Q}_{ij}\right]
\label{CosmoGWAmplitude} \\
& & - 2H\left[ {\cal L}^2_T{Q}_{ij} + H{\cal L}_T{Q}_{ij} + H{\cal
L}_T{\bar{Q}}_{ij} + H^2 \bar{Q}_{ij}\right] \nonumber 
\end{eqnarray}
The terms on the right hand side are evaluated at $t = t_{ret}$.  {Both
terms have the same derivatives of moments appearing in them and on
combining, lead to a coefficient of the form $((r\bar{a})^{-1} - H)$.
Thus in each order in $H$, the effect of the tail is to reduce the
amplitude.  The equation} (\ref{CosmoGWAmplitude}) matches with the
solution given by Ashtekar et al. \cite{AshtekarBongaKesavanIII} and the
$\Lambda \to 0$ limit of the solution goes over to the Minkowski
background solution.

In the next section we {define the gauge invariant deviation
scalar} to compare the computations done in the two charts.

\section{Tidal distortions}\label{TidalDistortion} 
The two solutions presented above were obtained in two different
gauges.  With a further choice of synchronous gauge, we could restrict
the solutions to the spatial components alone. While these conditions
fix the gauge completely, these spatial components  still have to
satisfy certain `spatial transversality and trace free' conditions.  The
solutions obtained above do {\em not} satisfy these conditions and hence
do not represent solutions of the original linearized Einstein equation.
Their dependence on the retarded time and the `radial' coordinate
however, offers an easy way to construct solutions which {\em do}
satisfy these spatial-TT conditions \cite{WillPoisson}. In flat
background, this is achieved by the {\em algebraic} TT-projector
(defined below) and the method extends to the de Sitter background as
well.

For $\chi_{ij}$, the spatial-TT conditions have the form:
$\partial^j\chi_{ji} = 0 = \delta^{ij}\chi_{ij}$ which have exactly the
same form as in the case of the Minkowski background. To deduce their
form for the $\tilde{h}^{ij}$ consider $\tilde{h}^{\mu\nu}$ satisfying
the TT gauge condition and the synchronous gauge condition:
$\bar{\nabla}_{\mu}\tilde{h}^{\mu\nu} = 0 =
\tilde{h}^{\mu\nu}\bar{g}_{\mu\nu}~,~ \tilde{h}^{\alpha 0} =
\Case{\Lambda}{3}(\tau - \tau_0)\tilde{h}^{\alpha i}\xi^j\delta_{ij}$.
These imply, $\tilde{h}^{\mu\nu} = h^{\mu\nu}$, 
\begin{equation}\label{FNCSynchronous}
h^{00} = o(\Lambda^2) ~ , ~ h^{0i} = \frac{\Lambda}{3}(\tau -
\tau_0)h^{ij}\xi^k\delta_{jk} ~~~;~~~ h^{ij}\delta_{ij} = -
\frac{\Lambda}{9}h^{ij}\xi_i\xi_j~. 
\end{equation}
Furthermore, 
\begin{eqnarray}
\bar{\nabla}_{\mu}h^{\mu\nu} & = &  \partial_{\mu}h^{\mu\nu} +
\bar{\Gamma}^{\mu}_{~\mu\lambda}h^{\lambda\nu} +
\bar{\Gamma}^{\nu}_{~\mu\lambda}h^{\mu\lambda}  ~~ = ~~ 0 \hspace{3.3cm}
\Rightarrow \nonumber \\
%
%
\partial_{\mu}h^{\mu\nu} & = & \frac{5\Lambda}{9}h^{i\nu}\xi_i +
\frac{2\Lambda}{3}\delta^{\nu}_0h^{0i}\xi_i +
\frac{2\Lambda}{9}\delta^{\nu}_i \left(h^{ij} -
h^{kl}\delta_{kl}\delta^{ij}\right)\xi_j \hspace{1.0cm} \Rightarrow \\
%
\partial_j{(h^{ji}\xi_i)} & = & - \frac{\Lambda}{3}(\tau -
\tau_0)\partial_{\tau}(h^{ij}{\xi_i}\xi_j) +
\frac{{5}\Lambda}{{9}}\xi_i\xi_jh^{ji}
\hspace{3.0cm}\mbox{($\nu = 0$)} \label{FNCSpatialTT0} \\
\partial_jh^{ji} & = & - \frac{\Lambda}{3}(\tau -
\tau_0)\partial_{\tau}(h^{ij}\xi_j) + \frac{4\Lambda}{9}\xi_jh^{ji}
\hspace{3.3cm} \mbox{($\nu = i$)} \label{FNCSpatialTT1}
%
%
%
\end{eqnarray}
{ Multiplying (\ref{FNCSpatialTT1}) by $\xi_i$, subtracting from
(\ref{FNCSpatialTT0}) and using (\ref{FNCSynchronous}) implies that
$h^{ij}\xi_i\xi_j = 0 = h^{ij}\delta_{ij}$. {The
$\xi_i\times$(\ref{FNCSpatialTT0}) then} implies that
$\partial_j(h^{ij}\xi_i) = 0$, satisfying eqn. (\ref{FNCSpatialTT0})
identically. {\em Provided $h^{ij}\xi_j = 0$,} the spatial
transversality condition, $\partial_jh^{ij} = 0$ will be satisfied.  The
TT-projector defined below will ensure $h^{ij}\xi_j = 0$ to the {\em
leading order in $s^{-1}$}. Hence the spatial transversality will also
hold for the projected $h^{ij}$ to the leading order in $s^{-1}$. The
projector being local (algebraic) in space-time while the spatial TT
conditions being non-local (differential), the projector ensures the
condition only for large $s$. Elsewhere, the condition must be satisfied
by adding solutions of the homogeneous wave equation. However, we need
the explicit forms of the solution only in the large $s$ regions for
which the projector suffices.}

As in the case of the Minkowski background, corresponding to each
spatial, unit vector $\hat{n}$, define the projectors,
\begin{equation}
P^i_{~j}(\hat{n}) ~ := ~ \delta^i_{~j} - \hat{n}^i\hat{n}_j ~~,~~
\Lambda^{ij}_{~~kl} ~ := ~ \frac{1}{2}\left(P^i_{~k}P^j_{~l} +
P^i_{~l}P^j_{~k} - P^{ij}P_{kl}\right) \ .
\end{equation}
Contraction with $\hat{n}$ gives zero and the trace of
$\Lambda$-projector in either pair of indices vanishes. From any
$X^{kl}$, the $\Lambda$-projector gives $X^{ij}_{TT} :=
\Lambda^{ij}_{~~kl}X^{kl}$ which is trace free and is transverse to the
unit vector $\hat{n}$. For the FNC fields we choose $\hat{n}^i :=
\vec{\xi}^i/s$ and for the conformal chart fields we choose $\hat{n}^i =
-H\eta\,\vec{x}^i/r$. When $\tilde{h}^{ij}_{TT}$ is substituted in the
eqn.  (\ref{FNCSpatialTT1}) the condition reduces to the `spatial
transversality', $\partial_j\tilde{h}^{ij}_{TT} = 0$.  The
$\chi^{TT}_{ij}$ also satisfies the same condition:
$\partial^j\chi_{ij}^{TT} = 0$.

Since $\hat{n}$ is a radial unit vector, It follows that,
$\partial_j\Lambda^{ij}_{~~kl} = \Case{1}{2r}(P^i_{~k}\hat{n}_l +
P^i_{~l}\hat{n}_k)$ which is down by a power of $r$ (or $s$ for FNC).
Therefore to the leading order in $r^{-1}$,
$\partial_j\tilde{h}^{ij}_{TT} =
\Lambda^{ij}_{~~kl}\partial_j\tilde{h}^{kl}$.

Noting that the retarded solutions have a form $\sim f^{ij}(\tau -
s)/s$, we get,
\[
\partial_j\left[\frac{f^{ij}(\tau - s)}{s}\right]  =  - \frac{1}{s^2}
\xi_j\left(\partial_{\tau}{f}^{ij} + s^{-1}f^{ij}\right) \approx -
\hat{\xi}_j\partial_{\tau}\left[\frac{f^{ij}(\tau - s)}{s}\right] +
o(\frac{1}{s^2}) ~,~ \hat{\xi}_j := s^{-1}\xi_j \ .
\]

It follows immediately that to the leading order in $s^{-1}$ (or
$r^{-1}$), $\partial_j\tilde{h}^{ij}_{TT} \approx - \partial_{\tau}
(\hat{\xi}_j\tilde{h}^{ij}_{TT}) = 0$ (and likewise
$\partial^j\chi^{TT}_{ij} = 0$). Note that although to begin with, the
spatial TT conditions in FNC look different from those of the conformal
chart, {they have the same form after the corresponding
$\Lambda$-projections.} {Thus, for the $\Lambda-$projected $h^{ij}$
too,} $h^{\alpha 0} = 0, \forall \alpha$.  There is no plane wave
assumption or spatial Fourier transform needed for this projection.  Of
course, the $\Lambda-$projector only ensures that the gauge conditions
are satisfied to the {\em leading order} in $r^{-1} (s^{-1})$.  These
$\Lambda$-projected field represent physical perturbations and gauge
invariant observables of interest can be computed using these.

From now on, the solutions will be in the synchronous gauge and with TT
projection implicit: $\tilde{h}^{\tau\beta} = 0\ , \tilde{h}^{ij}
\leftrightarrow \tilde{h}^{ij}_{TT}$ and $\chi_{\eta\alpha} = 0\ ,
\chi_{ij} \leftrightarrow \chi^{TT}_{ij}$. In particular $\tilde{h}^{ij}
= h^{ij}$. 

As an illustration, we consider the deviation induced in the nearby
geodesics, as tracked by a freely falling observer. Thus we consider a
congruence of time like geodesics of the {\em background space-time} and
consider the tidal effects of a transient gravitational wave. 

We begin with the observation that for {\em all} space-times satisfying
$R_{\mu\nu} = \Lambda g_{\mu\nu}$ (which include the de Sitter
background as well as its linearized perturbations {in source free
regions}) and for vectors $u, Z, Z'$ satisfying $u\cdot Z = u\cdot Z' =
Z\cdot Z' = 0$, the definition of the Weyl tensor implies,
\begin{eqnarray}
C_{\alpha\beta\mu\nu} - R_{\alpha\beta\mu\nu} & = & -
\frac{\Lambda}{3}(g_{\alpha\mu}g_{\beta\nu} - g_{\alpha\nu}g_{\beta\mu})
~~~ \Rightarrow \\
R_{\alpha\beta\mu\nu}Z'^{\alpha}u^{\beta}Z^{\mu}u^{\nu} & = &
C_{\alpha\beta\mu\nu}Z'^{\alpha}u^{\beta}Z^{\mu}u^{\nu} +
\frac{\Lambda}{3}\left\{(u\cdot u)(Z'\cdot Z) - (u\cdot Z')(u\cdot Z)
\right\} \\
\therefore {D(u, Z, Z')} & := & -
R_{\alpha\beta\mu\nu}Z'^{\alpha}u^{\beta}Z^{\mu}u^{\nu} ~ = ~ -
C_{\alpha\beta\mu\nu}Z'^{\alpha}u^{\beta}Z^{\mu}u^{\nu}
\label{WeylComponents}
%
\end{eqnarray}
The last equation shows the gauge invariance of $D({u}, Z', Z)$.
This is because the gauge transform of the Weyl tensor for the
background is zero and the gauge transform of the $Z' u Z u$ factor (it
depends on the perturbation through the normalizations) does not
contribute since the Weyl tensor of the de Sitter background itself is
zero. {Notice that $D(u, Z, Z')$ is symmetric in $Z \leftrightarrow
Z'$ and is the component of acceleration of one deviation vector $Z$,
along another orthogonal deviation vector.} 

A suitably chosen congruence of time-like geodesics,
{$u^{\alpha}\partial_{\alpha}$,} provides {a} required pair of
orthogonal deviation vectors for the gauge invariant observable $D(u,
Z', Z)$ which we now refer to as {\em deviation scalar}. {Since
deviation vectors are always defined with respect to a geodesic
congruence, we leave the argument $u$ implicit and restore it in the
final expressions. The deviation scalar} is related to Weyl scalars as
noted in \cite{Bishop}.  We compute this for the $\Lambda-$projected
solutions given in (\ref{FNCSoln}, \ref{NonCovariantGaugeGWAmplitude1}).
Note that the dot products in the above equations, involve the perturbed
metric, $(\bar{g} + h)_{\mu\nu}$. For the explicit choices that we will
make below, we denote the observer and the deviation vectors in the
form: $u = \bar{u} + \delta u, Z = \bar{Z} + \delta Z, Z' = \bar{Z}' +
\delta Z'$ with the `barred' quantities normalised using the background
metric while the `delta' quantities are treated as of the same order as
the perturbed field. Thus,
\begin{eqnarray}
\bar{u}\cdot\bar{\nabla}\bar{u}^{\alpha} = 0 & , & \bar{u}\cdot\partial
\bar{Z}^{\alpha} = \bar{Z}\cdot\partial \bar{u}^{\alpha} ~~,~~
\bar{u}\cdot\partial \bar{Z}^{'\alpha} = \bar{Z'}\cdot\partial
\bar{u}^{\alpha} ~; \nonumber \\
\bar{u}\cdot\bar{u} = -1 & , & \bar{u}\cdot \bar{Z} = \bar{u}\cdot
\bar{Z}' = \bar{Z}'\cdot \bar{Z} = 0 ~.
\end{eqnarray}
The `delta' quantities have to satisfy conditions so that the full
quantities satisfy the requisite orthogonality relations with respect to
the perturbed metric.

The deviation scalar is then given by, 
\begin{eqnarray}
-~D(Z', Z) & := & \left(\bar{g}_{\alpha\beta} + h_{\alpha\beta}\right)
\left( \bar{R}^{\alpha}_{~\lambda\mu\nu} +
R^{(1)\alpha}_{~~~~\lambda\mu\nu}\right)
\left(\bar{Z}^{'\beta} \bar{u}^{\lambda}\bar{Z}^{\mu}\bar{u}^{\nu} +
\delta(Z'^{\beta} u^{\lambda}Z^{\mu}u^{\nu}) \right) \nonumber \\
& = & \bar{g}_{\alpha\beta}\bar{R}^{\alpha}_{~\lambda\mu\nu}
\bar{Z}'^{\beta} \bar{u}^{\lambda}\bar{Z}^{\mu}\bar{u}^{\nu} +
\bar{g}_{\alpha\beta}R^{(1)\alpha}_{~~~~\lambda\mu\nu}\ \bar{Z}'^{\beta}
\bar{u}^{\lambda}\bar{Z}^{\mu}\bar{u}^{\nu} \nonumber \\
& & ~~ + \bar{g}_{\alpha\beta}\bar{R}^{\alpha}_{~\lambda\mu\nu}\
\delta\left(Z'^{\beta} u^{\lambda}Z^{\mu}u^{\nu}\right) +
h_{\alpha\beta}\bar{R}^{\alpha}_{~\lambda\mu\nu} \bar{Z}'^{\beta}
\bar{u}^{\lambda}\bar{Z}^{\mu}\bar{u}^{\nu} \label{DevScalarEqn0}\\
\therefore D(Z', Z) & = & - R^{(1)\alpha}_{~~~~\lambda\mu\nu}\
\bar{Z}'_{\alpha} \bar{u}^{\lambda}\bar{Z}^{\mu}\bar{u}^{\nu} +
\frac{\Lambda}{3}\bar{g}_{\alpha\beta} (\bar{Z}'^{\beta}\delta
Z^{\alpha} + \delta Z'^{\beta}\bar{Z}^{\alpha}) +
\frac{\Lambda}{3}h_{\alpha\beta}\bar{Z}'^{\alpha}\bar{Z}^{\beta}
\nonumber \\
& = & - R^{(1)\alpha}_{~~~~\lambda\mu\nu}\ \bar{Z}'_{\alpha}
\bar{u}^{\lambda}\bar{Z}^{\mu}\bar{u}^{\nu} +
\frac{\Lambda}{3}\delta\left(g_{\alpha\beta}Z'^{\alpha}Z^{\beta}\right)
\label{DevScalarEqn}
\end{eqnarray}
%
%
Here, $R^{(1)}$ refers to the Riemann tensor linear in $h_{\mu\nu}$. In
eqn. (\ref{DevScalarEqn0}), the first term vanishes thanks to the
properties of the barred quantities, while in the third term, only one
factor has a delta-quantity.  The only contributions that survive in the
third and the fourth terms are the ones with $\bar{u}^2 = -1$. These
terms combine (note the full metric in the last term) and
eqn. (\ref{DevScalarEqn}) reflects this. Next,
\begin{eqnarray}
R^{(1)\alpha}_{~~~~\lambda\mu\nu} & = &
\bar{\nabla}_{\mu}\Gamma^{(1)\alpha}_{~~~~\nu\lambda} -
\bar{\nabla}_{\nu}\Gamma^{(1)\alpha}_{~~~~\mu\lambda} \\
\Gamma^{(1)\alpha}_{~~\nu\lambda} & = &
\frac{1}{2}\bar{g}^{\alpha\beta}\left(
\bar{\nabla}_{\lambda}h_{\beta\nu} + \bar{\nabla}_{\nu}h_{\beta\lambda}
-~\bar{\nabla}_{\beta}h_{\nu\lambda}\right) ; \\
\therefore D(u, Z', Z) & = &
-\frac{1}{2}\left[\bar{Z}^{'\alpha}\bar{u}^{\lambda}\bar{Z}^{\mu}\bar{u}^{\nu}\left(
\bar{\nabla}_{\mu}\bar{\nabla}_{\lambda}h_{\alpha\nu} -
\bar{\nabla}_{\mu}\bar{\nabla}_{\alpha}h_{\nu\lambda} -
\bar{\nabla}_{\nu}\bar{\nabla}_{\lambda}h_{\alpha\mu} +
\bar{\nabla}_{\nu}\bar{\nabla}_{\alpha}h_{\mu\lambda}
\right.\right.\nonumber \\
& & \left.\left.~~~ + [\bar{\nabla}_{\mu},
\bar{\nabla}_{\nu}]h_{\alpha\lambda}\right)\right] +
\frac{\Lambda}{3}\delta\left(g_{\alpha\beta}Z'^{\alpha}Z^{\beta}\right)
\ . \label{DeviationScalarEqn}
\end{eqnarray}
Evaluating the commutator in the last term within the square brackets,
we write it as,
$\Case{\Lambda}{3}h_{\alpha\beta}\bar{Z}^{'\alpha}\bar{Z}^{\beta}$.  To
proceed further, we need to make choice of the congruence, the deviation
vectors and the delta-quantities. This is done in the respective charts.

A natural class of time-like geodesics of the background geometry, is
suggested in the conformal chart. From the appendix
eqn (\ref{SpatialGeodesicEqn}), we know that the curves $x^i = x^i_0$ are
time-like geodesics. The corresponding, normalised velocity is given by
$\bar{u}^{\alpha}(\eta, x^i) := -H\eta(1, \vec{0})$. The same family of
geodesics is given in FNC as, $\bar{u}^{\alpha} = (1 + \Lambda s^2/3,
\sqrt{\Lambda/3}\ \vec{\xi}\ )$.  From now on, we will use $\Lambda/3 =:
H^2$ for ease of comparison. 

{\em FNC chart of the Static patch}: 

From the explicit choice of the freely falling observer, we get the
following consequences:
\begin{eqnarray}
\bar{u}\cdot\bar{Z} = 0 & \Rightarrow & \bar{Z}^0 = H
\vec{\xi}\cdot\vec{\bar{Z}} ~~,~~ \bar{u}\cdot\bar{Z}' = 0 ~ \Rightarrow
~ \bar{Z}^{'0} = H \vec{\xi}\cdot\vec{\bar{Z}}' \ ;\\
\vec{\xi}\cdot\vec{\bar{Z}} = 0 = \vec{\xi}\cdot\vec{\bar{Z}}' &
\Rightarrow & \bar{Z}^0 = 0 = \bar{Z}^{'0} \ ; \\
\bar{Z}\cdot \bar{Z}' = 0 & \Rightarrow &
\vec{\bar{Z}}'\cdot\vec{\bar{Z}} = 0 \ ; \\
\bar{u}\cdot\partial\bar{Z}^{\alpha} =
\bar{Z}\cdot\partial\bar{u}^{\alpha} & \Rightarrow &
\bar{u}\cdot\partial \bar{Z}^i ~ = ~ H \bar{Z}^i ~~,~~ \nonumber \\
\bar{u}\cdot\partial\bar{Z}^{'\alpha} =
\bar{Z}'\cdot\partial\bar{u}^{\alpha} & \Rightarrow &
\bar{u}\cdot\partial \bar{Z}^{'i} ~ = ~ H \bar{Z}^{'i} \ ; \\
\partial_{\tau}\bar{Z}^i & = & H(\bar{Z}^i -
\vec{\xi}\cdot\vec{\partial}\bar{Z}^i) ~~,~~ \partial_{\tau}\bar{Z}^{'i}
~ = ~ H(\bar{Z}^{'i} - \vec{\xi}\cdot\vec{\partial}\bar{Z}^{'i}) \ ;
\end{eqnarray}
{In the second equation above, we have made the further choice
namely, the spatial parts of $\bar{Z}, \bar{Z}'$ are orthogonal to the
radial direction $\vec{\xi}$ as well.}

The idea is to bring the deviation vectors across the derivatives. Using
the 
properties above and $\tilde{h}_{ij}\bar{u}^i = 0$ which holds thanks to
the TT-projection, the equation (\ref{DeviationScalarEqn}) gives,
\begin{equation} \label{TidalScalarFNC}
D(Z', Z) - \frac{\Lambda}{3}\delta
\left(g_{\alpha\beta}Z'^{\alpha}Z^{\beta}\right) ~ = ~
\left[\frac{1}{2}(\bar{u}\cdot\bar{\nabla})^2 -
H(\bar{u}\cdot\bar{\nabla}) \right]
(\tilde{h}^{TT}_{ij}\bar{Z}^{'i}\bar{Z}^j)\ .
\end{equation}
The second term on the left hand side of the above equation vanishes. 

To see this, we collect the equations satisfied by the
$\delta-$quantities.
\begin{eqnarray}
g_{\alpha\beta}u^{\alpha}u^{\beta} = - 1 & \Rightarrow &
\bar{u}_{\alpha}\delta u^{\alpha} = 0 \label{OrthogonalityEqn}\\
g_{\alpha\beta}u^{\alpha}Z^{\beta} = 0 & \Rightarrow &
\bar{u}_{\alpha}\delta Z^{\alpha} + \bar{Z}_i\delta u^{i} = 0 \nonumber \\
g_{\alpha\beta}u^{\alpha}Z^{'\beta} = 0 & \Rightarrow &
\bar{u}_{\alpha}\delta Z^{'\alpha} + \bar{Z}'_i\delta u^{i} = 0 \nonumber \\
u\cdot\nabla u^{\alpha} = 0 & \Rightarrow &
\bar{u}\cdot\bar{\nabla}\delta u^{\alpha} = - \delta
u\cdot\bar{\nabla}\bar{u}^{\alpha} \label{EvolutionEqn}\\
u\cdot\partial Z^{\alpha} - Z\cdot\partial u^{\alpha} = 0 & \Rightarrow
& \bar{u}\cdot\bar{\nabla}\delta {Z}^{\alpha} =
\bar{Z}^i\bar{\nabla}_i\delta u^{\alpha} + \delta
Z\cdot\bar{\nabla}\bar{u}^{\alpha} - \delta
u\cdot\bar{\nabla}\bar{Z}^{\alpha} \nonumber \\
u\cdot\partial Z^{'\alpha} - Z'\cdot\partial u^{\alpha} = 0 &
\Rightarrow & \bar{u}\cdot\bar{\nabla}\delta {Z'}^{\alpha} =
\bar{Z}^{'i}\bar{\nabla}_i\delta u^{\alpha} + \delta
Z'\cdot\bar{\nabla}\bar{u}^{\alpha} - \delta
u\cdot\bar{\nabla}\bar{Z}^{'\alpha} \nonumber \\
g_{\alpha\beta}Z^{\alpha}Z^{'\beta} = 0 & \Rightarrow &
\bar{Z}_{\alpha}\delta Z^{'\alpha} + \bar{Z}'_i\delta Z^{i} +
\tilde{h}^{TT}_{ij}\bar{Z}^{'i}\bar{Z}^j = 0 \label{AuxilliaryEqn}
%
%
%
\end{eqnarray}

The equations (\ref{OrthogonalityEqn}) serve to give the zeroth
components of the $\delta-$vectors in terms of their spatial components.
The equations (\ref{EvolutionEqn}) are evolution equations along the
geodesic for the $\delta-$vectors and preserve the previous three
equations. The last equation (\ref{AuxilliaryEqn}), is needed for the
gauge invariance of the deviation scalar. The spatial components of
$\delta-$vectors are still free.  Demanding that the
(\ref{AuxilliaryEqn}) is preserved along the observer geodesic leads to, 
\begin{eqnarray} \label{DeltaUEqnFNC}
(\bar{Z}'_i\bar{Z}^j + \bar{Z}_i\bar{Z}^{'j})\bar{\nabla}_j\delta u^i ~
= ~ - (\bar{u}\cdot\bar{\nabla} - 2
H)(\tilde{h}^{TT}_{ij}\bar{Z}^{'i}\bar{Z}^j)
\end{eqnarray}
Here, we have used the evolution equations for $\delta Z, \delta Z'$,
equation (\ref{AuxilliaryEqn}), as well as $\bar{\nabla}_j\bar{u}^i =
H\delta^i_{~j} + o(H^3)$. This equation together with the evolution
equation for $\delta u^i$, can be taken to restrict $\delta u^i$ and we
are still left free with the $\delta Z^i, \delta Z^{'i}$ subject only to
the (\ref{AuxilliaryEqn}). This equation precisely sets the second term
on the left hand side of eq. (\ref{TidalScalarFNC}) to zero.

Thus we obtain the deviation scalar as a simple expression,
\begin{eqnarray}
D(u, Z', Z) & = & \left[\frac{1}{2}(\bar{u}\cdot\bar{\nabla})^2 -
H(\bar{u}\cdot\bar{\nabla})\right] Q ~~~,~~~ Q :=
(\tilde{h}^{TT}_{ij}\bar{Z}^{'i}\bar{Z}^j)\ ~~~ \mbox{with,} \\
\bar{u}\cdot\bar{\nabla} Q & = & \bar{u}\cdot\partial Q ~ = ~ \left((1 +
H^2s^2)\partial_{\tau} + H\xi^i\partial_i\right)Q \ . \nonumber
\end{eqnarray}

For subsequent comparison, it is more convenient to take the deviation
vectors across the derivatives, using $\bar{u}\cdot\bar{\nabla}\bar{Z}^i
= H\bar{Z}^i$ etc. The deviation scalar is then given by,
\begin{eqnarray}
D(u, Z', Z) & = &
\bar{Z}^{'i}\bar{Z}^j\left[\frac{1}{2}(\bar{u}\cdot\bar{\nabla})^2 +
H(\bar{u}\cdot\bar{\nabla}) \right]\tilde{h}^{TT}_{ij}\ ~~~ \mbox{with,}
\\
\bar{u}\cdot\bar{\nabla}\tilde{h}^{TT}_{ij}  & = & \bar{u}\cdot\partial
\tilde{h}^{TT}_{ij} + o(H^3) \ .
\end{eqnarray}
Substituting the solution (\ref{FNCSoln}) gives, 
\begin{eqnarray}\label{DeviationScalarFNC}
D(u, Z', Z) & = & \frac{1}{s}\left[\left(1 - 2Hs +
\frac{7}{2}H^2s^2\right)\partial^4_{\tau}{\cal M}_{ij}^{TT} - H^2 s
\partial^3_{\tau}{\cal M}_{ij}^{TT} - H^2\partial^2_{\tau}{\cal
M}_{ij}^{TT} \right. \nonumber \\
& & \left. \hspace{0.5cm} - \frac{3H^2}{4}\partial_{\tau}^4{\cal
M}^{TT}_{ijkl}\delta^{kl}\right]\bar{Z}^{'i}\bar{Z}^j
\end{eqnarray}
The $\tau$ derivatives are evaluated at the retarded time, $(\tau -
\bar{s}(s))$, defined in eqn. (\ref{RetTauDefn}).

{\em The Conformal chart of the Poincare patch}: For the solution
in the generalized transverse gauge, the full metric has the form
$g_{\mu\nu} = \Omega^2(\eta_{\mu\nu} + \chi_{\mu\nu}), ~\Omega^2 =
3\Lambda^{-1}\eta^{-2} = H^{-2}\eta^{-2}$. We can then use the Weyl
transformation property of the Riemann tensor and obtain the full
curvature in terms of the curvature of $(\eta + \chi)$ metric plus extra
terms depending on {\em derivatives} of $\ln(\Omega)$.  From these
derivatives, $\Lambda$ drops out and the full curvature (and hence the
relative acceleration) is completely independent of $\Lambda$.
Explicitly, 
\begin{eqnarray}
R_{\alpha\lambda\mu\nu}[\Omega^2(\eta + \chi)] & = & \Omega^2\left[
R_{\alpha\lambda\mu\nu}[\eta + \chi] 
+ \frac{1}{\eta^2}\left\{ \hat{g}_{\alpha\mu} \hat{g}_{\nu\lambda}
-~\hat{g}_{\alpha\nu} \hat{g}_{\mu\lambda}\right\}\right. \\
& & \left. \hspace{0.0cm} +
\frac{1}{\eta}\left\{\hat{g}_{\alpha\nu}\hat{\Gamma}^0_{~\mu\lambda}
-~\hat{g}_{\alpha\mu}\hat{\Gamma}^0_{~\nu\lambda} +
\hat{g}_{\mu\lambda}\hat{g}_{\alpha\beta}\hat{\Gamma}^{\beta}_{~0\nu} -
\hat{g}_{\nu\lambda}\hat{g}_{\alpha\beta}\hat{\Gamma}^{\beta}_{~0\mu}
\right\} \right] 
\hspace{0.0cm}\mbox{where,} \nonumber \\
\hat{g}_{\mu\nu} & = & \eta_{\mu\nu} + \chi_{\mu\nu} ~~~,~~~
\left.\hat{\Gamma}^{\alpha}_{~\mu\nu}\right|_{o(\chi)} ~ = ~
\frac{1}{2}\left(\partial_{\nu}\chi^{\alpha}_{~\mu} +
\partial_{\mu}\chi^{\alpha}_{~\nu} -
\partial^{\alpha}\chi_{\mu\nu}\right) \nonumber
\end{eqnarray}

The definition of the deviation scalar and its invariance remains the
same. We also choose the same geodesic congruence in the background
space-time so that $u^{\alpha} = -H\eta\delta^{\alpha}_{~0}$. As before
we choose two mutually orthogonal deviation vectors, $Z, Z'$ and write
$D(Z', Z) = - R_{\alpha\beta\mu\nu}Z^{'\alpha}u^{\beta}Z^{\mu}u^{\nu}$.
Using the Weyl transformation given above, we write,
\begin{eqnarray}\label{ScaledDeviationScalar}
\hat{D}(\hat{Z}', \hat{Z}) & = & \left[ R_{\alpha\lambda\mu\nu}[\hat{g}] 
+ \frac{1}{\eta^2}\left\{ \hat{g}_{\alpha\mu} \hat{g}_{\nu\lambda}
-~\hat{g}_{\alpha\nu} \hat{g}_{\mu\lambda}\right\}\right. \\
& & \left. \hspace{0.0cm} +
\frac{1}{\eta}\left\{\hat{g}_{\alpha\nu}\hat{\Gamma}^0_{~\mu\lambda}
-~\hat{g}_{\alpha\mu}\hat{\Gamma}^0_{~\nu\lambda} +
\hat{g}_{\mu\lambda}\hat{g}_{\alpha\beta}\hat{\Gamma}^{\beta}_{~0\nu} -
\hat{g}_{\nu\lambda}\hat{g}_{\alpha\beta}\hat{\Gamma}^{\beta}_{~0\mu}
\right\} \right]
\hat{Z}^{'\alpha}\hat{u}^{\lambda}\hat{Z}^{\mu}\hat{u}^{\nu}\ ,
\nonumber
\end{eqnarray}
where we have defined new {\em scaled variables} as: $u^{\alpha} :=
|\Omega|^{-1}\hat{u}^{\alpha}\ ,\ Z^{\alpha} := |\Omega|^{-1}
\hat{Z}^{\alpha}\ ,\  Z^{'\alpha} := |\Omega|^{-1} \hat{Z}^{'\alpha}$
and $D(Z', Z) := \Omega^{-2} \hat{D}(\hat{Z}', \hat{Z})$. This removes
all the explicit factors of $\Omega^{2}$ and we get an expression for
the scaled deviation scalar, defined by perturbations about {\em
Minkowski} background, with explicit additional terms. 

{\em For notational simplicity, we will suppress the `hat's in the
following, and restore them in the final equation.} The background
quantities, denoted by `overbars' refer to the Minkowski metric and the
corresponding $\delta-$quantities are treated as the of the same order
as the perturbation $\chi^{TT}_{ij}$.  In particular, $\bar{u}^{\alpha}
= \delta^{\alpha}_{~0}\ ,\ \bar{u}\cdot\partial\bar{u}^{\alpha} = 0\ ,\
\bar{u}\cdot\partial\bar{Z}^{\alpha} =
\bar{Z}\cdot\partial\bar{u}^{\alpha}$ and similarly for
$\bar{Z}^{'\alpha}$.  Proceeding exactly as before, we deduce:
\begin{eqnarray}
\bar{u}^i ~=~ \bar{Z^0} = \bar{Z}^{'0} =
\bar{Z}^i\bar{Z}^{'j}\delta_{ij} = 0 & , & \partial_{\eta}\bar{Z}^i =
\partial_{\eta}\bar{Z}^{'i} = 0 \ ; \\
\delta u^0 = 0 = \delta Z^0 - \bar{Z}_i\delta u^i = \delta Z^{'0} -
\bar{Z}'_i\delta u^i & , & \bar{Z}'_i\delta Z^i + \bar{Z}_i\delta Z^{'i}
+ \chi_{ij}\bar{Z}^{'i}\bar{Z}^j = 0 \ ; \label{DeltaNormEqn}\\
\partial_{\eta}\delta Z^{\alpha} = \bar{Z}^i\partial_i\delta u^{\alpha} -
\delta u^i\partial_i\bar{Z}^i  & , &  \partial_{\eta}\delta Z^{'\alpha} =
\bar{Z}^{'i}\partial_i\delta u^{\alpha} - \delta
u^i\partial_i\bar{Z}^{'i} \ ; \\
(\bar{Z}'_i\bar{Z}^j + \bar{Z}_i\bar{Z}^{'j})\partial_j\delta u^i & = &
- \partial_{\eta}(\chi_{ij}\bar{Z}^{'i}\bar{Z}^j) ~~ , ~~
\partial_{\eta}\delta u^i = 0 \ . \label{DeltaUEqnConf}
\end{eqnarray}
As before, demanding preservation of the last of the normalization
conditions in (\ref{DeltaNormEqn}) under $\eta$ evolution, gives
conditions on $\delta u^i$ given in equation (\ref{DeltaUEqnConf}).
These are used in simplifying eqn. (\ref{ScaledDeviationScalar}). The
$\eta^{-2}$ term of this equation vanishes as before while the
$\eta^{-1}$ coefficient gives only one contribution. In the first term,
$R(\hat{g})$ gets replaced by $R^{(1)}$ which is linear in
$\chi^{TT}_{ij}$. This leads to (restoring the `hats'),
\begin{equation}
\hat{D}(\hat{Z}', \hat{Z}) ~ = ~
\frac{1}{2}\left(\partial^2_{\eta}\chi^{TT}_{ij} -
\frac{1}{\eta}\partial_{\eta}\chi^{TT}_{ij}\right)
\bar{\hat{Z}}^{'i}\bar{\hat{Z}}^j \ .
%
\end{equation}

Noting that $\chi_{ij}$ is a function of $\eta$ only through
$\eta_{\mathrm{ret}} = \eta - r$, we can replace $\partial_{\eta}$ by
$\partial_{\eta_{\mathrm{ret}}} =: \partial_{\bar{\eta}}$. Going to the
cosmological chart via the definitions $\eta = - H^{-1}e^{-Ht} =:
-H^{-1}a^{-1}$ and $\bar{\eta} := \eta - r := - H^{-1}\bar{a}^{-1}$
which defines the retarded time $\bar{t}$ through $\bar{a} =
a(\bar{t})$, we replace $\bar{\partial}_{\eta} =
\bar{a}\partial_{\bar{t}}$. This leads to,
\[
\hat{D}(\hat{Z}', \hat{Z}) ~ = ~ \frac{
\bar{\hat{Z}}^{'i}\bar{\hat{Z}}^j}{2} \
\bar{a}^2\left(\partial_{\bar{t}}^2 + H\left(1 +
\frac{a}{\bar{a}}\right)\partial_{\bar{t}}\right)\chi^{TT}_{ij} ~~~
\mbox{with $~\chi_{ij}~$ from equation (\ref{CosmoGWAmplitude0})} \ .
\]

To express the deviation scalar in terms of the Killing time $\tau$, we
observe that on scalars, ${\cal L}_Tf = T\cdot\partial f$ while on
tensorial functions of the retarded time, 
\[
{\cal L}_TQ_{ij}(\bar{t}) = \left((\partial_t -
Hx^i\partial_i)(\bar{t})\right)(\partial_{\bar{t}} Q_{ij}(\bar{t})) -~2H
Q_{ij}(\bar{t})\ , ~~~ \mbox{with}~~~ (\partial_t -
Hx^i\partial_i)(\bar{t}) = 1.
\]

After a straightforward computation, we get,
\begin{eqnarray}\label{DeviationScalarConf}
\hat{D}(u, \hat{Z}', \hat{Z}) & = &
\left(\bar{\hat{Z}}^{'i}\bar{\hat{Z}}^j\right)\left(\frac{\bar{a}^2}{ra}\right)
\Big[{\cal L}_T^4Q_{ij} +~6 H {\cal L}_T^3Q_{ij} + 11H^2 {\cal
L}_T^2Q_{ij} + 6 H^3 {\cal L}_TQ_{ij}  \nonumber \\
& & \hspace{2.5cm}  + H{\cal L}_T^3\bar{Q}_{ij} + 6 H^2 {\cal
L}_T^2\bar{Q}_{ij} + 11H^3 {\cal
L}_T\bar{Q}_{ij} + 6 H^4 \bar{Q}_{ij} \Big] . 
\end{eqnarray}

To compare the deviation scalars computed above, we need to ensure that
we use the `same' deviation vectors. Since the same observer is used,
the deviation vectors are defined the same way with the only exception
of their normalization. So let us use\footnote{We now suppress the
overbars on the deviation vectors to avoid cluttering.} {\em normalised
deviation vectors}: $Z^i := \gamma \hat{Z}^i$ where
$\hat{Z}^i\hat{Z}^j\delta_{ij} = 1$. Then $Z^2 = 1$ determines $\gamma$.
In the FNC, $\gamma = (1 + H^2s^2/6)$ whereas in the conformal chart
$\gamma = |\Omega|^{-1}$.  Thus, in the conformal chart, the hatted
deviation vectors are already normalized. In the FNC, we need to replace
the deviation vectors by $(1 + H^2s^2/6)\times\hat{Z}$ and in the
conformal chart, we write $\hat{D}(\hat{Z}', \hat{Z}) = \Omega^2D(Z',
Z)$. In the conformal chart, we retain terms up to order $H^2$ only and
since the FNC calculation uses traceless stress tensor, we take
$\bar{Q}$ moments to equal the $Q$ moments. The two expressions are
given below. (Recall that in FNC, ${\cal L}_T$ on all tensors reduces to
$\partial_{\tau}$.)
\begin{eqnarray}
D^{FNC}(u, Z', Z) & = & \frac{1}{s}\left(1 +
\frac{H^2s^2}{3}\right)\left[\left(1 - 2Hs +
\frac{7}{2}H^2s^2\right)\partial^4_{\tau}{\cal M}_{ij}^{TT} - H^2 s
\partial^3_{\tau}{\cal M}_{ij}^{TT} \right. \nonumber \\
& & \left. \hspace{1.0cm} - H^2\partial^2_{\tau}{\cal M}_{ij}^{TT} -
\frac{3H^2}{4}\partial_{\tau}^4{\cal
M}^{TT}_{ijkl}\delta^{kl}\right]\hat{Z}^{'i}\hat{Z}^j \nonumber \\
& = & \frac{1}{s}\left(1 - 2Hs +
\frac{23}{6}H^2s^2\right)\Big[\partial^4_{\tau}{\cal M}^{TT}_{ij} -
H^2s\partial^3_{\tau}{\cal M}^{TT}_{ij}  \nonumber \\
& & \left. \hspace{1.0cm} - H^2\partial^2_{\tau}{\cal M}_{ij}^{TT} -
\frac{3H^2}{4}\partial_{\tau}^4{\cal
M}^{TT}_{ijkl}\delta^{kl}\right]\hat{Z}^{'i}\hat{Z}^j
\label{FNCScalarFinal}\\
\left.D^{Conf}(u, Z', Z)\right|_{o(H^2)} & = &
\left.\left(\frac{\bar{a}^2}{a^2}\frac{1}{ra}\right)\right|_{o(H^2)}
\left[{\cal L}_T^4Q_{ij} +~7 H {\cal L}_T^3Q_{ij} +~17H^2 {\cal
L}_T^2Q_{ij} \right. \nonumber \\ 
& & \left. \hspace{3.0cm} + 17 H^3{\cal L}_TQ_{ij} + 6
H^4Q_{ij}\right]\hat{Z}^{'i}\hat{Z}^j 
\nonumber \\
& = & \frac{1}{s}\left(1 - 2Hs + \frac{19}{6}H^2s^2\right)\Big[{\cal
L}_T^4Q^{TT}_{ij} + 7H{\cal L}_T^3Q^{TT}_{ij} 
\nonumber \\ 
& &  \hspace{1.0cm} + 17 H^2{\cal L}_T^2Q^{TT}_{ij} \Big]
\hat{Z}^{'i}\hat{Z}^j \label{ConfScalarFinal}
\end{eqnarray}

Equations (\ref{DeviationScalarFNC}) and (\ref{DeviationScalarConf})
give the deviation scalars in the two charts. The comparable expressions
are given in (\ref{FNCScalarFinal}, \ref{ConfScalarFinal}). {These are
obtained for the specific choice of the congruence of the de Sitter
background: $\bar{u}^{\alpha}(\eta, x^i) := - H\eta(1, \vec{0})$.}

We have obtained two different looking expressions for the same, gauge
invariant deviation scalar. The difference can be attributed to the
definition of moments.  They have been defined on two different spatial
hypersurfaces - the $\tau = $constant in FNC and the $\eta =$ constant
in the conformal chart. In the conformal chart solution there is no
truncation of powers of $H$ (in the leading $`r'$ approximation) and it
includes the contribution of both the sharp and the tail terms. By
contrast, the FNC chart computation is obtained as an expansion in $H$
only up to the quadratic order.  Furthermore, it includes only the
contribution of the sharp term. While it is possible to relate the frame
components of the stress tensor in the two charts, the relation among
the moments is non-trivial and is not obtained here.

{ We have defined a gauge invariant quantity and illustrated how to
compute it. It depends on a time-like geodesic congruence and two
mutually orthogonal deviation vectors. At the linearised level, it also
depends on the $\hat{n}$ direction used in the TT projection. What
information about the wave does it contain? To see this, consider the
simpler case of Minkowski background, choose the congruence so that
$\bar{u}^{\alpha} = (1, \vec{0})$. It follows that at the linearised
level, the quantity
\begin{equation}
A_{\alpha\beta}(\eta_{\mu\nu} + h_{\mu\nu}) := -
R_{\alpha\mu\beta\nu}(\eta_{\mu\nu} + h_{\mu\nu}) u^{\mu}u^{\nu} \approx
-~R^{(1)}_{\alpha\mu\beta\nu}(h)\bar{u}^{\mu}\bar{u}^{\nu} =
-~R^{(1)}_{\alpha 0 \beta 0}(h)\ 
\end{equation}
is symmetric in $\alpha \leftrightarrow \beta$ and spatial i.e. $A_{00}
= 0 = A_{0i}$. When the transient wave $h_{\mu\nu}$ is in synchronous
gauge and TT projected, the matrix $A_{ij}(h^{TT}_{kl})$ is also
transverse. This is because, the $\Lambda-$projector can be taken across
the derivatives up to terms down by powers of $r$.  Explicitly,
\begin{equation}
A_{ij}(h^{TT}) \approx
\frac{1}{2}\Lambda_{ij}^{~~~kl}(\hat{n})\partial_0^2h_{kl}\ , ~~~
A_{ij}\delta^{ij} = 0 \ .
\end{equation}
Since the deviation vectors too are taken to be transverse, in effect
the deviation scalar reduces to $D(u, Z', Z) \approx \hat{\bar{Z}}'^a
A_{ab}(h)\hat{\bar{Z}}^b$ where $a, b$ take two values and the real,
symmetric matrix $A_{ab}$ is traceless. With respect to an arbitrarily
chosen basis, $\{\hat{e}_1, \hat{e}_2\}$ in the plane transverse to the
wave direction, $\hat{n}$, we can define the `+' and the `$\times$'
polarizations by setting the matrix $A := h_+\sigma_3 +
h_{\times}\sigma_1$. If $\hat{\bar{Z}}$ makes an angle $\phi$ with
$\hat{e}_1$, then the unit deviation vectors are given by,
$\hat{\bar{Z}} = (cos(\phi), sin(\phi))\ , \hat{\bar{Z}}' = (-
sin(\phi), cos(\phi))$. It follows that,
\begin{equation}
D(u, Z', Z) = - h_{+} sin(2\phi) + h_{\times} cos(2\phi) \ .
\end{equation}
Thus, for a pair of bases $(\hat{e}_1, \hat{e}_2)$ and $(\hat{\bar{Z}},
\hat{\bar{Z}}')$, determination of the deviation scalar gives one
relation between the amplitudes of the two polarizations. A similar
determination at another detector location gives a second relation,
thereby providing amplitudes of individual polarizations.

A natural choice for $\hat{e}_1, \hat{e}_2$ would be the unit vectors
provided by the RA/Dec coordinate system used by astronomers, at the
$\hat{n}$ direction. The basis of unit deviation vectors could be
constructed in {many ways. For instance,} using the wave direction
$\hat{n}$ and one of the arms of the interferometer which form a plane.
Its unit normal may be taken as $\hat{\bar{Z}}$ and then,
$\hat{n}\times\hat{\bar{Z}}$ can be taken as $\hat{\bar{Z}}'$.  To avoid
the exceptional case where the wave is incident along the chosen arm of
the interferometer, one could repeat the procedure with the other arm.
The construction gives $\phi$ at the detector location. Suffice it to
say that measurement of deviation scalar for appropriate deviation
vectors, at two or more detectors would constitute a measurement of the
amplitudes of individual polarizations of a gravitational wave.

To be useful in observations, the deviation scalar must be computed for
congruence related to specific interferometer (earth based ones are not
in free fall, the space based ones would be) and related to the
waveform. These details are beyond the scope of the present work.
}

\section{Summary and discussion}\label{FinalSec} 
Let us begin by recalling the main motivation for this work. The
concordance model of cosmology favours dark energy modelled conveniently
in terms of a positive cosmological constant which is about $10^{-29}
gm/cc$ or about $10^{-52} m^{-2}$ in the geometrized units with $G = 1 =
c$. In the vicinity of any astrophysical sources, this density is
extremely small and only over vast distances of matter free regions, we
may expect its effects to be felt. Over distances of typical, detectable
compact sources of gravitational waves - about mega-parsecs - its effect
may be estimated to be of order $\sqrt{\Lambda} r \sim 10^{-4}$. (This
is comparable to the 4th order PN corrections for a $v/c \sim 0.2 - 0.3$
and is relevant for direct detection of gravitational waves.) On the
other hand, the asymptotic structure of ${\cal J}^+$ - the final
destination for all massless radiation - is qualitatively different for
{\em arbitrarily small} values of $\Lambda$ and has significant impact
on asymptotic symmetry groups and the fluxes associated with them. Does
this affect {\em indirect detection} of gravitational waves, eg in the
orbital decays of binary pulsars? To the extent that the Hulse-Taylor
pulsar observations have already vindicated the quadrupole formula
computed in Minkowski background, one does not expect the radically
different nature of ${\cal J}^+$ to play any significant role in
such indirect detections. A physically relevant question then is: how
are the effects of positive $\Lambda$ to be estimated quantitatively?
Our main motivation has been to address this question. 

In the introduction, we noted the different features and issues that
arises: multiple charts, gauges, identification of physical
perturbations, source multipole moments and energy measures. We
considered two different charts (FNC and conformal) and two different
gauge choices (TT and Generalized-TT), defined the corresponding
synchronous gauges to identify the physical components and these were
expressed in terms of the appropriately defined source moments.

{A strategy to determine of the waveform of a transient
gravitational waves, eg using an interferometer, always selects a
frequency window of sensitivity and corresponding class of sources. For
the class of sources we have assumed (rapidly varying and distant), it
seems sufficient to confine attention to a region maximally up to the
cosmological horizon.  The physical distance (eg luminosity distance)
from the source to the cosmological horizon, eg $\eta = -r$ in the
conformal chart, is $\sqrt{3/\Lambda\eta^2}\ r = \sqrt{3/\Lambda}$.
This contains typical, currently detectable sources and thus should
suffice for estimation. We obtained the corresponding fields, to order
$\Lambda$, using Fermi Normal coordinates based near the compact source
and is given in eqn.(\ref{FNCSoln}). For a subsequent comparison, we
also computed the field in the conformal/cosmological charts. It is
given in equation (\ref{NonCovariantGaugeGWAmplitude1}).}

{By contrast, an indirect detection via observation of orbital
decays of binary systems, is premised on the energy lost due to
gravitational radiation. This is typically the in-spiral phase of the
binary system and has much lower frequencies (about $10^{-5}$ Hz for
Hulse-Taylor).  This is beyond the capabilities of earth based
interferometers and one has to appeal to the energy carried away by
gravitational waves.  The energy flux calculations are ideally done at
infinity}.  For these, done in the conformal/cosmological chart, we
refer the reader to \cite{AshtekarBongaKesavanII,
AshtekarBongaKesavanIII}.  As mentioned in the introduction, there are
two distinct prescriptions and it would be useful to compare them. Flux
computations and comparisons will be dealt with in a separate
publication. 

In the Minkowski background analysis, tail terms appear at higher orders
of perturbations and these are understood to be due to scattering off
the curvature generated at the lower orders. In the de Sitter
background, curvature effects are felt by the perturbations at the
linear order itself. This is manifested in both the gauges. In the
generalized transverse gauge, the tail term is explicitly available and
plays a crucial role at the null infinity
\cite{AshtekarBongaKesavanIII}. In the TT gauge however, the tail term
itself is order $\Lambda^2$ and within the FNC patch does not seem
likely to give significant contribution by cumulative effects. However
this remains to be computed explicitly.

As a by product of expressing the retarded solution in terms of the
source moments, we also saw (not surprisingly) that the `mass' (zeroth
moment) and the `momentum' (first moment) are not conserved, thanks to
the curvature of the de Sitter background. More generally, it also
implied that static (test) sources cannot exist in curved background.
This is just a consequence of the conservation equation in a curved
background, quite independent of any gravitational waves. 

{
In the Minkowski background geodesic deviation acceleration, to the
linearised order, is gauge invariant and is used to infer the wave form.
In a general curved background, its gauge invariance is lost. However,
for a conformally flat background, component of a deviation vector along
another, orthogonal deviation vector defines a gauge invariant function,
$D(u, Z, Z')$, which we termed as deviation scalar.  In the simpler
context of flat background, we saw that its measurement at two or more
detetctors would give the amplitudes of individual polarizations. Its
determination could provide useful information on polarization of
gravitational waves even for non-zero cosmological constant.
}

We computed the deviation scalar, for the solutions given in the two
charts. This is a new result.  The expressions obtained
(\ref{FNCScalarFinal},\ref{ConfScalarFinal}) are different. The
comparison is expected to be possible when both charts overlap and only
up to order $\Lambda \sim H^2$. In FNC, we have computed only the sharp
term. However, it is not clear if the `sharp' contribution can be
identified in a chart independent manner. So in the conformal chart, we
took the full field and restricted its contribution to order $\Lambda$.
While we compute the same observable, a chart dependence, or more
precisely a dependence on the spatial hypersurface, enters through the
definition of source moments. There is also a choice of moment variable
involved ($\zeta^{\U{i}}$ in FNC).  Thus, the solutions are given in
terms of source moments which are defined on {\em different spatial
hypersurfaces}. As such they cannot be compared immediately.  An
explicit model system for which the two different moments are computed,
should help clarify some of these aspects and show the equality of the
deviation scalar computed in two ways. This needs to be checked. 

Lastly we comment on the lessons from these computations. Even a
smallest cosmological constant (positive or negative), immediately
brings up the more than one `natural' choices of charts in a given
patch.  Quite apart from the qualitatively distinct structure of the
respective ${\cal J}^+$, even the local (near source) analysis reveals
different issues to be faced.  The FNC is very natural to the local
analysis and goes through the same way for AdS as well. It naturally
gives the answer as corrections to the corresponding Minkowski answer,
in { $powers~of~ \mathit{\Lambda}$}. This is also seen the Bondi-Sachs
chart \cite{Bishop}.  From the intuition from Minkowski background
analysis, neighbourhood of infinity is the natural place for
characterising {\em radiation} in a gauge invariant manner.  Then the
conformal chart (for de Sitter) is a natural choice. And here the
corrections to the Minkowski answer are obtained in
$powers~of~\sqrt{\mathit{\Lambda}}$.  {This difference in the
powers of $\Lambda$, was seen in the solutions obtained in equations
(\ref{FNCSoln}, \ref{NonCovariantGaugeGWAmplitude1}).  However it is
meaningless to compare the gauge fixed fields. For this purpose the
gauge invariant deviation scalar was computed and compared.} The
manifest dependence of the corrections on $\Lambda$ {does}
distinguish a local (neighbourhood of source) form from the one in the
asymptotic region.

{
To conclude, linearization about the de Sitter background provides a
simplified arena for an extension of the computational steps from a flat
background to a curved background. The weak gravitational waves can be
computed as corrections in powers of the cosmological constant. There is
a gauge invariant observable that could provide information about the
amplitudes of the two polarizations. More precise computations at least
for a model source are needed for a quantitative estimate of corrections
to the waveforms. If the $\Lambda-$corrections could be identified from
the signal, it could provide an independent measurement of the
cosmological constant. 
}

\acknowledgments We would like to thank Abhay Ashtekar for helpful
discussions during his visit to Chennai and useful suggestions. We also
acknowledge discussions with Alok Laddha and K G Arun. {We would
like to thank Yi-Zen Chu for correspondence and for pointing out an
erroneous boundary term in equation (79) which has been corrected.}

\appendix

\section{Triangle law for World function} \label{TriangleLaw}
We sketch the steps that go in the computation of the world function
between the observation event and a source event, $\sigma(P', P)$, given
in the equation (\ref{LeadingOrderDistance}). In reference to the figure
(\ref{RNCfig}), we want to compute: $\phi := \Case{1}{3}\int_0^1 dv (1 -
v)^3 \frac{D^4 \sigma(q', q)}{Dv^4}$ \cite{Synge}.

Let $u$  denote the parameter along the geodesics connecting $q', q$ as
they vary along the geodesics $P_0P'$ and $P_0P$. These geodesics are
all parameterized such that they begin at $q'(u_1, v)$ and end at $q(u_2,
v)$. In general $\sigma(q', q)$ is a function of $u_1, u_2, v$. But
since $u_1, u_2$ are the same for all such pairs, we have $\sigma(q', q)
= \sigma(v)$. Therefore,
\begin{equation}
\frac{D\sigma(v)}{Dv} ~ = ~ \frac{dx'^{\alpha}}{dv}\frac{\partial
\sigma}{\partial x'^{\alpha}} + \frac{dx^{\alpha}}{dv}\frac{\partial
\sigma}{\partial x^{\alpha}} ~ := ~ \sigma_{\alpha'}V^{\alpha'} +
\sigma_{\alpha}V^{\alpha} \ .
\end{equation}
The $V$'s denote the tangent vectors at the respective end points while
the prime on the component labels indicate which end point is implied.
The suffix on the $\sigma$ denote the covariant derivative at the
corresponding point.  Since $\sigma$ is a (bi-)scalar, its covariant
derivative equals the partial derivative. 

The second and higher derivatives of $\sigma$ with respect to $v$ are
computed similarly, noting that $\Case{DV^{\alpha}}{Dv} =
\Case{DV^{\alpha'}}{Dv} = 0$ since $P_0\to P', P_0\to P$ are both
geodesics and $v$ is the affine parameter along them. We also note the
property of the world function \cite{Synge}, $\sigma_{\alpha'\beta} =
\sigma_{\beta\alpha'}$. This leads to,
\begin{eqnarray}
\frac{D^2\sigma}{Dv^2} & = & \sigma_{\alpha'\beta'}V^{\alpha'}V^{\beta'}
+ \sigma_{\alpha\beta}V^{\alpha}V^{\beta} +
2\sigma_{\alpha'\beta}V^{\alpha'}V^{\beta} \\
\frac{D^4\sigma}{Dv^4} & = &
\sigma_{\alpha'\beta'\mu'\nu'}V^{\alpha'}V^{\beta'}V^{\mu'}V^{\nu'} + 4
\sigma_{\alpha'\beta'\mu'\nu}V^{\alpha'}V^{\beta'}V^{\mu'}V^{\nu}
\nonumber \\ & & + 6
\sigma_{\alpha'\beta'\mu\nu}V^{\alpha'}V^{\beta'}V^{\mu}V^{\nu} + 4
\sigma_{\alpha'\beta\mu\nu}V^{\alpha'}V^{\beta}V^{\mu}V^{\nu} \nonumber
\\ & & + \sigma_{\alpha\beta\mu\nu}V^{\alpha}V^{\beta}V^{\mu}V^{\nu} 
\end{eqnarray}
We have not written the third derivative as we do not need it.

The desired world function is, using Taylor expansion with a remainder,
about $P_0$ ($v = 0$),
\begin{eqnarray}\label{TaylorExp}
\sigma(P', P) = \sigma(\bar{v}) & = & \sigma(0) +
\bar{v}\left.\frac{D\sigma}{D v}\right|_0 + \frac{1}{2}
\bar{v}^2\left.\frac{D^2\sigma}{D v^2}\right|_0 + \frac{1}{6}
\bar{v}^3\left.\frac{D^3\sigma}{D v^3}\right|_0 \nonumber \\ & &
\hspace{2.0cm} + \frac{1}{6}\int_0^1 dv (1 - v)^3 \frac{D^4 \sigma(q',
q)}{Dv^4} 
\end{eqnarray}

It is known that $\sigma(0) = 0 = \Case{D\sigma}{Dv}(0) =
\Case{D^3\sigma}{Dv^3}(0)$. The coincidence limits of the second
derivatives of $\sigma$ are given by, $[\sigma_{\alpha'\beta'}] =
[\sigma_{\alpha\beta}] = g_{\alpha\beta}$ and $[\sigma_{\alpha'\beta}] =
[\sigma_{\alpha\beta'}] = - g_{\alpha'\beta} = - g_{\alpha\beta'}$ and
$\bar{v}V^{\alpha'} = - g^{\alpha'\beta'}\sigma_{\beta'}$ and
$\bar{v}V^{\alpha} = + g^{\alpha\beta}\sigma_{\beta}$ \cite{Synge}. This
leads to,
\begin{eqnarray}
\left.\bar{v}^2\frac{D\sigma}{Dv^2}\right|_0 & = &
g_{\alpha\beta}(\bar{v}V^{\alpha})(\bar{v}V^{\beta}) +
g_{\alpha'\beta'}(\bar{v}V^{\alpha'})(\bar{v}V^{\beta'}) -
2g_{\alpha'\beta}(\bar{v}V^{\alpha'})(\bar{v}V^{\beta}) \nonumber \\
& = & g^{\alpha\beta}\sigma_{\alpha}\sigma_{\beta} +
g^{\alpha'\beta'}\sigma_{\alpha'}\sigma_{\beta'} +
2g^{\alpha'\beta}\sigma_{\alpha'}\sigma_{\beta} \nonumber \\
& = & 2\sigma(P_0, P) + 2\sigma(P_0, P') - 2\sigma_{\alpha}(P_0,
P')\sigma^{\alpha}(P_0, P) 
\end{eqnarray}
In the last line, we have used $2\sigma =
g^{\alpha\beta}\sigma_{\alpha}\sigma_{\beta}$. Substituting in eqn.
(\ref{TaylorExp}), we get,
\begin{eqnarray} 
\sigma(P',P) & = & \sigma(P_0, P') + \sigma(P_0, P) - \left. \left(
g^{\alpha\beta}\frac{\partial \sigma(y, P')}{\partial y^{\alpha}}
\frac{\partial \sigma(y, P)}{\partial y^{\beta}}\right)\right|_{P_0}
\nonumber \\
& &  + \frac{1}{6}\int_0^1 dv (1 - v)^3 \frac{D^4 \sigma(q', q)}{Dv^4}
\end{eqnarray}

To compare with the triangle law, we denote, $\overrightarrow{PQ}^2 := 2
\sigma(P, Q)$. Then the above equation can be written as,
\begin{equation}
\overrightarrow{P'P}^2 ~ = ~ \overrightarrow{P_0P'}^2 ~ + ~
\overrightarrow{P_0P}^2 ~ - ~ 2
\overrightarrow{P_0P'}\cdot\overrightarrow{P_0P} + \phi
\end{equation}

To evaluate $\phi$, we need to evaluate the fourth order covariant
derivatives of the world function. These are obtained in terms of the
parallel propagator and integrals of curvature. To state the result, we
introduce the notation:
\begin{eqnarray}
\mbox{Parallel propagator:} \hspace{0.5cm} & & X_{\parallel}^{\alpha}(p)
:= g^{\alpha}_{~\beta'}(p', p) X_{\parallel}^{\beta'}(p') ~~
\mbox{where} ~~ V^{\gamma}\nabla_{\gamma}X^{\alpha}_{\parallel} = 0 \ .
\\
\mbox{Symmetrized Riemann:} \hspace{0.5cm} & & S_{\alpha\beta\mu\nu} ~
:= ~ - \frac{1}{3}\left(R_{\alpha\mu\beta\nu} +
R_{\alpha\nu\beta\mu}\right) \ .
\end{eqnarray}
The {\em parallel propagator}, $g^{\alpha}_{~\beta'}(p', p)$ is a
bi-tensor and its indices are raised/lowered by the metric at the
respective points. 

It is convenient to introduce a tetrad basis, $E_{~a}^{\alpha},
E_{~\alpha}^{a}$, at $p'$ and define it at $p$ by parallel transporting
it along the geodesic from $p'$ to $p$. The parallel propagator is then
given by $g^{\alpha}_{~\beta'}(p', p) = E_{~a}^{\alpha}(p)
E^{a}_{~\beta'}(p')$. Denoting the components with respect to these
parallelly transported tetrad by Latin indices, the second order
covariant derivatives of the world function are given by (equation 97 of
\cite{Synge}),
\begin{eqnarray}
\sigma_{a'b'}(q', q) & = & g_{a'b'}(q') + \frac{3}{2}\frac{1}{u_2 -
u_1}\int_{u_1}^{u_2}du (u_2 - u)^2 S_{abcd}(u) U^cU^d(u) \\
\sigma_{a'b}(q', q) & = & g_{a'b}(q') + \frac{3}{2}\frac{1}{u_2 -
u_1}\int_{u_1}^{u_2}du (u_2 - u)(u - u_1) S_{abcd}(u) U^cU^d(u) \\
\sigma_{ab}(q', q) & = & g_{ab}(q') + \frac{3}{2}\frac{1}{u_2 -
u_1}\int_{u_1}^{u_2}du (u - u_1)^2 S_{abcd}(u) U^cU^d(u) 
\end{eqnarray}
Note that the tetrad components of the parallel propagator are just
$\eta_{ab}$ while the tetrad components of the geodesic tangent vectors,
$U^{a}$ are constant along the geodesics and may be taken out of the
integration. These expression have corrections at the second order in
curvature.

The fourth covariant derivatives have a similar form but now involve
covariant derivatives of the symmetrized Riemann tensor. In our context
of maximally symmetric background, all these covariant derivatives of
the Riemann tensor vanish and the expressions simplify drastically. In
particular, the third covariant derivatives are all absent as they
involve the covariant derivatives of the Riemann tensor and the index
distribution also gets restricted thanks to the symmetries of the
Riemann tensor.  This leads to (equation 117 of \cite{Synge}),
\begin{eqnarray}
\sigma_{a'b'c'd'}(q', q) & = & \frac{3}{(u_2 - u_1)^3}\int_{u_1}^{u_2}du
(u_2 - u)^2 S_{abcd}(u) \ , \\
\sigma_{a'b'c'd}(q', q) & = & - \frac{3}{(u_2 -
u_1)^3}\int_{u_1}^{u_2}du (u_2 - u)^2 S_{abcd}(u) \ , \\
\sigma_{a'b'cd}(q', q) & = & \frac{3}{(u_2 - u_1)^3}\int_{u_1}^{u_2}du
(u_2 - u)^2 S_{abcd}(u) \ . 
\end{eqnarray}
These again have correction at the second order in curvature. Note that
the tetrad components refer to the tetrad derived from an arbitrary
choice at $q'$, by parallel transport along the geodesic $q'\to q$.

In section \ref{RNC}, we choose a tetrad at the base point of the RNC,
$P_0$ and set it up elsewhere by parallel transporting along the
geodesics emanating from $P_0$. This gives the tetrad
$E^{\alpha'}_{~a'}$ at $q'$.  However the tetrad at $q$,
$E^{\alpha}_{~a}$, is not equal to $\tilde{E}^{\alpha}_{~a}$ - the one
obtained from $E^{\alpha'}_{~a'}$ by parallel transport along $q'\to q$
geodesic. They are related through the holonomy group element along the
closed curve $q\to P_0\to q'\to q$: $\tilde{E}^a_{~\alpha} =
H_{\alpha}^{~\beta}E^a_{~\beta}$. Because of the smallness of the
curvature, $H_{\alpha}^{~\beta}$ differs from the identity element by a
term of order $\Lambda$. In short, the error committed in replacing the
tetrad components of curvature relative to the $q'\to q$ parallelly
transported tetrad, by those derived from tetrad at $P_0$, will be of
second order in the curvature, i.e. order $\Lambda^2$.

With this understood, we regard all the tetrad components in the fourth
covariant derivatives to be relative to the tetrad derived from $P_0$.
The equation (136) of \cite{Synge} then gives,
\begin{eqnarray}
\phi = \phi_0 & = & \frac{3}{(u_2 - u_1)^3}\int_0^1 dw (1 -
w)^3\int_{u_1}^{u_2}du \nonumber \\
& & \hspace{2.0cm} \left\{(u_2 - u)^2 + (u - u_1)^2\right\}\times
\left\{S_{a'b'cd}\bar{v}^4V^{a'}V^{b'}V^c V^d\right\}(u, w).
%
\end{eqnarray}

The tetrad components of the symmetrized Riemann tensor simplify further
thanks to the maximal symmetry.
\begin{eqnarray}
S_{abcd}(u,w) & = & - \frac{1}{3}\left(R_{\alpha\mu\beta\nu} +
R_{\alpha\nu\beta\mu}\right)E^{\alpha}_{~a}E^{\beta}_{~b}E^{\mu}_{~c}E^{\nu}_{~d}(u,w)
\ . \\
& = & - \frac{\Lambda}{9}\left(g_{\alpha\beta}g_{\mu\nu} -
g_{\alpha\nu}g_{\mu\beta} + g_{\alpha\beta}g_{\nu\mu} -
g_{\alpha\mu}g_{\nu\beta}\right)E^{\alpha}_{~a}E^{\beta}_{~b}E^{\mu}_{~c}E^{\nu}_{~d}(u,w)
\\
& = & - \frac{\Lambda}{9}\left[2 (E_a\cdot E_b)(E_c\cdot E_d) -
(E_a\cdot E_d)(E_b\cdot E_c) - (E_a\cdot E_c)(E_b\cdot E_d)\right]
\nonumber \\
& = & - \frac{\Lambda}{9}\left[2 \eta_{ab}\eta_{cd} - \eta_{ad}\eta_{bc}
-~\eta_{ac}\eta_{bd}\right] ~~ \because \mbox{(orthonormality of the
tetrad.)}~~~~
\end{eqnarray}
Consequently, the symmetrized Riemann tensor comes out of the integrals.
The vectors $V^a, V^{a'}$ are independent of $u$ because they come from
the expansion of $\sigma(v)$ and are independent of $v$ since they are
geodesic tangents and refer to the parallelly transported tetrad. The
terms enclosed in the second pair of braces, come out of the integration
and we get,
\begin{eqnarray}
\phi & = & \frac{3}{(u_2 - u_1)^3}\left[\int_0^1 dw (1 -
w)^3\int_{u_1}^{u_2}du \left\{(u_2 - u)^2 + (u - u_1)^2\right\}\right]
\times \nonumber \\
& & \hspace{4.0cm}\left\{S_{a'b'cd}\bar{v}^4V^{a'}V^{b'}V^c V^d\right\}
\\
& = &
\left[\frac{1}{2}\right]\left\{S_{a'b'cd}X^{a'}X^{b'}X^cX^d\right\}
\hspace{1.0cm},~ \bar{v}V^* =: X^* ( = \mbox{~corresponding RNC~}) \\ &
= & - \frac{\Lambda}{9}\left(X^2 X'^2 - (X\cdot X')^2\right) \ .
\end{eqnarray}

Notice that the reference to the choice of the tetrad, $E^{\alpha}_{~a}$
has disappeared.

\section{Calculation of the parallel
propagator}\label{ParallelPropagatorCaln}
In the main text we needed the parallel propagator
$g^{\mu}_{~\alpha'}(x, x')$ along the null geodesic from the observation
point $P$ to a source point $P'$. To this end, introduce an arbitrary
tetrad $e^{\mu}_a(P)$ and its inverse co-tetrad $e^a_{\alpha}(P)$ which
is parallel transported along the null geodesic. These will drop out at
the end. The parallel propagator is then given by,
\[
g^{\mu}_{~\alpha}(x, x') ~ = ~ e^{\mu}_{~a}(x) e^a_{~\alpha}(x') \ .
\]

The geodesic satisfies the equation,
\[
\frac{d^2 x^{\mu}}{d\lambda} +
\Gamma^{\mu}_{~\alpha\beta}(x(\lambda))\frac{dx^{\alpha}}{d\lambda}\frac{dx^{\beta}}{d\lambda}
~ = ~ 0 ~~;~~ x^{\mu}(0) = x^{\mu}(P) := \hat{x}^{\mu} ~~,~~
\dot{x}^{\mu}(0) = \hat{t}^{\mu} \ .
\]
The parallel transported co-tetrad satisfies the equation,
\[
\frac{d e^a_{\alpha}}{d\lambda} -
\Gamma^{\gamma}_{~\alpha\beta}\frac{dx^{\beta}}{d\lambda}e^a_{\gamma} ~
= ~ 0 ~~;~~ e^a_{~\alpha}(0) ~ = ~ e^a_{~\alpha}(P) :=
\hat{e}^a_{~\alpha}\ .
\]

These are solved by Taylor expanding in the affine parameter $\lambda$
and determining the coefficients. Denoting the evaluations at $\lambda =
0$ by hatted quantities, we write,
\begin{eqnarray}\label{TaylorExpansion}
e^a_{~\alpha}(\lambda) & = & \hat{e}^a_{~\alpha} + \lambda
\dot{e}^a_{~\alpha}(0) + \frac{\lambda^2}{2}\ddot{e}^a_{~\alpha}(0)
\cdots \\
x^{\mu}(\lambda) & = & \hat{x}^{\mu} + \lambda\hat{t}^{\mu} +
\frac{\lambda^2}{2}\left(-\hat{\Gamma}^{\mu}_{~\alpha\beta}\hat{t}^{\alpha}
\hat{t}^{\beta}\right) + \frac{\lambda^3}{6}\left( -
{\partial_{\gamma}\hat{\Gamma}^{\mu}_{~\alpha\beta}}
\hat{t}^{\alpha}\hat{t}^{\beta}\hat{t}^{\gamma}\right) + \cdots
\end{eqnarray}
In the last equation we have used the geodesic equation. By
differentiating the geodesic equation, the higher order terms in
$x^{\mu}(\lambda)$ are determined. We note that the connection is order
$\Lambda$ and linear in coordinates. So more than the first derivative
of the connection is not needed. In the Taylor expansion of $x^{\mu}$, we
have shown only the terms to order $\Lambda$. Substituting these
expansions in the parallel transport equation, determines the solution
as,
\begin{equation}\label{ParallelTetrad}
e^a_{~\alpha}(\lambda) ~ = ~ \hat{e}^a_{~\mu}\left[
\delta^{\mu}_{\alpha} + (\lambda\hat{t}^{\beta})
\hat{\Gamma}^{\mu}_{~\alpha\beta} +
\frac{1}{2}(\lambda\hat{t}^{\gamma})(\lambda\hat{t}^{\beta})
\partial_{\gamma}\hat{\Gamma}^{\mu}_{~\alpha\beta}\right]
\end{equation}

From the Taylor expansions of $x^{\mu}$ and $e^a_{~\alpha}$, we
eliminate $\lambda\hat{t}$ and obtain the parallel tetrad in terms of
the coordinates. To the linear order in $\Lambda$, this simply replaces
$\lambda\hat{t}^{\beta}$ by $(x' - x)^{\beta}$. The parallel propagator
is then given by,
\begin{equation} \label{ParallelPropagator}
g^{\mu}_{~\alpha'}(P, P') ~ = ~ \hat{\delta}^{\mu}_{~\alpha'} +
\hat{\Gamma}^{\mu}_{~\alpha'\beta'}(x' - x)^{\beta'} +
\frac{1}{2}\partial_{\gamma'}\hat{\Gamma}^{\mu}_{~\alpha'\beta'}(x' -
x)^{\gamma'}(x' - x)^{\beta'}  + o(\Lambda^2)
\end{equation}
We have used primed indices for notational consistency for bi-tensors.
The hatted quantities are the coincidence limits.

Notice that the arbitrary tetrad introduced at the beginning has
disappeared. We have not used any specific property of the Fermi or
Riemann normal coordinates, except for the order $\Lambda$. In the main
text, we have used the Fermi normal coordinates and the connection
together with its derivative, are given in equations
(\ref{ConnectionFNC}, \ref{ConnectionDerivativeFNC}). 

\section{FNC $\leftrightarrow$ Conformal chart transformations}
\label{FNC2Conformal}

We used two different charts in presenting the quadrupole field, the FNC
restricted to the static patch and the conformal coordinates covering
the Poincare patch which overlaps with the static patch. To relate these
two sets of coordinates, $(\tau, \xi^i)$ and $(\eta, x^i)$, consider the
geodesic equation in the conformal coordinates. In conformal
coordinates, 
\begin{eqnarray}
ds^2 & = & \frac{\alpha^2}{\eta^2}\left[ -d\eta^2 + \sum_i
(dx^i)^2\right] ~,~~~ \alpha^2 ~ = ~~~~ \frac{3}{\Lambda} \ ;\\
\Gamma^0_{~00} & = & -\frac{1}{\eta} ~~,~~ \Gamma^0_{~0j} ~ = ~~~~~~ 0
~~~~,~~ \Gamma^0_{~ij} ~ = ~ -\frac{\delta_{ij}}{\eta} ~, \\
\Gamma^i_{~00} & = & ~~~ 0 ~~,~~ \Gamma^i_{~0j} ~ = ~
-\frac{1}{\eta}\delta^i_j ~~,~~ \Gamma^k_{~ij} ~ = ~~~~~ 0 ~~~. 
\end{eqnarray}

The geodesic equation splits as,
\begin{eqnarray}
0 ~ & = & \frac{d^2\eta}{d\lambda^2} -
\frac{1}{\eta}\left(\frac{d\eta}{d\lambda}\right)^2 -
\frac{\delta_{ij}}{\eta}\frac{dx^i}{d\lambda}\frac{dx^j}{d\lambda} , \\
0 ~ & = & \frac{d^2x^i}{d\lambda^2} -
\frac{2}{\eta}\frac{d\eta}{d\lambda}\frac{dx^i}{d\lambda} \ ;
\hspace{1.0cm} \Rightarrow ~~ \frac{d\vec{x}}{d\lambda} ~ = ~ \eta^2
\vec{C} ~~,\nonumber \\
\therefore \vec{x}(\lambda) & = & \vec{C}\int_0^{\lambda}d\lambda'
\eta^2(\lambda') ~ + ~ \vec{x}_0 ~\hspace{1.0cm} \mbox{where $\vec{C}$
is a constant vector, and} \label{SpatialGeodesicEqn} \\
0 ~ & = & \frac{d^2\eta}{d\lambda^2} -
\frac{1}{\eta}\left(\frac{d\eta}{d\lambda}\right)^2 -
\vec{C}^2 \eta^3  \ . \label{TemporalGeodesicEqn} 
\end{eqnarray}
The choice $\vec{x}_0 = \vec{0}$ corresponds to `radial' geodesics. 

To define FNC, we have to choose one time-like geodesic whose proper
time provides the time coordinate, $\tau$. We choose this to be the line
AD in figure \ref{GlobalChartFig}. This corresponds to the choice
$\vec{x}_0 = 0$ and $\vec{C} = 0$. The $\eta$ equation can be
immediately integrated to give the {\em reference geodesic} as:
\begin{equation} \label{RefGeodesic}
\eta_*(\tau) ~ = ~ -\sqrt{\frac{3}{\Lambda}} e^{- \tau \sqrt{\Lambda/3}}
~~,~~ \vec{x}_*(\tau) ~ = ~ \vec{0} \ .
\end{equation}
For future convenience, we have chosen an integration constant to be $-
\sqrt{3/\Lambda}$ while the integration constant in the exponent is
determined by the proper time condition (norm = -1) which makes $\tau$
to be one of the FNC.

To determine $\xi^i$ coordinates, we consider spatial geodesics,
emanating orthogonally from the reference geodesic. Clearly, we consider
a radial geodesic, $\vec{x}_0 = \vec{0}$ and defining $\vec{x}(\sigma)
:= \hat{C} r(\sigma)$ where $\hat{C} := \vec{C}/|\vec{C}|$. The geodesic
is determined by solving the equation for $\eta(\sigma)$ with initial
conditions reflecting the orthogonality, $-d_{\tau}\eta_*\
d_{\sigma}\eta + d_{\tau}r_*\ d_{\sigma}r = 0$, 
\begin{equation}
d^2_{\sigma}\eta - \frac{(d_{\sigma}\eta)^2}{\eta} - \vec{C}^2\eta^3 ~ =
0 ~~,~~ \eta(0) ~=~ \eta_*(\tau) ~~,~~ d_{\sigma}\eta(0) = 0 ~~,~~ r(0)
~ = ~ 0 ~~,~~ d_{\sigma}r(0) ~ = ~ \gamma \ . 
\end{equation}

Let $P$ be the point with conformal coordinates $(\eta_P, r_P)$ and FNC
$(\tau, s)$. Taking the norm of the initial tangent vector to be $s^2$, the pairs of coordinates are related as:
\[
\eta_P ~:=~ \eta(\sigma = 1) ~~,~~ r_P ~:=~ r(\sigma = 1) ~~,~~ s^2 =
\frac{3}{\Lambda\eta^2(0)}\gamma^2 .
\]

Using the first integral of the r-equation, we get 
\begin{equation}
d_{\sigma}r(0) = |\vec{C}|\eta^2(0) = \gamma =
\sqrt{\frac{\Lambda}{3}}|\eta(0)| s ~ = ~ s
e^{-\tau\sqrt{\Lambda/3}}\ ~~\Rightarrow ~~ |\vec{C}| ~=~ s
\frac{\Lambda}{3}e^{\tau\sqrt{\Lambda/3}} \ .
\end{equation}
To obtain $(\eta_P, r_P)$, we need to solve the $\eta-$equation.

For this, we first take out a scale $\zeta$ by defining $\eta(\sigma) :=
\zeta y(\sigma)$ which gives $y'' - y'^2/y - |\vec{C}|^2\zeta^2 y^3 = 0$ and
choosing $\zeta = \eta_*(\tau)$, we get $|\vec{C}|^2\zeta^2 = \Lambda
s^2/3 =: \epsilon$. The desired coordinates are then given by,
\begin{eqnarray}
r_P & := & r(\sigma = 1) ~ = s e^{-\tau \sqrt{\Lambda/3}} \int_0^1
d\sigma' y^2(\sigma') \\
\eta_P & := & \eta(\sigma = 1) ~ = ~ - \sqrt{\frac{3}{\Lambda}} e^{-
\tau \sqrt{\Lambda/3}} y(\sigma = 1) ~~~ \mbox{with,} \\
0 & = & y'' - \frac{y'^2}{y} - \epsilon y^3 ~~~,~~~ y(0) = 1~~,~~ y'(0) ~
= ~ 0 ~~~,~~~ \epsilon ~:=~ \frac{\Lambda}{3}s^2
\end{eqnarray}

To order $\epsilon$, the solution for $y(\sigma) := y_0(\sigma) +
\epsilon y_1(\sigma)$ is obtained as, $y(\sigma) = 1 + \epsilon
\sigma^2/2$ which leads to the coordinate transformation,
\begin{equation}\label{ConformalToFNC}
r(\tau, s) ~ = ~ s e^{-\tau\sqrt{\Lambda/3}}\left( 1 + \frac{\Lambda
s^2}{9}\right) ~~~,~~~ \eta(\tau, s) ~ = ~ - \sqrt{\frac{3}{\Lambda}}
e^{-\tau\sqrt{\Lambda/3}}\left( 1 + \frac{\Lambda s^2}{6}\right)
\end{equation}

For inverting the transformation, it is more convenient to use the
combinations: $a(\eta) := - \sqrt{3/\Lambda} \eta^{-1}~,~ A(\tau) :=
e^{\tau\sqrt{\Lambda/3}}$ so that,
\begin{eqnarray}
r(A, s) = \frac{s}{A}\left(1 + \frac{\Lambda}{9}s^2\right) & , & a(A, s)
~ = ~ A\left(1 - \frac{\Lambda}{6}s^2\right) \\
s(a, r) ~ = ~ (ra)\left(1 + \frac{\Lambda}{18}(ra)^2\right) & , & A(a,
r) ~ = ~ a\left(1 + \frac{\Lambda}{6}(ra)^2\right) \label{CosmoToFNC}
\end{eqnarray}
Note that $ra$ is {\em a physical distance} such as the commonly used
luminosity distance in cosmology while $s$ is also a physical distance
but along a spatial geodesic. The equation (\ref{CosmoToFNC}) gives the
relation between them.

From these relations, it is easy to verify that the stationary Killing
vector field,
\begin{equation}
-\sqrt{3/\Lambda} T ~:=~ \eta\partial_{\eta} + x^i\partial_i ~ = ~
\eta\partial_{\eta} + r\partial_r ~ = ~ \partial_{\tau}
\end{equation}

For completeness, we list the transformations between the conformal
chart and the FNC chart in the static patch, up to order $H^2$..
\begin{eqnarray}
\eta(\tau, \xi^i) & := & - \frac{e^{-H\tau}}{H}\left(1 + \frac{H^2s^2}{2}\right)
~ , ~
x^i(\tau, \xi^i) ~ := ~ \xi^i e^{-H\tau}\left(1 +
\frac{H^2s^2}{3}\right) \\
e^{-H\tau}(\eta, x^i) & := & - \eta H \left(1 - \frac{r^2}{2\eta^2}\right)
\hspace{0.85cm}  , ~
\xi^i(\eta, x^i) ~ := ~ - \frac{x^i}{\eta H} \left(1 +
\frac{r^2}{6\eta^2}\right) 
\end{eqnarray}
With these, it can be checked that the two metrics go into each other.
%


\end{document}